\documentclass[12pt]{article}

\usepackage[dvipsname]{xcolor}

\usepackage{graphicx}
\usepackage{amsmath}        
\usepackage{amssymb}        
\usepackage{slashed}

\def\mydate{June 21, 2019 (corrected typos.)}
\def\ignore#1{{}}

\def\go{\rightarrow}
\def\dd{\partial}
\def\ep{{\epsilon}}

\def\eff{{\rm eff}}
\def\SM{{\rm SM}}
\def\KK{{\rm KK}}
\def\EM{{\rm EM}}
\def\onehalf{\hbox{$\frac{1}{2}$}}
\def\la{\langle}
\def\ra{\rangle}

\def\mybig{\displaystyle \strut }
\def\mbig{\displaystyle }
\def\myfrac#1#2{\frac{\mybig #1}{\mybig #2}}
\def\mfrac#1#2{\frac{\mbig #1}{\mbig #2}}

\def\mynoalign{\noalign{\kern 4pt}}

\makeatletter
\@addtoreset{equation}{section}
\makeatother

\begin{document}

\thispagestyle{empty}

{\small \noindent \mydate    \hfill OU-HET 989}

{\small \noindent Published in PRD}


\vskip 3.cm

\baselineskip=30pt plus 1pt minus 1pt

\begin{center}
{\Large \bf  GUT inspired}

{\Large \bf   $SO(5)\times U(1) \times SU(3)$ gauge-Higgs unification}

\end{center}


\baselineskip=22pt plus 1pt minus 1pt

\vskip 1.5cm

\begin{center}
{\bf Shuichiro Funatsu$^1$, Hisaki Hatanaka$^2$,  Yutaka Hosotani$^3$,}

{\bf Yuta Orikasa$^4$ and Naoki Yamatsu$^5$}

\baselineskip=18pt plus 1pt minus 1pt

\vskip 10pt
{\small \it $^1$Institute of Particle Physics and Key Laboratory of Quark and Lepton 
Physics (MOE), Central China Normal University, Wuhan, Hubei 430079, China} \\
{\small \it $^2$Osaka, Osaka 536-0014, Japan} \\
{\small \it $^3$Department of Physics, Osaka University, 
Toyonaka, Osaka 560-0043, Japan} \\
{\small \it $^4$Czech Technical University, Prague 12800, Czech Republic} \\
\small \it $^5$Department of Physics, Kyoto University, Kyoto 606-8502, Japan \\

\end{center}

\vskip 1.5cm
\baselineskip=18pt plus 1pt minus 1pt

\begin{abstract}
$SO(5)\times U(1) \times SU(3)$ gauge-Higgs unification model inspired by
$SO(11)$ gauge-Higgs grand unification is constructed in the Randall-Sundrum
warped space. 
The 4D Higgs boson is identified with the Aharonov-Bohm phase in the fifth dimension.
Fermion multiplets are introduced in the bulk in the spinor, vector and singlet 
representations of $SO(5)$ such that they are implemented in the spinor and
vector representations of $SO(11)$. 
The mass spectrum of quarks and leptons in three generations is reproduced
except for the down quark mass.
The small neutrino masses are explained by the gauge-Higgs seesaw mechanism 
which takes the same form as in the inverse seesaw mechanism in grand unified theories
in four dimensions.  
\end{abstract}


\newpage

\baselineskip=20pt plus 1pt minus 1pt
\parskip=0pt

\section{Introduction} 

The existence of the Higgs boson of a mass  125$\,$GeV has been
firmly established at LHC.\cite{HiggsLHC2012}
It supports the unification scenario of electromagnetic and weak forces.  
So far almost all of the experimental results and observations have been
consistent with  the standard model (SM) based on
the gauge group ${\cal G}_\SM = SU(3)_C \times SU(2)_L \times U(1)_Y$.
Yet it is not clear whether or not the observed Higgs boson is precisely what the SM
assumes.  All of the  Higgs couplings to other fields and to itself  need to be 
determined with better accuracy.
Furthermore, the SM is afflicted with the gauge hierarchy problem
which becomes apparent when the model is generalized to incorporate
grand unification.  
The fundamental problem is the lack of a principle which regulates
the Higgs sector, in quite contrast to the gauge sector which is
controlled by the gauge principle.

There are several attempts to overcome these difficulties.
Supersymmetric theory is one of them which has been extensively investigated.
An alternative approach is gauge-Higgs unification in which the Higgs boson is 
identified with the zero mode of the fifth dimensional component of the
gauge potential.  It appears as a fluctuation mode of the Aharonov-Bohm  (AB) phase 
$\theta_H$  in  the fifth dimension.\cite{Hosotani1983}-\cite{Kubo2002}   
Already a realistic gauge-Higgs unification (GHU)
model has been constructed.
It is the $SO(5) \times U(1)_X$ gauge theory in the Randall-Sundrum (RS) warped space
with quark and lepton multiplets in the vector representation of $SO(5)$.\cite{ACP2005}-\cite{Yoon2018}
It has been shown that the $SO(5) \times U(1)_X$ GHU yields nearly the same
phenomenology at low energies  as the SM.  
Deviations of the gauge couplings of quarks and leptons from the SM values are
less than $10^{-3}$ for $\theta_H \sim 0.1$. 
Higgs couplings of quarks, leptons, $W$ and $Z$ are approximately 
the SM values times $\cos \theta_H$, the deviation being about 1{\%}.
The Kaluza-Klein (KK) mass scale is about $m_\KK \sim 8\,$TeV  for $\theta_H \sim 0.1$. 
Implications of GHU to dark matter and Majorana neutrino masses are also under intensive
study.\cite{FHHOS-DM2014}-\cite{Lim2018}

The model predicts $Z'$ bosons, which are the first KK modes of $\gamma$, $Z$, and
$Z_R$ ($SU(2)_R$ gauge boson), in the $7 \sim 9\,$TeV range for $\theta_H=0.1 \sim 0.07$.  
They have broad widths and can be produced at 14$\,$TeV LHC.\cite{Funatsu:2014fda, FHHO2017} 
The current non-observation of $Z'$ signals puts the limit $\theta_H < 0.11$.
Right-handed quarks and charged leptons have rather large couplings to $Z'$.
It has been pointed out recently that the interference effects of $Z'$ bosons can be
clearly observed at 250$\,$GeV $e^+ e^-$ linear collider (ILC).\cite{FHHO2017ILC, Yoon2018}
For instance, in the process $e^+ e^- \go \mu^+ \mu^-$ the deviation from the SM
amounts to $-4${\%} with the electron beam polarized in the right-handed mode by 80{\%}
$(P_{e^-} = 0.8)$ for $\theta_H \sim 0.09$, whereas there appears negligible deviation 
with the electron beam polarized in the left-handed mode by 80{\%} $(P_{e^-} = -0.8)$.
In the forward-backward asymmetry $A_{FB} (\mu^+ \mu^-)$
the deviation from the SM becomes $-2${\%} for $P_{e^-} = 0.8$.
These deviations can be seen at 250$\,$GeV ILC with 250$\,$fb$^{-1}$ data,
namely in the early stage of the ILC project.\cite{Bilokin2017, Richard2018, ILC2019}

At this point one may pause to ask a question.  
Is there an alternative way of introducing quark-lepton multiplets in the 
$SO(5) \times U(1)_X \times SU(3)_C$ GHU?
A different choice may lead to different predictions for the $Z'$ couplings.

In this paper we present an alternative way of introducing fermions in the 
$SO(5) \times U(1)_X \times SU(3)_C$ GHU based on the compatibility 
with grand unification of forces.  Many gauge-Higgs grand unification models have
been proposed.\cite{Burdman2003}-\cite{Frigerio2011}
Among them the $SO(11)$ GHU  generalizes  the gauge structure of the  previous 
$SO(5) \times U(1)_X \times SU(3)_C$ model, yielding the 4D Higgs boson as an AB 
phase.\cite{HosotaniYamatsu2015}-\cite{HosotaniYamatsu2018}
Fermions are introduced in the spinor and vector representations of $SO(11)$.
The current $SO(11)$ GHU models in either 5D or 6D warped space are
not completely satisfactory, however.   The  models yield exotic light fermions in addition to
quarks and leptons at low energies.

In the framework of grand unification,  the representation in $SO(5)$ and $U(1)_X$ charge
are not independent.   Only certain combinations are allowed.
For instance, fields with quantum numbers of up-type quarks are contained
in an $SO(11)$ spinor, but not in an $SO(11)$ vector.
This fact immediately implies that the fermion content in the previous
$SO(5) \times U(1)_X \times SU(3)_C$ model, in which all quark multiplets are
introduced in the vector representation of $SO(5)$, need to be modified
to be consistent with the $SO(11)$ unification.
The purpose of the present paper is to formulate an  $SO(5) \times U(1)_X \times SU(3)_C$ 
GHU which is compatible with  the $SO(11)$ GHU scheme.
Models must yield phenomenology of the SM at low energies.
In particular, the  mass spectrum and gauge-couplings of quarks and leptons need to 
be reproduced within experimental errors.

In Section 2 we review the general structure of the group $SO(11)$ which is necessary 
to construct a model compatible with gauge-Higgs grand unification.
A new model of $SO(5) \times U(1)_X \times SU(3)_C$ GHU is introduced in Section 3.
In Section 4 the mass spectrum of gauge fields is determined.
In Section 5  the mass spectra of various fermion fields are determined.
Brane interactions become important for down-type quarks and neutral leptons.
$W$ couplings of quarks and leptons are also evaluated.
Section 6 is devoted to summary and discussions.
Appendix A summarizes generators of $SO(5)$.
Basis mode functions in the RS space are summarized in Appendix B.
In subsection B.3 modes functions for massive fermion fields are given.
In Appendix C notation for Majorana fermions is summarized.
In Appendix D the mass spectra and wave functions of additional dark fermion fields 
are derived.


\section{Structure of $SO(11)$}
\label{Sec:SO(11)}

We would like to formulate $SO(5) \times U(1)_X \times SU(3)_C$ GHU
inspired from $SO(11)$ GHU.  For that purpose it is
useful to review  branching rules of $SO(11)$ to its subgroups. 
We check  them for $SO(11)$ singlet, vector, spinor, and adjoint
representations ${\bf 1}$, ${\bf 11}$, ${\bf 32}$, ${\bf 55}$.
All the necessary information is found in Ref.~\cite{Yamatsu2015}.
First we note 
\begin{align}
SO(11) &\supset SO(6)_C\times SO(5)_W \simeq SU(4)_C\times USp(4)_W \cr
&\supset SU(3)_C\times U(1)_X \ \times\ SU(2)_L\times SU(2)_R \cr
&\supset SU(3)_C\times SU(2)_L\times U(1)_Y\times U(1)_Z ~.
\label{SO11branching}
\end{align}
Here  $U(1)_X$ represents   $U(1)$ in 
 $SO(6)_C\simeq SU(4)_C\supset SU(3)_C\times U(1)_X$,  whereas
 $U(1)_Z$ represents   $U(1)$ in  $SO(10)\supset SU(5)\times U(1)_Z$.

The branching rules of
$SO(11)\supset SO(6)_C\times SO(5)_W(\simeq SU(4)_C\times USp(4)_W)$
are given by 
\begin{align}
 {\bf 1}&=
 ({\bf 1,1}),\nonumber\\
{\bf 11}&=
({\bf 6,1})
\oplus({\bf 1,5}),\nonumber\\
{\bf 32}&=
({\bf 4,4})
\oplus({\bf \overline{4},4}),\nonumber\\
{\bf 55}&=
({\bf 15,1})
\oplus({\bf 6,5})
\oplus({\bf 1,10}).
\label{Eq:Branching-rules-SO11_SU4-USp4} 
\end{align}
The branching rules of
$SO(6)_C\simeq SU(4)_C\supset SU(3)_C\times U(1)_X$
are given by 
\begin{align}
 {\bf 1}&=
 ({\bf 1})_0 ~,\nonumber\\
{\bf 4}&=
({\bf 3})_{\frac{1}{6}}
\oplus({\bf 1})_{-\frac{1}{2}} ~,\nonumber\\
{\bf \overline{4}}&=
({\bf \overline{3}})_{-\frac{1}{6}}
\oplus({\bf 1})_{\frac{1}{2}} ~,\nonumber\\
{\bf 6}&=
({\bf 3})_{-\frac{1}{3}}
\oplus({\bf \overline{3}})_{\frac{1}{3}} ~,\nonumber\\
{\bf 15}&=
({\bf 8})_0
\oplus({\bf 3})_{\frac{2}{3}}
\oplus({\bf \overline{3}})_{-\frac{2}{3}}
\oplus({\bf 1})_0 ~.
\label{Eq:Branching-rules-SU4_SU3-U1} 
\end{align}
Here the subscript represents the $U(1)_X$ charge $Q_X$.
For later use $Q_X$ has been normalized such that the electric charge $Q_\EM$
is given by $Q_\EM = T^L_3 + T^R_3 + Q_X$ where  $T^L_a$ and  $T^R_a$
($a=1,2,3$)  are generators of $SU(2)_L$ and $SU(2)_R$.
From the  branching rules  (\ref{Eq:Branching-rules-SO11_SU4-USp4}) and
(\ref{Eq:Branching-rules-SU4_SU3-U1}),
one obtains the branching rules of
$SO(11)\supset SU(3)_C \times SO(5)_W \times U(1)_X$ as
\begin{align}
 {\bf 1}&=
 ({\bf 1,1})_0 ~,\nonumber\\
{\bf 11}&=
({\bf 3,1})_{-\frac{1}{3}}
\oplus({\bf \overline{3},1})_{\frac{1}{3}}
\oplus({\bf 1,5})_0 ~,\nonumber\\
{\bf 32}&=
({\bf 3,4})_{\frac{1}{6}}
\oplus({\bf 1,4})_{-\frac{1}{2}}
\oplus({\bf \overline{3}, 4})_{-\frac{1}{6}}
\oplus({\bf 1, 4})_{\frac{1}{2}} ~,\nonumber\\
{\bf 55}&=
({\bf 8,1})_0 
\oplus({\bf 3,1})_{\frac{2}{3}}
\oplus({\bf \overline{3},1})_{-\frac{2}{3}}
\oplus({\bf 1,1})_0
\oplus({\bf 3,5})_{-\frac{1}{3}}
\oplus({\bf \overline{3},5})_{\frac{1}{3}}
\oplus({\bf 1,10})_0 ~.
\label{Eq:Branching-rules-SO11_SU3-USp4-U1} 
\end{align}
The branching rules of
$SO(5)_W \simeq USp(4)\supset SU(2)_L\times SU(2)_R$
are given by 
\begin{align}
 {\bf 1}&=
 ({\bf 1,1}),\nonumber\\
{\bf 4}&=
({\bf 2,1})
\oplus({\bf 1,2}),\nonumber\\
{\bf 5}&=
({\bf 2,2})
\oplus({\bf 1,1}),\nonumber\\
{\bf 10}&=
({\bf 3,1})
\oplus({\bf 2,2})
\oplus({\bf 1,3}).
\label{Eq:Branching-rules-USp4_SU2-SU2} 
\end{align}
(For more information, see Table~471 in Ref.~\cite{Yamatsu2015}.)

It has been shown \cite{HosotaniYamatsu2017, HosotaniYamatsu2018} that 
in 6D $SO(11)$ gauge-Higgs grand unification in the hybrid warped space
4D SM chiral fermions and other vectorlike fermions can be extracted from
6D Weyl fermions without 6D and 4D gauge anomalies.
With appropriate boundary conditions imposed, only 
$({\bf 3,4})_{\frac{1}{6}} \oplus ({\bf 1,4})_{-\frac{1}{2}}$ of
$SU(3)_C\times SO(5)_W\times U(1)_X$ have 
zero modes of  6D $SO(11)$ ${\bf 32}$ Weyl fermions.
Also, only either ${\bf (1,5)}_0$ or 
$({\bf 3,1})_{-\frac{1}{3}} \oplus ({\bf \overline{3},1})_{\frac{1}{3}}$ 
have zero modes of  6D $SO(11)$ ${\bf 11}$ Weyl fermions.

The gauge symmetry breaking takes place in three steps;
\begin{align}
SU(3)_C\times &SO(5)_W\times U(1)_X \cr
\noalign{\kern 5pt}
\underset{\rm BCs}{\longrightarrow}
\hskip .6cm
&SU(3)_C\times SU(2)_L\times SU(2)_R\times U(1)_X \cr
\noalign{\kern 5pt}
\underset{\la \Phi_{({\bf 1,4})_{1/2}} \ra \not=0}{\longrightarrow}
&SU(3)_C\times SU(2)_L\times U(1)_Y=G_{\rm SM} \cr
\noalign{\kern 3pt}
\underset{\theta_H \not=0}{\longrightarrow}
\hskip .5cm
&SU(3)_C \times  U(1)_\EM ~.
\label{SymBreaking1}
\end{align}
In the first step $SO(5)_W$ is broken to $SO(4) \simeq SU(2)_L\times SU(2)_R$
by orbifold boundary conditions.  In the second step $SU(2)_R \times U(1)_X$
is spontaneously broken to $U(1)_Y$ by nonvanishing vacuum expectation value (VEV)
of a brane scalar field
$\Phi_{({\bf 1,4})_{1/2}}$.  In the third step $SU(2)_L\times U(1)_Y$ is
broken to $U(1)_\EM$ by the Hosotani mechanism $\theta_H \not= 0$.
At the moment we need to introduce an elementary brane scalar field
$\Phi_{({\bf 1,4})_{1/2}}$ on the UV brane, which is not completely in harmony with
the philosophy of gauge-Higgs unification.
The $\Phi_{({\bf 1,4})_{1/2}}$ field not only reduces the gauge symmetry to 
$G_{\rm SM} $ in the second step in (\ref{SymBreaking1}), but also plays
a crucial role in realizing the mass spectrum of quarks and leptons through brane interactions.
The origin of the brane scalar field remains to be clarified.

\section{$SU(3)_C\times SO(5)_W\times U(1)_X$ GHU --- new model}
\label{Sec:5D-GHEWU}

A new model of $SU(3)_C\times SO(5)_W\times U(1)_X$ GHU is defined 
in the Randall-Sundrum warped space.  The construction is guided by 
the $SO(11)$ gauge-Higgs grand unified model\cite{HosotaniYamatsu2015}-\cite{HosotaniYamatsu2018}
The metric $g_{MN}$ of the Randall-Sundrum (RS)  warped space \cite{RS1999} is given by
\begin{align}
ds^2= g_{MN} dx^M dx^N =
e^{-2\sigma(y)} \eta_{\mu\nu}dx^\mu dx^\nu+dy^2,
\label{Eq:5D-metric}
\end{align}
where $M,N=0,1,2,3,5$, $\mu,\nu=0,1,2,3$, $y=x^5$,
$\eta_{\mu\nu}=\mbox{diag}(-1,+1,+1,+1)$,
$\sigma(y)=\sigma(y+ 2L)=\sigma(-y)$,
and $\sigma(y)=ky$ for $0 \le y \le L$.
The topological structure of the  RS space is $S_1/ \mathbb{Z}_2$.
In terms of the conformal coordinate $z=e^{ky}$
($1\leq z\leq z_L=e^{kL}$) in the region $0 \leq y \leq L$ 
\begin{align}
ds^2=  \frac{1}{z^2} \bigg(\eta_{\mu\nu}dx^{\mu} dx^{\nu} + \frac{dz^2}{k^2}\bigg) .
\label{Eq:5D-metric-2}
\end{align}
The bulk region $0<y<L$ ($1<z<z_L$) is anti-de Sitter (AdS) spacetime 
with a cosmological constant $\Lambda=-6k^2$, which is sandwiched by the
UV brane at $y=0$ ($z=1$) and the IR brane at $y=L$ ($z=z_L$).  
The KK mass scale is $m_{\rm KK}=\pi k/(z_L-1) \simeq \pi kz_L^{-1}$
for $z_L\gg 1$.

Parity transformations  around the two fixed points  $(y_0, y_1) = (0, L)$ are
defined as $(x^\mu,y_j+y) \to (x^\mu,y_j-y)$. 
We choose orbifold boundary conditions (BCs) such that 
they break $SO(5)_W$ to
$SO(4)\simeq SU(2)_L\times SU(2)_R$ as described below.

\subsection{Gauge fields and orbifold boundary conditions}

The structure of the gauge field part is the same as in the previous
$SU(3)_C\times SO(5)_W\times U(1)_X$ GHU model.
We have  $SU(3)_C\times SO(5)_W\times U(1)_X$
$({\bf 8,1})_0$, $({\bf 1,10})_0$, and $({\bf 1,1})_0$
gauge bosons  denoted by $A_M^{SU(3)_C}$, $A_M^{SO(5)_W}$, and $A_M^{U(1)_X}$.
The orbifold BCs are given by
\begin{align}
&\begin{pmatrix} A_\mu \cr  A_{y} \end{pmatrix} (x,y_j-y) =
P_{j} \begin{pmatrix} A_\mu \cr  - A_{y} \end{pmatrix} (x,y_j+y)P_{j}^{-1}
\label{Eq:BC-gauge}
\end{align}
for each gauge field.  In terms of  
\begin{align}
P_{\bf 3}^{SU(3)}&=I_3,\nonumber\\
P_{\bf 4}^{SO(5)}&=\mbox{diag}\left(I_{2},-I_{2}\right),\nonumber\\
P_{\bf 5}^{SO(5)}&=\mbox{diag}\left(I_{4},-I_{1}\right) , 
\label{Eq:SO5-BCs}
\end{align}
$P_0=P_1= P_{\bf 3}^{SU(3)}$ for  $A_M^{SU(3)_C}$  and 
$P_0=P_1= 1$ for $A_M^{U(1)_X}$.  
$P_0=P_1 = P_{\bf 5}^{SO(5)}$ for $A_M^{SO(5)_W}$ in the vector
representation and $P_{\bf 4}^{SO(5)}$ in the spinor representation, respectively.
$P_{\bf 4}^{SO(5)}$ and $P_{\bf 5}^{SO(5)}$ break $SO(5)_W$ to $SO(4)$.
The parity assignments of 
$A_\mu$ and $A_y$ are summarized in Table~\ref{Tab:BC-gauge}.
Note that the 4D Higgs field is contained in the $({\bf 1,2,2})_0$ part 
of $A_y$.
\begin{table}[tbh]
{
\renewcommand{\arraystretch}{1.1}
\begin{center}
\caption{ Parity assignment $P_0=P_1$ of $A_\mu$ and $A_y$ in 
 $SU(3)_C\times SU(2)_L\times SU(2)_R\times U(1)_X$.
$G_{3221}:=SU(3)_C\times SU(2)_L\times SU(2)_R\times U(1)_X$.
 }
\vskip 5pt
\begin{tabular}{|c|cc|}\hline
$G_{3221}$ &$A_\mu$&$A_y$  \\\hline
$({\bf 8,1,1})_0$ &$(+,+)$&$(-,-)$\\
$({\bf 1,3,1})_0$ &$(+,+)$&$(-,-)$\\
$({\bf 1,1,3})_0$ &$(+,+)$&$(-,-)$\\
$({\bf 1,2,2})_0$ &$(-,-)$&$(+,+)$\\
$({\bf 1,1,1})_0$ &$(+,+)$&$(-,-)$\\\hline
\end{tabular}
\label{Tab:BC-gauge}
\end{center}
}
\end{table}

\subsection{Matter fields and orbifold boundary conditions}

Matter fields are introduced both in 5D bulk and on the UV brane.
They are listed in Table~\ref{Tab:matter}.
Quark multiplets $({\bf 3}, {\bf 4})_{\frac{1}{6}}$ and $({\bf 3}, {\bf 1})_{-\frac{1}{3}}^\pm$
are introduced in the 5D bulk in three generations.  They are denoted as
$\Psi_{({\bf 3,4})}^{\alpha}(x,y)$ and $\Psi_{({\bf 3,1})}^{\pm \alpha}(x,y)$
$(\alpha=1,2,3)$.  
All $\Psi_{({\bf 3,4})}^{\alpha}$ and $\Psi_{({\bf 3,1})}^{\pm \alpha}$ intertwine 
with each other.  
Lepton multiplets in the bulk are introduced in $({\bf 1}, {\bf 4})_{-\frac{1}{2}}$, being denoted 
as $\Psi_{({\bf 1,4})}^{\alpha}(x,y)$.  In addition brane fermions $\chi_{({\bf 1}, {\bf 1})}^\alpha (x)$
in the singlet $({\bf 1}, {\bf 1})_{0}$ are introduced on the UV brane, which satisfy
the Majorana condition $\chi (x)^c = \chi (x)$.
$\chi_{({\bf 1}, {\bf 1})}^\alpha$ and $\Psi_{({\bf 1,4})}^{\alpha}$ intertwine with each other 
to induce the seesaw mechanism for neutrino masses.
Two types of dark fermion multiplets, 
$\Psi_{({\bf 3,4})}^{\alpha=4}(x,y)$  in $({\bf 3}, {\bf 4})_{\frac{1}{6}}$ and
 $\Psi^{\pm \beta}_{({\bf 1}, {\bf 5})} (x,y)$ ($\beta=1, \cdots, n_F$) in $({\bf 1}, {\bf 5})_0^\pm$, 
 are introduced in the bulk,  which is necessary to have desired electroweak (EW) symmetry breaking 
with $0 < \theta_H < \onehalf \pi$. 
$\Psi_{({\bf 3,4})}^{\alpha=4}$ obeys orbifold boundary conditions such that no zero modes arise.
Zero modes of $\Psi^{\pm \beta}_{({\bf 1}, {\bf 5})}$ appear, but $\Psi^{+ \beta}_{({\bf 1}, {\bf 5})}$
and $\Psi^{- \beta}_{({\bf 1}, {\bf 5})}$ intertwine to have large Dirac masses.
The brane scalar field $\Phi_{({\bf 1}, {\bf 4})} (x)$ is introduced in $({\bf 1}, {\bf 4})_{\frac{1}{2}}$ 
on the UV brane.  
All of these fields can be implemented in the representations
{\bf 1}, {\bf 11}, and {\bf 32} of $SO(11)$ as seen from (\ref{Eq:Branching-rules-SO11_SU3-USp4-U1}).
$SU(3)_C\times SO(5) \times U(1)_X$  gauge symmetry is preserved on the UV brane, 
which should be contrasted to the previous model in which only 
$SU(3)_C\times SO(4) \times U(1)_X$  symmetry is preserved on the UV brane.
$(\bar{\bf 3}, {\bf 1})_{+\frac{1}{3}}^\pm$ fermion fields accompany with
$({\bf 3}, {\bf 1})_{-\frac{1}{3}}^\pm$ fermion fields when they are implemented in the 
{\bf 11} representation in $SO(11)$ GHU.  
Zero modes of $(\bar{\bf 3}, {\bf 1})_{+\frac{1}{3}}^+$ and 
$(\bar{\bf 3}, {\bf 1})_{+\frac{1}{3}}^-$ couple to have large Dirac masses
so that they may be ignored here.
One can confirm that anomalies are cancelled in the present model.

\begin{table}[tbh]
{
\renewcommand{\arraystretch}{1.2}
\begin{center}
\caption{Matter fields.   $SU(3)_C\times SO(5) \times U(1)_X$ content
is shown. For comparison the matter content in the  previous model is
listed in the last column.  In the  previous model only $SU(3)_C\times SO(4) \times U(1)_X$
symmetry is preserved on the UV brane so that the $SU(2)_L \times SU(2)_R$ content
is shown for brane fields.}
\vskip 10pt
\begin{tabular}{|c|c|c|}
\hline
&$\begin{matrix} \hbox{Present model} \cr \hbox{Type B}\end{matrix}$
&$\begin{matrix} \hbox{Previous model} \cr \hbox{Type A} \end{matrix}$\\
\hline \hline
quark
&$({\bf 3}, {\bf 4})_{\frac{1}{6}} ~ ({\bf 3}, {\bf 1})_{-\frac{1}{3}}^+ 
    ~ ({\bf 3}, {\bf 1})_{-\frac{1}{3}}^-$
&$({\bf 3}, {\bf 5})_{\frac{2}{3}} ~ ({\bf 3}, {\bf 5})_{-\frac{1}{3}}$ \\
lepton
&$\strut ({\bf 1}, {\bf 4})_{-\frac{1}{2}}$ 
&$({\bf 1}, {\bf 5})_{0} ~ ({\bf 1}, {\bf 5})_{-1}$  \\
\hline
dark fermion & $({\bf 3}, {\bf 4})_{\frac{1}{6}} ~ ({\bf 1}, {\bf 5})_{0}^+ ~ ({\bf 1}, {\bf 5})_{0}^-$ 
&$({\bf 1}, {\bf 4})_{\frac{1}{2}}$ \\
\hline \hline
brane fermion &$({\bf 1}, {\bf 1})_{0} $ 
&$\begin{matrix} ({\bf 3}, [{\bf 2,1}])_{\frac{7}{6}, \frac{1}{6}, -\frac{5}{6}} \cr
({\bf 1}, [{\bf 2,1}])_{\frac{1}{2}, -\frac{1}{2}, -\frac{3}{2}} \end{matrix}$\\
\hline
brane scalar &$({\bf 1}, {\bf 4})_{\frac{1}{2}} $ 
&$({\bf 1}, [{\bf 1,2}])_{\frac{1}{2}}$ \\
\hline
$\begin{matrix} {\rm symmetry ~of} \cr {\rm brane ~interactions} \end{matrix}$
&$SU(3)_C \times SO(5) \times U(1)_X$ &$SU(3)_C \times SO(4) \times U(1)_X$ \\
\hline
\end{tabular}
\label{Tab:matter}
\end{center}
}
\end{table}

Orbifold boundary conditions for bulk fermions are specified in the following manner.

\noindent
\underline{(i) Quark multiplets:}  
$\Psi_{({\bf 3,4})}^{\alpha}$, $\Psi_{({\bf 3,1})}^{\pm \alpha}$
\begin{align}
&\Psi_{({\bf 3,4})}^{\alpha} (x, y_j - y) = 
- P_{\bf 4}^{SO(5)} \gamma^5 \Psi_{({\bf 3,4})}^{\alpha} (x, y_j + y) ~, \cr
&\Psi_{({\bf 3,1})}^{\pm \alpha}  (x, y_j - y) =
\mp \gamma^5 \Psi_{({\bf 3,1})}^{\pm \alpha}  (x, y_j + y) ~.
\label{quarkBC1}
\end{align}
Here 5D Dirac matrices $\gamma^a$ $(a=0,1,2,3,5)$ satisfy 
$\{\gamma^a,\gamma^b\}=2\eta^{ab}$
$(\eta^{ab}=\mbox{diag}(-I_1,I_4))$, and 
$\gamma^5  =\mbox{diag} (1,1,-1,-1)$.

\noindent
\underline{(ii) Lepton multiplets:}  $\Psi_{({\bf 1,4})}^{\alpha}$
\begin{align}
&\Psi_{({\bf 1,4})}^{\alpha} (x, y_j - y) = 
- P_{\bf 4}^{SO(5)} \gamma^5 \Psi_{({\bf 1,4})}^{\alpha} (x, y_j + y) ~.
\label{leptonBC1}
\end{align}

\noindent
\underline{(iii) Dark fermions:}  $\Psi_{({\bf 1,5})}^{\pm \beta}$
\begin{align}
&\Psi_{({\bf 1,5})}^{\pm \beta} (x, y_j - y) = 
\pm P_{\bf 5}^{SO(5)} \gamma^5 \Psi_{({\bf 1,5})}^{\pm \beta} (x, y_j + y) ~.
\label{darkFBC1}
\end{align}
Alternatively one may adopt the parity assignment $\pm (-1)^j P_{\bf 5}^{SO(5)}$ 
instead of $\pm P_{\bf 5}^{SO(5)}$ in  (\ref{darkFBC1}).

\noindent
\underline{(iv) Dark fermion:}  $\Psi_{({\bf 3,4})} \equiv \Psi_F$
\begin{align}
&\Psi_F(x, y_j - y) = 
(-1)^j  P_{\bf 4}^{SO(5)} \gamma^5 \Psi_F (x, y_j + y) ~.
\label{darkFBC2}
\end{align}

The parity assignment  of 4D left- and right-handed component of each fermion field
is summarized in Table~\ref{Tab:parity}.
$\Psi_{({\bf 3,4})}^{\alpha}$  and $\Psi_{({\bf 1,4})}^{\alpha}$
$(\alpha=1,2,3)$ has zero modes,
corresponding to one generation of quarks and leptons for each $\alpha$.

\begin{table}[tbh]
\renewcommand{\arraystretch}{1.2}
\begin{center}
\caption{Parity assignment $(P_0, P_1)$ of fermion fields in the bulk.
The corresponding names adopted in Ref.~\cite{Furui2016} are 
listed in the last column for the first generation.
Brane fermion and scalar fields are  listed at the bottom for convenience.
}
\vskip 10pt
\begin{tabular}{|c|c|c|c|c|}
\hline
Field & $G_{3221}$ &Left &Right &Name\\
\hline
$\Psi_{({\bf 3,4})}^{\alpha}$ &$({\bf 3,2,1})_{\frac{1}{6}}$
&$(+,+)$ &$(-,-)$ &$\begin{matrix} u_j \cr d_j \end{matrix}$\\
\cline{2-5}
&$({\bf 3,1,2})_{\frac{1}{6}}$ 
&$(-,-)$ &$(+,+)$ &$\begin{matrix} u_j' \cr d_j' \end{matrix}$\\
\hline
$\Psi_{({\bf 3,1})}^{\pm \alpha}$ &$({\bf 3,1,1})_{-\frac{1}{3}}$
&$(\pm ,\pm )$ &$(\mp , \mp )$ &$D^{\pm}_j$\\
\hline
$\Psi_{({\bf 1,4})}^{\alpha}$ &$({\bf 1,2,1})_{-\frac{1}{2}}$
&$(+,+)$ &$(-,-)$ &$\begin{matrix} \nu_e \cr e \end{matrix}$\\
\cline{2-5}
&$({\bf 1,1,2})_{-\frac{1}{2}}$ 
&$(-,-)$ &$(+,+)$ &$\begin{matrix} \nu_e' \cr e' \end{matrix}$\\
\hline
$\Psi_F$ &$({\bf 3,2,1})_{\frac{1}{6}}$
&$(-,+)$ &$(+,-)$ &$\begin{matrix} F_{1j }\cr F_{2j} \end{matrix}$\\
\cline{2-5}
&$({\bf 3,1,2})_{\frac{1}{6}}$ 
&$(+,-)$ &$(-,+)$ &$\begin{matrix} F_{1j}' \cr F_{2j}' \end{matrix}$\\
\hline
$\Psi_{({\bf 1,5})}^{\pm \beta}$ &$({\bf 1, 2,2})_0$
&$(\pm ,\pm )$ &$(\mp , \mp )$ 
&$\begin{matrix} N^{\pm} &\hat E^{\pm } \cr
E^{\pm} & \hat N^{\pm }\end{matrix}$\\
\cline{2-5}
&$({\bf 1,1,1})_0$
&$(\mp , \mp )$  &$(\pm ,\pm )$ &$S^{\pm }$ \\
\hline \hline
$\chi^\alpha$ & $({\bf 1,1,1})_0$ & --- & --- &$\chi$ \\
\hline
$\Phi_{({\bf 1}, {\bf 4})}$ &$({\bf 1,2,1})_{\frac{1}{2}}$ & --- & --- &$\Phi_{[{\bf 2}, {\bf 1}]}$\\
\cline{2-5}
&$({\bf 1,1,2})_{\frac{1}{2}}$ & --- & --- &$\Phi_{[{\bf 1}, {\bf 2}]}$\\
\hline
\end{tabular}
\label{Tab:parity}
\end{center}
\end{table}

\subsection{Action}
\label{Sec:action}

The action consists of the 5D bulk action and 4D brane action.

\subsubsection{Bulk action}
\label{Sec:bulk-action}

The bulk part of the action is given by 
\begin{align}
S_{\rm bulk}&=
S_{\rm bulk}^{\rm gauge}+S_{\rm bulk}^{\rm fermion},
\label{Eq:Action-bulk}
\end{align}
where $S_{\rm bulk}^{\rm gauge}$ and $S_{\rm bulk}^{\rm fermion}$ are
bulk actions of gauge and fermion fields, respectively.
The action of each gauge field, $A_M^{SU(3)_C}$, $A_M^{SO(5)_W}$, or $A_M^{U(1)_X}$,
is given in the form 
\begin{align}
S_{\rm bulk}^{\rm gauge}&=
\int d^5x\sqrt{-\det G}\, \bigg[-\mbox{tr}\left(
\frac{1}{4}F_{}^{MN}F_{MN}
+\frac{1}{2\xi}(f_{\rm gf})^2+{\cal L}_{\rm gh}\right)\bigg],
\label{Eq:Action-bulk-gauge}
\end{align}
where $\sqrt{-\det G}=1/k z^5$, $z=e^{ky}$, $M,N=0,1,2,3,5$,
$\mbox{tr}$ is a trace over all group generators for each group.
Field strength $F_{MN}$ is defined by  
\begin{align}
F_{MN}&:=
\partial_MA_N-\partial_NA_M-i g[A_M,A_N]
\end{align}
with each 5D gauge coupling constant $g$.
For the gauge fixing and ghost terms we take
\begin{align}
f_{\rm gf}&=
 z^2\left\{ \eta^{\mu\nu} {\cal D}^{\rm c}_\mu A^{\rm q}_\nu
 +\xi k^2 z{\cal D}^{\rm c}_z \Big( \frac{1}{z} A^{\rm q}_z\Big) \right\}, \cr
\noalign{\kern 10pt} 
{\cal L}_{\rm gh}& =
\bar{c} \bigg\{ \eta^{\mu\nu} {\cal D}_\mu^{\rm c} {\cal D}_\nu
+ \xi k^2 z {\cal D}_z^{\rm c} \frac{1}{z} {\cal D}_z \bigg\} c,
\end{align}
where $\mu,\nu=0,1,2,3$, 
$\eta^{\mu\nu}=\eta_{\mu\nu}=\mbox{diag}(-1,1,1,1)$, and
$A_M=A_M^c+A_M^q$.
${\cal D}_M^{c}B=\partial_M B-ig[A_M^c,B]$ and
${\cal D}_M^{c+q}B=\partial_M B-ig[A_M,B]$ 
where $B=A_\mu^q$, ${A_z^q/z}$ and $c$.
In the present paper  only $A_z$ component of $A_M^{SO(5)}$ has
non-vanishing classical background $A_z^c$.

Each fermion multiplet $\Psi (x,y)$ in the bulk has its own bulk-mass parameter $c$.
The covariant derivative is given by
\begin{align}
&{\cal D}(c)= \gamma^A {e_A}^M
\bigg( D_M+\frac{1}{8}\omega_{MBC}[\gamma^B,\gamma^C]  \bigg) -c\sigma'(y) ~, \cr
\noalign{\kern 5pt}
&D_M =  \dd_M - ig_S A_M^{SU(3)} -i g_A A_M^{SO(5)}
 -i g_B Q_X A_M ^{U(1)} ~. 
\label{covariantD}
\end{align}
Here $\sigma'(y):=d\sigma(y)/dy$ and $\sigma'(y) =k$ for $0< y < L$. 
$g_S$, $g_A$, $g_B$ are $SU(3)_C$, $SO(5)_W$, $U(1)_X$ gauge
coupling constants.
Let $\Psi^J$ collectively denote all fermion fields in the bulk.   Then the action
in the bulk becomes
\begin{align}
&S_{\rm bulk}^{\rm fermion} =  \int d^5x\sqrt{-\det G} \,
\bigg\{ \sum_J  \overline{\Psi^J}  {\cal D} (c_J) \Psi^J \cr
\noalign{\kern 5pt}
&\quad 
- \sum_\alpha \Big( m_D^\alpha \overline{\Psi}{}_{({\bf 3},{\bf 1})}^{+ \alpha}
\Psi_{({\bf 3},{\bf 1})}^{- \alpha}+ \mbox{h.c.} \Big)
- \sum_\beta \Big( m_V^\beta \overline{\Psi}{}_{({\bf 1},{\bf 5})}^{+ \beta}
\Psi_{({\bf 1},{\bf 5})}^{- \beta}+ \mbox{h.c.} \Big) \bigg\} ,
\label{fermionAction1}
\end{align} 
where $\overline{\Psi} = i \Psi^\dagger \gamma^0$.
$m_D^{\alpha}$ and  $m_V^{\beta}$  are ``pseudo-Dirac'' bulk mass terms.

In terms of  $\check{\Psi}$ defined by 
\begin{align}
\check{\Psi}:=\frac{1}{z^{2}}\Psi ~,~~
\Big(\partial_z-\frac{2}{z}\Big)\Psi =z^{2}\partial_z\check{\Psi} ~, 
\label{checkedPsi}
\end{align}
the bulk part of the fermion action becomes 
\begin{align}
&S_{\rm bulk}^{\rm fermion}=
\int d^4x\int_1^{z_L}\frac{dz}{k} \, \bigg\{ 
\sum_J \overline{\check \Psi} {}^J \Big[ \gamma^\mu D_\mu
+ k \Big( \gamma^5 D_z - \frac{c_J}{z}\Big) \Big]  \check \Psi^J \cr
\noalign{\kern 10pt}
&
- \sum_\alpha \Big( \frac{m_D^\alpha}{z}  \overline{\check \Psi}{}_{({\bf 3},{\bf 1})}^{+ \alpha}
\check \Psi_{({\bf 3},{\bf 1})}^{- \alpha}+ \mbox{h.c.} \Big)
- \sum_\beta \Big( \frac{m_V^\beta}{z} \overline{\check \Psi}{}_{({\bf 1},{\bf 5})}^{+ \beta}
\check\Psi_{({\bf 1},{\bf 5})}^{- \beta}+ \mbox{h.c.} \Big) \bigg\} .
\label{fermionAction2}
\end{align}

\subsubsection{Action for the brane scalar $\Phi_{({\bf 1,4})}$}

The action for the brane scalar field $\Phi_{({\bf 1,4})}  (x)$ in $({\bf 1, 4})_{ \frac{1}{2}}$ is given by
\begin{align}
S_{\rm brane}^{\Phi} & = 
\int d^5x\sqrt{-\det G} \,  \delta(y) \cr
\noalign{\kern 5pt}
&
\times \Big\{ 
-(D_\mu\Phi_{({\bf 1,4})})^{\dag}D^\mu\Phi_{({\bf 1,4})}
-\lambda_{\Phi_{({\bf 1,4})}}
\big(\Phi_{({\bf 1,4})}^\dag\Phi_{({\bf 1,4})} - |w|^2  \big)^2 \Big\} ,
\label{Eq:Action-brane-scalar}
\end{align}
where 
\begin{align}
D_\mu\Phi_{({\bf 1,4})}&=
\bigg\{\partial_\mu- ig_A 
 \sum_{\alpha=1}^{10}   A_{\mu}^{\alpha} T^{\alpha}  
 -ig_B Q_X B_\mu  \bigg\}\Phi_{({\bf 1,4})} ~.
\end{align}
Here $SO(5)_W$ generators $\{ T^\alpha \}$ consist of
$SU(2)_L$, $SU(2)_R$ generators $\{ T^{a_L}, T^{a_R} \}$ ($a=1,2,3$) and
$SO(5)/SO(4)$ generators $\{ T^{\hat p} = T^{p5}/\sqrt{2} \}$ ($p=1 \sim 4$).
The corresponding canonically normalized gauge fields are 
\begin{align}
A_{M}^{a_L} &= \frac{1}{\sqrt{2}}
\Big(\frac{1}{2} \epsilon^{abc}A_M^{bc}+A_M^{a4}\Big) ~,  \cr
\noalign{\kern 5pt}
A_{M}^{a_R} &= \frac{1}{\sqrt{2}}
\Big(\frac{1}{2} \epsilon^{abc}A_M^{bc} -A_M^{a4}\Big) ~,  \cr
\noalign{\kern 5pt}
A_M^{\hat p} &= A_M^{p5} ~.
\end{align}
$B_M$ represents the $U(1)_X$ gauge field.

The brane scalar field $\Phi_{({\bf 1,4})}$ is decomposed as 
\begin{align}
\Phi_{({\bf 1,4})} =
\begin{pmatrix} \Phi_{[{\bf 2,1}]} \cr \Phi_{[{\bf 1,2}]} \end{pmatrix}
\end{align}
where $[{\bf 2,1}]$ and $[{\bf 1,2}]$ represent $SU(2)_L \times SU(2)_R$ content.
$\Phi_{({\bf 1,4})}$ develops a nonvanishing VEV
\begin{align}
\la \Phi_{({\bf 1,4}) }\ra = \begin{pmatrix} 0_2 \cr v_2 \end{pmatrix} , ~~
v_2 = \begin{pmatrix} 0 \cr w \end{pmatrix} ~.
\label{Svev1}
\end{align}
The nonvanishing VEV  breaks
$SU(3)_C\times SO(5)\times U(1)_X$ to
$SU(3)_C\times SU(2)_L\times U(1)_Y$.
As shown in Appendix~\ref{Sec:SO5}, one can define  the conjugate scalar field
$\widetilde{\Phi}_{({\bf 1,4})}$  in $({\bf 1, 4})_{- \frac{1}{2}}$ by  
\begin{align}
\widetilde{\Phi}_{({\bf 1,4})}= \begin{pmatrix}
i\sigma^2\Phi_{[{\bf 2,1}]}^* \cr \noalign{\kern 5pt}
-i\sigma^2\Phi_{[{\bf 1,2}]}^*  \end{pmatrix} .
\end{align}
Its VEV is given by
\begin{align}
\la \widetilde \Phi_{({\bf 1,4}) }\ra = \begin{pmatrix} 0_2 \cr \tilde v_2 \end{pmatrix} , ~~
\tilde v_2 = \begin{pmatrix} - w^*  \cr 0 \end{pmatrix} ~.
\label{Svev2}
\end{align}

The combination of the nonvanishing VEV
$\langle\Phi_{({\bf 4,1})(3)}\rangle$ on
the UV brane (at $y=0$) and the orbifold BCs $P_j (j=0,1)$
reduces $SU(3)_C\times SO(5) \times U(1)_X$ to the SM gauge group 
$G_{\rm SM}=SU(3)_C\times SU(2)_L\times U(1)_Y$.

\subsubsection{Action for the  brane fermion $\chi^\alpha$}

The action for the gauge-singlet brane fermion $\chi^\alpha (x)$ is
\begin{align}
S_{\rm brane}^\chi = &
\int d^5x\sqrt{-\det G} \, \delta(y) \bigg\{  
\frac{1}{2}\overline{\chi}^\alpha \gamma^\mu\partial_\mu \chi^\alpha
 - \frac{1}{2} M^{\alpha \beta}  \overline{\chi}^\alpha \chi^{\beta} \bigg\} ~.
\label{Eq:Action-brane-fermion-chi}
\end{align}
$\chi^\alpha (x)$ satisfies the Majorana condition
$\chi^c=\chi$;
\begin{align}
\chi = \begin{pmatrix} \xi \cr \eta \end{pmatrix} , ~~
\chi^c = \begin{pmatrix} + \eta^c \cr - \xi^c \end{pmatrix} 
=e^{i\delta_C} \begin{pmatrix} + \sigma^2 \eta^* \cr - \sigma^2 \xi^* \end{pmatrix} .
\label{Majorana1}
\end{align}

\subsubsection{Brane  interactions and mass terms for fermions}

On the UV brane there can be
$SU(3)_C\times SO(5) \times U(1)_X$-invariant
brane interactions among the bulk fermion,   brane fermion,
and brane scalar fields. We consider  
\begin{align}
&
 S_{\rm brane}^{\rm int}=
\int d^5x\sqrt{-\det G} \, \delta(y)
\big({\cal L}_1+{\cal L}_2+{\cal L}_3\big) ~, \cr
\noalign{\kern 5pt}
&{\cal L}_1 = 
- \Big\{ \kappa^{\alpha\beta} \,
\overline{\Psi}{}_{({\bf 3,4})}^{\alpha} \Phi_{({\bf 1,4})}
\cdot \Psi_{({\bf 3,1})}^{+\beta}  + {\rm h.c.} \Big\} ~, \cr
\noalign{\kern 5pt}
&{\cal L}_2 =
- \Big\{ \widetilde{\kappa}^{\prime\alpha\beta} \,
\overline{\Psi}{}_{({\bf 1,4})}^{\alpha} \, \Gamma^a \,
\widetilde{\Phi}_{({\bf 1,4})} \cdot \big(\Psi_{({\bf 1,5})}^{-\beta} \big)_a
 + {\rm h.c.} \Big\} ~, \cr
\noalign{\kern 5pt}
&{\cal L}_3 =
- \Big\{ \widetilde{\kappa}_{\bf 1}^{\alpha \beta} \,
\overline{\chi}^\beta 
\widetilde{\Phi}_{({\bf 1,4})}^\dag \Psi_{({\bf 1,4})}^{\alpha}   + {\rm h.c.} \Big\} ~, 
\label{Eq:Action-brane-fermion}
\end{align}
where $\kappa$'s are coupling constants.

$\langle\Phi_{({\bf 1,4})}\rangle\not=0$ generates mass terms on the UV brane
from  the interaction in (\ref{Eq:Action-brane-fermion}).
Together with the inherent Majorana masses in (\ref{Eq:Action-brane-fermion-chi})
brane fermion masses are given by
\begin{align}
&S_{\rm brane\ mass}^{\rm fermion}=
\int d^5x\sqrt{-\det G} \, \delta(y)
 \Big(
 {\cal L}_1^m+{\cal L}_2^m+{\cal L}_3^m
+{\cal L}_{\chi}^m\Big) ~, \cr
\noalign{\kern 5pt}
&{\cal L}_1^m =
 2\mu_1^{\alpha\beta} \, 
\overline{\check{d}\,}{}_{R}^{\prime\alpha} \check{D}_{L}^{+\beta}   +\mbox{h.c.}  ~, \cr
 \noalign{\kern 5pt}%
&{\cal L}_2^m = 
-\widetilde{\mu}_2^{\alpha\beta}
\left\{i2
(\overline{\check{e}_{L}^{\alpha}}\check{E}_{R}^{-\beta}
+\overline{\check{\nu}_{L}^{\alpha}}\check{N}_{R}^{-\beta})
 +\sqrt{2}\ \overline{\check{\nu}_{R}^{\prime\alpha}}\check{S}_{L}^{-\beta}
\right\}
+\mbox{h.c.} ~ , \cr
\noalign{\kern 5pt}
&{\cal L}_3^m = 
-\frac{m_B^{\alpha \beta}}{\sqrt{k}} \, 
 ( \overline{\chi}^\beta  \check{\nu}_{R}^{\prime\alpha}
 +\overline{\check{\nu}} {}_{R}^{\prime\alpha}  \chi^\beta ) ~, \cr
\noalign{\kern 5pt}
&{\cal L}_{\chi_{\bf 1}}^m =
-\frac{1}{2}  M^{\alpha \beta}\overline{\chi}^\alpha \chi^{\beta}.
\label{braneFmass1}
\end{align}
Here 
$2\mu_{1}^{\alpha\beta}=\sqrt{2}\kappa^{\alpha\beta}w$,
$2\widetilde{\mu}_{2}^{\alpha\beta}=
\sqrt{2}\widetilde{\kappa}^{\prime\alpha\beta}w$, and
$m_B^{\alpha\beta}=\widetilde{\kappa}_{\bf 1}^{\alpha\beta}w\sqrt{k}$.
$\mu_{1}^{\alpha\beta}$ and $\widetilde{\mu}_{2}^{\alpha\beta}$ are dimensionless, whereas
$m_B^{\alpha \beta}$ and $M^{\alpha \beta}$ have dimension of mass.

\subsubsection{Brane mass terms for gauge bosons}

$\langle\Phi_{({\bf 1,4})}\rangle\not=0$ also yields additional brane mass terms for 
the 4D components of the $SO(5)\times U(1)_X$ gauge fields.
It follows from (\ref{Eq:Action-brane-scalar}) that 
\begin{align}
&S_{\rm brane}^{\rm gauge}=
\int d^5x\sqrt{-\det G} \, \delta(y) \, \cr 
\noalign{\kern 10pt}
& 
\times \bigg\{ -\frac{g_A^2|w|^2}{4}
 \big( A_{\mu}^{1_R} A^{1_R \mu}+A_{\mu}^{2_R} A^{2_R \mu} \big)
 -\frac{(g_A^2+g_B^2)|w|^2}{4}
 A_{\mu}^{3_R'} A^{3_R' \mu} \bigg\} ,
 \label{gaugeBranemass1}
\end{align}
where
\begin{align}
&\begin{pmatrix}  A_M^{3_R '} \cr \noalign{\kern 3pt} B_M^{Y} \end{pmatrix} 
=\begin{pmatrix}  c_\phi& -s_\phi \cr \noalign{\kern 3pt}s_\phi &c_\phi \end{pmatrix} 
\begin{pmatrix}  A_M^{3_R} \cr \noalign{\kern 3pt} B_M \end{pmatrix} , \cr
\noalign{\kern 5pt}
& c_\phi =\cos\phi = \frac{g_A}{\sqrt{g_A^2+g_B^2}} ~, ~~
s_\phi =\sin\phi = \frac{g_B}{\sqrt{g_A^2+g_B^2}} ~.  
\label{gaugeBranemass2}
\end{align}
The 5D gauge coupling $g_Y^{\rm 5D}$ of $U(1)_{Y}$  is given by
\begin{align}
 g_Y^{\rm 5D} =\frac{g_Ag_B}{\sqrt{g_A^2+g_B^2}}=g_A s_\phi ~.
\label{Eq:g_Y-g_A-g_B}
\end{align}
$A_{\mu}^{1_R}, A_{\mu}^{2_R}$ and $A_{\mu}^{3_R '}$ obtain
large brane masses, which effectively change the BCs on the UV brane for the
corresponding fields.

Note that the 4D $SU(2)_L$ gauge coupling constant is related to $g_A$ by
\begin{align}
g_w = \frac{g_A}{\sqrt{L}} ~.
\label{gaugecoupling1}
\end{align}
The three 4D SM gauge coupling constants $g_s, g_w, g_Y$ of $SU(3)_C$, $SU(2)_L$, $U(1)_Y$
at the $m_Z$ scale are $\alpha_s = g_s^2/4\pi = 0.1184 \pm 0.0007$, 
$\alpha_w = g_s^2/4\pi = \alpha_{\rm EM}/\sin^2\theta_W$, and
$\alpha_Y = g_Y^2/4\pi = \alpha_{\rm EM}/\cos^2\theta_W$
where $\alpha_{\rm EM}^{-1}=127.916\pm 0.015$ and 
$\sin^2\theta_W =0.23116\pm 0.00013$.\cite{PDG2014}
In the $SU(3)_C\times SO(5) \times U(1)_X$ GHU,
the $SU(2)_R$ gauge coupling constants are the same as
the SM $SU(2)_L$ gauge coupling constants.  With the relation (\ref{Eq:g_Y-g_A-g_B})
one finds that
\begin{align}
&\frac{4 \pi L}{g_A^2} = \alpha_w^{-1} \simeq 29.56 ~, \cr
\noalign{\kern 5pt}
&\frac{4 \pi L}{g_B^2} =  \alpha_Y^{-1} -  \alpha_w^{-1} \simeq 68.78 ~,
\label{gaugecoupling2}
\end{align}
at the $m_Z$ scale.

\subsection{Higgs boson and the twisted gauge}

4D Higgs boson is contained in the ({\bf 1,2,2}) component of $A_y^{SO(5)}$ as tabulated in
Table~\ref{Tab:BC-gauge}.  In the $z$ coordinate $A_z = (kz)^{-1} A_y$ ($1 \le z \le z_L$), and 
\begin{align}
A_z^{(j5)} (x, z) &= \frac{1}{\sqrt{k}} \, \phi_j (x) u_H (z) + \cdots , \cr
\noalign{\kern 5pt}
u_H (z) &= \sqrt{ \frac{2}{z_L^2 -1} } \, z ~,  \cr
\noalign{\kern 5pt}
\Phi(x) &= \frac{1}{\sqrt{2}} \begin{pmatrix} \phi_2 + i \phi_1 \cr \phi_4 - i\phi_3 \end{pmatrix} .
\label{4dHiggs}
\end{align}
$\Phi(x) $ corresponds to the doublet Higgs field in the SM.

At the quantum level $\Phi$ develops a nonvanishing expectation value.  Without loss of generality
we assume $\la \phi_1 \ra , \la \phi_2 \ra , \la \phi_3 \ra  =0$ and  $\la \phi_4 \ra \not= 0$, 
which is related to the Aharonov-Bohm (AB) phase $\theta_H$ in the fifth dimension.  Eigenvalues of 
\begin{align}
\hat W &= P \exp \bigg\{ i g_A \int_{-L}^L dy \, A_y \bigg\}  \cdot P_1 P_0 
\label{ABphase1}
\end{align}
are gauge invariant.  For $A_y = (2k)^{-1/2} \phi_4 (x)v_H (y) T^{(45)}$,  where
$v_H (y) = k e^{ky} u_H (z)$ for $0 \le y \le L$ and $v_H (-y) =  v_H (y) =  v_H (y + 2L)$,
one finds
\begin{align}
&\hat W = \exp \Big\{ i \hat \theta_H (x) \cdot 2 T^{(45)} \Big\} ~, \cr
\noalign{\kern 5pt}
&\hat \theta_H (x)  = \frac{g_A}{2} \sqrt{ \frac{z_L^2 - 1}{k}} \, \phi_4 (x) ~.
\label{ABphase2}
\end{align}
The eigenvalues of $2 T^{(45)} $ in the spinor representation are $\pm 1$, and $\hat \theta_H (x)$
is the AB phase.  We denote $\la \hat \theta_H \ra = \theta_H$.
4D neutral Higgs field $H(x)$ is the fluctuation mode of $\phi_4(x)$ around $\la \phi_4 \ra$.
Hence one finds
\begin{align}
&A_z^{(45)} (x, z) = \frac{1}{\sqrt{k}} \big\{ \theta_H f_H + H(x) \big\} \, u_H(z) + \cdots , \cr
\noalign{\kern 5pt}
&f_H = \frac{2}{g_A} \sqrt{ \frac{k}{z_L^2 -1}} = \frac{2}{g_w} \sqrt{ \frac{k}{L(z_L^2 -1)}} ~.
\label{ABphase3}
\end{align}

Under an $SO(5)$ gauge transformation
\begin{align}
\Omega (y; \alpha) = \exp \bigg\{ - i \frac{g_A \alpha}{\sqrt{2k}} 
\int_y^L dy \, v_H(y) T^{(45)} \bigg\} ~, 
\label{largeGT1}
\end{align}
orbifold boundary conditions  $\{ P_0, P_1 \}$  are changed to
\begin{align}
&P_0' = \Omega (0; 2\alpha) P_0  = \exp \bigg\{ - i \frac{\alpha}{f_H} \cdot 2 T^{(45)} \bigg\} \cdot P_0 ~, \cr
\noalign{\kern 5pt}
&P_1' = P_1 ~,
\label{twistedBC1}
\end{align}
and $\hat \theta_H (x)$ is transformed  to $\hat \theta_H' (x) = \hat \theta_H (x) + (\alpha/f_H)$.
For $\alpha/f_H = 2\pi n$ ($n$: an integer), the boundary conditions remain unchanged whereas
$\theta_H$ changes to $\theta_H ' = \theta_H + 2\pi n$.   This property reflects the gauge-invariant nature
of the AB phase $e^{i\theta_H}$.

Now we go to a new gauge by adopting $\alpha = - \theta_H f_H$ so that $\la \hat \theta_H' \ra = \theta_H' = 0$,
which  is called the twisted gauge.
It is most convenient to evaluate various physical quantities in this gauge.
The twisted gauge was originally introduced in Refs.\ \cite{Falkowski2007, HS2007}, 
and has been extensively employed
in the analysis of GHU. (See, e.g. Refs.~\cite{HOOS2008, Furui2016}.)
Note that the gauge transformation  in (\ref{largeGT1})
becomes, for $0 \le y \le L$, 
\begin{align}
\Omega (z) &= \Omega (y; - \theta_H f_H) = \exp \big\{ i \theta (z) T^{(45)} \big\}  ~, \cr
\noalign{\kern 5pt}
\theta (z) &= \theta_H \, \frac{z_L^2 - z^2}{z_L^2 - 1} ~.
\label{largeGT2}
\end{align}
Quantities in the twisted gauge are denoted with tildes below.
In the twisted gauge the background field vanishes ($\tilde \theta_H = 0$), whereas
the boundary conditions change as (\ref{twistedBC1}). 
For the $SO(5)$ vector representation ${\bf 5}$, the boundary condition
matrices $\tilde{P}_j^{\rm vec}$ $(j=0,1)$ are 
\begin{align}
 \tilde{P}_0^{SO(5)}=\Omega(0)^2P_0^{SO(5)}=
 e^{2i \theta_H T^{(45)}}P_0^{SO(5)},\ \ \tilde{P}_1^{SO(5)}=P_1^{SO(5)}.
\end{align}
For the $SO(5)$ vector representation ${\bf 5}$, the boundary condition
matrices $\tilde{P}_j^{\rm vec}$ $(j=0,1)$ become
\begin{align}
 \tilde{P}_0^{\rm vec}= \begin{pmatrix} I_3 && \cr &\cos2\theta_H &-\sin2\theta_H \cr
&-\sin2\theta_H &-\cos2\theta_H \end{pmatrix} , ~~
\tilde{P}_1^{\rm vec}= \begin{pmatrix} I_4 & \cr & -1 \end{pmatrix} , 
\label{twistedBC2}
\end{align}
and  for the $SO(5)$ spinor representation ${\bf 4}$
\begin{align}
\tilde{P}_0^{\rm sp} &= \sigma^0 \otimes 
\begin{pmatrix}  \cos\theta_H &-i\sin\theta_H \cr  i\sin\theta_H &-\cos\theta_H \end{pmatrix} , ~~
\tilde{P}_1^{\rm sp} = \begin{pmatrix} I_2 & \cr & - I_2 \end{pmatrix} .
\label{twistedBC3}
\end{align}
Here $T^{(45)}_{\rm sp}=  \onehalf \sigma^0 \otimes \sigma^1$ has been used.

\section{Spectrum of gauge fields}

The spectrum of gauge fields in the present model (Type B) is the same
as the spectrum in the previous model (Type A).
We here quote the result for completeness.
The bilinear part of the action of  gauge fields in
(\ref{Eq:Action-bulk-gauge}) takes the form 
\begin{align}
&S' =  \int d^4x\frac{dz}{kz} \sum_{j<k}\bigg[\frac{1}{2}A_\mu^{(jk)}
\Big\{ \eta^{\mu\nu} \big(\Box+k^2{\cal P}_4 \big)
-(1-\xi^{-1} )  \partial^\mu \partial^\nu \Big\}A_\nu^{(jk)} \cr
\noalign{\kern 5pt}
&\hskip 2.cm
+\frac{1}{2}k^2A_z^{(jk)} \big(\Box+\xi k^2{\cal P}_z\big) A_z^{(jk)}
+\bar c^{(jk)}\big(\Box+\xi k^2{\cal P}_4\big) c^{(jk)} \bigg], \cr
\noalign{\kern 5pt}
&\quad 
\Box = \eta^{\mu\nu} \partial_\mu \partial_\nu ~, ~~
 {\cal P}_4 = z \frac{\partial}{\partial z}\frac{1}{z}  \frac{\partial}{\partial z} ~,~~
 {\cal P}_z = \frac{\partial}{\partial z} z  \frac{\partial}{\partial z}\frac{1}{z} ~.
\label{actionGaugeBulk2}
\end{align}
Additional brane mass terms in  (\ref{gaugeBranemass1})
arise for the $A_\mu$ components of $(SU(2)_R \times U(1)_X)/U(1)_Y$.

Boundary conditions in the original gauge are given, in the absence of brane
interactions,  by
\begin{align}
&\begin{cases}
N: ~ \myfrac{\dd}{\dd z}  A_\mu = 0 &\mbox{for parity } + \cr
\noalign{\kern 5pt}
D: ~  A_\mu=0 &\mbox{for parity } -
\end{cases} \cr
\noalign{\kern 10pt}
&\begin{cases}
N: ~  \myfrac{\dd}{\dd z} \Big( \myfrac{1}{z} A_z \Big)=0 &\mbox{for parity }+ \cr
\noalign{\kern 5pt}
D: ~  A_z = 0 &\mbox{for parity } - 
\end{cases}
\label{gaugeBC2}
\end{align}
at $z=1$ ($y=0$) and $z=z_L$ ($y=L$).
Parity of each field is summarized in Table~\ref{Tab:BC-gauge}.
Because of the brane interaction (\ref{gaugeBranemass1}) 
boundary conditions of $A_\mu^{1_R, 2_R, 3_R'}$  at $z=1$ become
\begin{align}
D_\eff (\omega) &: ~ \Big( \frac{\dd}{\dd z} - \omega \Big) A_\mu^{1_R, 2_R} = 0 ~,~~
\omega = \frac{g_A^2 w^2}{4k} ~, \cr
\noalign{\kern 5pt}
D_\eff (\omega') &: ~ \Big( \frac{\dd}{\dd z} - \omega' \Big) A_\mu^{3_R'} = 0 ~,~~
\omega' = \frac{(g_A^2+ g_B^2) w^2}{4k} ~. 
\label{gaugeBC3}
\end{align}
For sufficiently large $w$, boundary conditions of $A_\mu^{1_R, 2_R, 3_R'}$ at $z=1$ are
modified from the Neunmann condition to the Dirichlet condition for low-lying modes
in their KK towers.
Boundary conditions of gauge fields are summarized in Table~\ref{Table:gaugeBC}.

\begin{table}
\caption{The boundary conditions for the  gauge fields at $z=1, z_L$ are summarized.
$N$ and $D$ stand for Neumann and Dirichlet conditions, respectively. 
$D_{\rm eff}$ stands for the effective Dirichlet condition specified in (\ref{gaugeBC3}).
}
\begin{center}
\begin{tabular}{|c|c|c|c|c|}\hline
&&$\begin{matrix}{\rm No.~of}\\ {\rm generators} \end{matrix}$ &$A_\mu$ &$A_z$ \\
\hline
(1)&$SU(3)_C$ &8 &$(N,N)$ &$(D,D)$\\
(2)&$SU(2)_L$ &3 &$(N,N)$ &$(D,D)$\\
(3)&$U(1)_Y$  &1 &$(N,N)$  &$(D,D)$\\
(4)&$(SU(2)_R\cup U(1)_X)/U(1)_Y$ &3 &$(D_{\rm eff},N)$ &$(D,D)$ \\
(5)&$SO(5)_W/(SU(2)_L\cup SU(2)_R)$ &4 &$(D,D)$ &$(N,N)$ \\
\hline
\end{tabular}
\end{center}
\label{Table:gaugeBC}
\end{table}

In the twisted gauge all fields obey free equations in the bulk $1 < z <  z_L$,
whereas boundary conditions at $z=1$ become $\theta_H$-dependent and nontrivial.
$SO(5)$ gauge fields in the twisted gauge are given by
$\tilde{A}_M= \Omega(z) A_M \Omega(z)^{-1} + (i/g_A) \Omega(z) \dd_M \Omega (z)^{-1}$
where $\Omega (z)$ is given by (\ref{largeGT2}).  
In particular one finds that
\begin{align}
 A_M^{a4} &= \cos \theta(z)  \tilde{A}_M^{a4} - \sin \theta(z) \tilde{A}_M^{a5} ~, ~(a=1,2,3),  \cr
 A_M^{a5} &= \sin \theta(z) \tilde{A}_M^{a4} + \cos \theta(z) \tilde{A}_M^{a5}~, \cr
 A_z^{45} &= \tilde{A}_z^{45} - \frac{\sqrt{2}}{g_A} \theta'(z)
 = \tilde{A}_z^{45} + \frac{2\sqrt{2}}{g_A}\,  \theta_H \, \frac{z}{z_L^2-1} ~, 
\label{Eq:Twisted-gauge-field}
\end{align}
while the other components are unchanged.

At $z=z_L$, $\theta(z_L)=0$, and 
$\tilde{A}_M$ satisfies the same boundary condition as $A_M$ at $z=z_L$.
Consequently wave functions  for $\tilde{A}_\mu$  and $\tilde{A}_z$ are given by
the  functions tabulated in Table \ref{Table:Twisted-Wave-Function-Gauge}.
The basis functions $C(z;\lambda)$ and $S(z; \lambda)$ there are defined
in e.g., Refs. \cite{HOOS2008} and \cite{Furui2016}, and are listed in Appendix B.

\begin{table}
\caption{Wave functions of the gauge fields in the twisted gauge.  
$N$ and $D$ stand for Neumann and Dirichlet conditions at $z=z_L$.
The basis functions $C(z;\lambda)$ and $S(z; \lambda)$ are given in Appendix B.
}
\renewcommand{\arraystretch}{1.3}
\begin{center}
\begin{tabular}{|c|c|c|}\hline
BC at $z=z_L$ &$N$ &$D$ \\
\hline
$\tilde{A}_\mu$ &$C(z;\lambda)$ & $S(z;\lambda)$\\
\hline
$\tilde{A}_z$ & $ S'(z;\lambda)$ & $C'(z;\lambda)$\\
\hline
\end{tabular}
\end{center}
\label{Table:Twisted-Wave-Function-Gauge}
\end{table}

\subsection{$A_\mu$ components}

The mass spectra of $A_\mu$ components are the following.

\vskip 5pt

\noindent
\underline{(i) $(\tilde{A}_{\mu}^{a_L},\ \tilde{A}_{\mu}^{a_R},\
\tilde{A}_\mu^{\hat a})$ $(a=1,2)$: $W$ and $W_R$ towers}

The boundary conditions at $z=1$ are
\begin{align}
&\frac{\partial}{\partial z} A_{\mu}^{a_L}=0~, ~~
\Big( \frac{\partial}{\partial z} - \omega \Big) A_{\mu}^{a_R}=0 ~, ~~
A_\mu^{\hat a}=0 ~.
\label{Eq:BCs-Gauge-W-1}
\end{align}
$\partial A_{\mu}^{a_R}/\partial z$ is evaluated at $z=1^+$.  
These conditions with (\ref{Eq:Twisted-gauge-field}) lead to
the equation which determine the mass spectrum $\{ m_n = k \lambda_n \}$:
\begin{align}
2C'(SC'+\lambda\sin^2\theta_H)-\omega C(2SC'+\lambda\sin^2\theta_H)=0 ~.
\label{Eq:Spectra-Gauge-W-1}
\end{align}
Here $C=C(1;\lambda)$, $S=S(1;\lambda)$,
$C'=C'(1;\lambda)$, and $S'=S'(1;\lambda)$.

For sufficiently large $\omega$,
the second term in Eq.~(\ref{Eq:Spectra-Gauge-W-1})
approximately determines the spectra  of low-lying KK modes.
This approximation is justified for $w \gg m_\KK$.
In this approximation  the spectra of $W$ and $W_R$ towers are determined by
\begin{align}
 W\ \mbox{tower:}\
 & 2S(1;\lambda)C'(1;\lambda)+\lambda\sin^2\theta_H =  0 ~, \cr
W_R\ \mbox{tower:}\
 & C(1;\lambda)= 0 ~.
\label{Eq:Spectra-Gauge-W-2}
\end{align}
It follows that the mass of $W$ boson $m_W=m_{W^{(0)}}$ is given by
\begin{align}
m_W \simeq \sqrt{\frac{k}{L}} \, z_L^{-1}\sin\theta_H 
\simeq\frac{\sin\theta_H}{\pi\sqrt{kL}} \, m_{\rm KK},
\label{Eq:W-mass-1}
\end{align}
where $m_{\rm KK}=\pi k/(z_L-1) \simeq \pi kz_L^{-1}$.

\vskip 5pt
\noindent
\underline{(ii) $(\tilde{A}_{\mu}^{3_L},\tilde{A}_{\mu}^{3_R'},\tilde{A}_\mu^{\hat 3},{B}_\mu^Y)$: 
$\gamma$, $Z$ and $Z_R$ towers}

The boundary conditions at $z=1$ are
\begin{align}
&\frac{\partial}{\partial z} A_{\mu}^{3_L}=0 ~,~~
\Big( \frac{\partial}{\partial z} - \omega' \Big) A_{\mu}^{3_R'}= 0 ~, ~~
A_\mu^{\hat 3}=0 ~, ~~
\frac{\partial}{\partial z}  B_\mu^Y=0 ~.
\label{Eq:BCs-Gauge-Z-1}
\end{align}
The spectrum is determined by 
\begin{align}
 C'\left[ 2C'(SC'+\lambda\sin^2\theta_H) 
 -\omega'C\left\{2SC'+(1+s_\phi^2)\lambda\sin^2\theta_H\right\}
 \right]=0 ~.
\label{Eq:Spectra-Gauge-Z-1}
\end{align}
For sufficiently large $\omega'$, the spectrum  of low-lying KK modes 
is approximately determined by the second term.  One finds that 
\begin{align}
 \gamma\ \mbox{tower:}\ &
 C'(1;\lambda)=0 ~, \cr
 Z\ \mbox{tower:}\ &
 2S(1;\lambda)C'(1;\lambda)+(1+s_\phi^2)\lambda\sin^2\theta_H =  0 ~, \cr
 Z_R\ \mbox{tower:}\ &
 C(1;\lambda) = 0 ~.
\label{Eq:Spectra-Gauge-Z-2}
\end{align}
The mass of the$Z$ boson $m_Z = m_{Z^{(0)}}$ is given by
\begin{align}
 m_Z\simeq\sqrt{1+s_\phi^2}
 \sqrt{\frac{k}{L}}z_L^{-1}\sin\theta_H
 \simeq \sqrt{1+s_\phi^2} \, \frac{\sin\theta_H}{\pi\sqrt{kL}}  \,  m_{\rm KK} ~.
\label{Eq:Z-mass-1}
\end{align}
We recall the relation\cite{Funatsu:2014fda}
\begin{align}
 \frac{1}{\sqrt{1+s_\phi^2}}\simeq \cos\theta_W ~,~~
 \sin\theta_W\simeq \frac{g_B'}{\sqrt{g_A^2+2g_B^{\prime 2}}} ~.
\end{align}
It follows from (\ref{Eq:W-mass-1}) and (\ref{Eq:Z-mass-1}) that 
\begin{align}
m_Z \simeq \frac{m_W}{\cos \theta_W} ~, 
\label{WZmass-relation}
\end{align}
which coincides with the relation in the SM.

\vskip 5pt
\noindent
\underline{(iii) $\tilde{A}_\mu^{\hat 4}$: $A^{\hat 4}$ tower}

$A_\mu^{\hat 4}$ obeys $(D,D)$ boundary condition and there is no zero mode.  Its spectrum is
determined by 
\begin{align}
\hat{A}^{4}\ \mbox{tower:} \ \  S(1;\lambda)=0.
\label{Eq:Spectra-Gauge-hatA}
\end{align}

\vskip 5pt
\noindent
\underline{(iv) $SU(3)_C$ gluons}

The boundary condition is $(N,N)$ so that
\begin{align}
\mbox{gluon tower:}\ \  C'(1;\lambda)=0.
\label{Eq:Spectra-Gauge-Gluon}
\end{align}

\subsection{$A_z$ components}

The mass spectra of $A_z$ components are the following.
Except for the zero modes, masses are given by $\{ m_n = \xi k \lambda_n \}$.

\vskip 5pt
\noindent
\underline{(i) $A_z^{ab} (1\leq a<b\leq 3), B_z$}

These components satisfy boundary conditions $(D,D)$ so that 
\begin{align}
C'(1;\lambda)=0.
\label{Eq:Spectra-GaugeZ-Z-1}
\end{align}

\vskip 5pt
\noindent
\underline{(ii) $ A_z^{a4}, A_z^{a5} (a=1,2,3)$}

The boundary conditions at $z=1$ are 
\begin{align}
&A_{z}^{a4}=0 ~, ~~
\frac{\partial}{\partial z} \Big( \frac{1}{z}A_{z}^{a5} \Big) =0 ~.
\end{align}
The spectrum is determined by
\begin{align}
S(1;\lambda)C'(1;\lambda)+\lambda\sin^2\theta_H=0 ~.
\label{Eq:Spectra-GaugeZ-Z-3}
\end{align}

\vskip 5pt
\noindent
\underline{(iii) $A_z^{45}$: Higgs tower}

The boundary conditions of $A_z^{45}$ is $(N,N)$ and the spectrum is determined by 
\begin{align}
\mbox{Higgs tower:}\ \ S(1;\lambda)=0.
\label{Eq:Spectra-Gauge-Higgs}
\end{align}
There is a zero mode, which will acquire a mass at the 1-loop level.

\vskip 5pt
\noindent
\underline{(iv) $SU(3)_C ~A_z $}

There are no zero modes. Their components satisfy boundary conditions
$(D,D)$. The mass spectrum is determined by 
\begin{align}
C'(1;\lambda)=0.
\end{align}

\section{Spectrum of fermion fields}

We determine the mass spectra of fermion fields.  It will be seen that
the mass spectrum of quarks and leptons in three generations is reproduced
except for the down quark mass which turns out smaller than the up quark mass ($m_d < m_u$).
To evaluate the effective potential $V_\eff (\theta_H)$ for the AB phase $\theta_H$
one needs to know the mass spectra of the dark fermion fields in (\ref{darkFBC1}) and 
 (\ref{darkFBC2}) as well.  We summarize the result for dark fermions 
in Appendix D for completeness.

In the original gauge the background gauge field in $SO(5)$ is
\begin{align}
 &g A_z^{cl}=
 \frac{g_A}{\sqrt{2}}A_z^{(45)}T^{45}
 =-\theta'(z) T^{45} 
 \label{classicalAz1}
\end{align}
where $\theta (z)$ is defined in (\ref{largeGT2}).
We introduce the following derivatives
\begin{align}
&D_\pm(c) =\pm\frac{\partial}{\partial z}+\frac{c}{z} ~,~~
\hat{D}_\pm(c) = D_\pm(c) \pm i \theta'(z) T^{45}  ~.
\label{derivativeD}
\end{align}
To simplify the notation the bulk mass parameters of  various fields are denoted as
\begin{align}
&c_{Q} = c_{\Psi_{({\bf 3,4})}^{\alpha}} ~,~~
c_{L} = c_{\Psi_{({\bf 1,4})}^{\alpha}} ~,~~
c_{D^\pm} = c_{\Psi_{({\bf 3,1})}^{\pm\alpha}} ~,~~
c_{V^ \pm} = c_{\Psi_{({\bf 1,5})}^{\pm \beta}} ~.
\label{bulkmass1}
\end{align}
We have suppressed generation indices $\alpha, \beta$.
In this paper we consider the cases ${c_{D^+} =\pm c_{D^-}}$ and
$c_{V^+}=\pm c_{V^-}$, for which exact solutions are available.

The components of $SO(5)$ spinor fermions $\Psi_{({\bf 3,4})}$ and $\Psi_{({\bf 1,4})}$  
in the original and twisted gauges are related to each other by
\begin{align}
&\chi = \begin{pmatrix} \cos \onehalf \theta (z) & -i \sin \onehalf \theta (z) \cr
\noalign{\kern 5pt}
-i \sin \onehalf \theta (z) & \cos \onehalf \theta (z)  \end{pmatrix}  \tilde{\chi} ~, 
\label{Eq:twisted-gauge-fermion}
\end{align}
where $\chi$ is given by
\begin{align}
\chi  = 
\begin{pmatrix} u \cr u' \end{pmatrix}, ~
\begin{pmatrix} d \cr d' \end{pmatrix}, ~
\begin{pmatrix} e \cr e' \end{pmatrix}, ~
\begin{pmatrix} \nu \cr \nu' \end{pmatrix}.
\end{align}
$T^{45} = \onehalf \sigma^1$ for these $\chi$'s.

\subsection{Up-type quarks}
\label{Sec:Up-type-quark}

\noindent
\underline{$Q_{\rm EM}=+\frac{2}{3}$: $u, u'$ $(\Psi_{({\bf 3,4})})$}

There are no brane mass terms.  The boundary conditions are given by
$D_+\check{u}_{L}=0$, $\check{u}_{R}=0$, $\check{u}_{L}'=0$, and
$D_-\check{u}_{R}'=0$ at $z=1,z_L$.
The equations of motion in the twisted gauge are 
\begin{align}
- i \delta
\begin{pmatrix}   u_{L}^{\dag} \cr  u_{L}^{\prime\dag} \end{pmatrix} : 
&-kD_-(c_{Q}) \begin{pmatrix}  \tilde{\check u}{}_{R} \cr  \tilde{\check u}{}_{R}' \end{pmatrix}
+\sigma^\mu\partial_\mu
\begin{pmatrix}  \tilde{\check u}{}_{L} \cr  \tilde{\check u}{}_{L}' \end{pmatrix} = 0 ~, \cr
\noalign{\kern 5pt}
 i \delta
\begin{pmatrix}   u_{R}^{\dag} \cr  u_{R}^{\prime\dag} \end{pmatrix} : 
&-kD_+(c_{Q}) \begin{pmatrix}  \tilde{\check u}{}_{L} \cr  \tilde{\check u}{}_{L}' \end{pmatrix}
+ \bar \sigma^\mu\partial_\mu
\begin{pmatrix}  \tilde{\check u}{}_{R} \cr  \tilde{\check u}{}_{R}' \end{pmatrix} = 0 ~.
\label{upEq1}
\end{align}
$(\tilde{\check u},\tilde{\check u}')$ satisfy the same boundary
conditions  at $z=z_L$ as $({\check u}, {\check u}')$ so that one can write,
in terms of basis functions summarized in Appendix B, as
\begin{align}
\begin{pmatrix}  \tilde{\check u}{}_{R} \cr  \tilde{\check u}{}_{R}' \end{pmatrix} 
= \begin{pmatrix}  \alpha_u S_R^Q \cr \alpha_{u'} C_R^Q \end{pmatrix} f_R(x) ~, ~~
\begin{pmatrix}  \tilde{\check u}{}_{L} \cr  \tilde{\check u}{}_{L}' \end{pmatrix}
= \begin{pmatrix}  \alpha_u C_L^Q \cr \alpha_{u'} S_L^Q \end{pmatrix} f_L(x) 
\label{upWave1}
\end{align}
where
$C_{L/R}^{Q} = C_{L/R}(z,\lambda,c_Q)$,
$S_{L/R}^{Q} = S_{L/R}(z,\lambda,c_Q)$,
$\bar{\sigma}\partial f_R(x)=k\lambda f_L(x)$ and
$\sigma\partial f_L(x)=k\lambda f_R(x)$.
Both right- and left-handed modes have the same coefficients
$\alpha_{u}$ and $ \alpha_{u'}$ as a consequence of the equations (\ref{upEq1}).

By making use of (\ref{Eq:twisted-gauge-fermion}) 
the boundary conditions at $z=1$ for the right-handed components
$\check{u}_{R}=0$ and $D_-\check{u}_{R}'=0$ become
\begin{align}
K_u \begin{pmatrix} \alpha_u \cr  \alpha_{u'} \end{pmatrix} = 
\begin{pmatrix} \cos \onehalf \theta_H S_R^{Q} & -i \sin \onehalf \theta_H  C_R^{Q} \cr
 -i \sin \onehalf \theta_H  C_L^{Q} & \cos \onehalf \theta_H S_L^{Q} \end{pmatrix} 
 \begin{pmatrix} \alpha_u \cr  \alpha_{u'} \end{pmatrix}  = 0 ~.
\end{align}
Here $S_{L/R}^{Q} = S_{L/R}(1,\lambda,c_Q)$ etc..
$\det K_u=0$ leads to the equation determining the spectrum;
\begin{align}
S_L^{Q}S_R^{Q}+\sin^2\frac{\theta_H}{2}=0 ~.
\label{Eq:Up-type-quark-mass-spectra}
\end{align}
The mass of the lowest mode (up-type quark)  $m=k\lambda$ is given by
\begin{align}
 m_u=
\begin{cases}
 \pi^{-1}\sqrt{1-4c_{Q}^2} \, \sin \onehalf \theta_H  \, m_{\rm KK}
&\ \mbox{for}\  |c_{Q}| < \frac{1}{2} ~, \cr
\noalign{\kern 5pt}
\pi^{-1}\sqrt{4c_{Q}^2-1} \, z_L^{-|c_{Q}|+0.5} \sin \onehalf \theta_H  \, m_{\rm KK}
&\ \mbox{for}\ |c_{Q}| > \frac{1}{2} ~.
\end{cases}
\label{Eq:Up-type-quark-mass-approximate}
\end{align}
Note that $S_L(z;\lambda, -c)=-S_R(z;\lambda, c)$, $C_L(z;\lambda, -c)=C_R(z;\lambda, c)$.
With given $m_u$, there are two solutions to (\ref{Eq:Up-type-quark-mass-spectra}); 
$c_Q >0$ and $c_Q <0$.

\subsection{Down-type quarks}
\label{Sec:Down-type-quark}

\underline{$Q_{\rm EM}=- \frac{1}{3}$: 
$d,d',D^{\pm}$ $(\Psi_{({\bf 3,4})},\Psi_{({\bf 3,1})}^{\pm})$}

As seen from Table~\ref{Tab:parity}, parity even modes at $y=0$ with $(P_0,P_1) = (+,+)$ are 
$d_{L}$, $d_{R}'$, $D_{L}^{+}$, and $D_{R}^{-}$.
From the action (\ref{fermionAction2}) and  the ${\cal L}_1^m$ term in (\ref{braneFmass1}), 
the equations of motion in the original gauge are given by
\begin{align}
 \begin{matrix}  (a) \cr  (b) \end{matrix}:\
-i\delta  \begin{pmatrix}  d_{L}^{\dag} \cr d_{L}^{\prime\dag} \end{pmatrix} :\
 &  -k\hat{D}_-(c_{Q}) \begin{pmatrix}  \check{d}_{R} \cr \check{d}_{R}^{\prime} \end{pmatrix}
 +\sigma^\mu\partial_\mu \begin{pmatrix}  \check{d}_{L} \cr \check{d}_{L}^{\prime} \end{pmatrix}
 =0 ~, \cr
 \noalign{\kern 5pt}
 \begin{matrix}  (c) \cr  (d) \end{matrix}:\ \hspace{0.75em}
 i\delta  \begin{pmatrix}  d_{R}^{\dag} \cr d_{R}^{\prime\dag} \end{pmatrix} :\
 & \overline{\sigma}^\mu\partial_\mu 
 \begin{pmatrix}  \check{d}_{R} \cr \check{d}_{R}^{\prime} \end{pmatrix}
 -k\hat{D}_+(c_{Q}) \begin{pmatrix}  \check{d}_{L} \cr \check{d}_{L}^{\prime} \end{pmatrix}
 = 2\mu_1\delta(y) \begin{pmatrix} 0 \cr \check{D}_{L}^{+} \end{pmatrix} , \cr
\noalign{\kern 5pt}
 (e):\ \hspace{1em}  -i\delta D_{L}^{+\dag}:\
 & -k\hat{D}_-(c_{D+}) \check{D}_{R}^{+}
 +\sigma^\mu\partial_\mu\check{D}_{L}^{+}
 -\frac{m_{D}^*}{z}\check{D}_{R}^{-}
=2\mu_1^*\delta(y)\check{d}_{R}^{\prime} ~, \cr
\noalign{\kern 5pt}
(f):\ \hspace{1.75em}  i\delta D_{R}^{+\dag}:\
 & \overline{\sigma}^\mu\partial_\mu\check{D}_{R}^{+}
 -k\hat{D}_+(c_{D+})\check{D}_{L}^{+}
 -\frac{m_{D}}{z}\check{D}_{L}^{-} = 0 ~,  \cr
 \noalign{\kern 5pt}
(g):\ \hspace{1em}  -i\delta D_{L}^{-\dag}:\
 &  -k\hat{D}_-(c_{D-})\check{D}_{R}^{-}
 +\sigma^\mu\partial_\mu\check{D}_{L}^{-}
 -\frac{m_{D}^*}{z}\check{D}_{R}^{+}  = 0 ~, \cr
\noalign{\kern 5pt}
(h):\ \hspace{1.75em}  i\delta D_{R}^{-\dag}:\
 & \overline{\sigma}^\mu\partial_\mu\check{D}_{R}^{-}
 -k\hat{D}_+(c_{D-})\check{D}_{L}^{-} -\frac{m_{D}}{z}\check{D}_{L}^{+}
=0 ~.
\label{downEq1}
\end{align}
Note that the mass dimension of each coupling constant and field is
e.g., $[\check{d}_{R/L}]=2$, $[k]=[m_D]=1$ and $[\mu_1]=0$.

The following arguments are parallel to  those in Ref.\ \cite{Furui2016}.
Under the parity transformation around $y=0$, 
$\Psi_+ = d_L, d'_R, D^+_L, D^-_R$ are parity even  whereas
$\Psi_- = d_R, d'_L, D^+_R, D^-_L$  are parity odd. 
Note that $\Psi_-(y) \big|_{-\epsilon}^{+\epsilon} = 2\Psi_-(+\epsilon)$ and
\begin{align}
D_\pm (c) = \frac{e^{-\sigma(y)}}{k} \bigg\{ \pm \frac{\dd}{\dd y} + c \sigma' (y) \bigg\}
\label{derivativeD2}
\end{align}
in the $y$ coordinate.
We integrate the equations for parity odd fields, $(a),(d),(e),(h)$ in (\ref{downEq1}), 
from $y=-\ep$ to $+\ep$ to find
\begin{align}
(a)\ \Rightarrow\ & 
\check{d}_{R}(\epsilon)=0 ~, \cr
(d)\ \Rightarrow\ & 
- 2 \check{d}'_{L}(\epsilon) - 2\mu_1 \check{D}^+_{L}(0)=0 ~, \cr
(e)\ \Rightarrow\ & 
2 \check{D}^+_{R}(\epsilon) - 2\mu_1^* \check{d}'_{R}(0)=0 ~, \cr
(h)\ \Rightarrow\ & 
\check{D}^-_{L}(\epsilon)=0 ~.
\label{Eq:BCs-down-type-quark-1}
\end{align}
For parity-even fields, we evaluate the equations at $y=+\epsilon$ by using 
the relations (\ref{Eq:BCs-down-type-quark-1}).
\begin{align}
(c)\ \Rightarrow\ &
\hat{D}_+(c_Q) \check{d}_L=0 ~, \cr
(b)\ \Rightarrow\ &
 \mu_1 \left[ \hat{D}_-(c_{D+}) \check{D}^+_R +
 \tilde{m}_D^* \check{D}^-_R \right] + \hat{D}_-(c_Q) \check{d}'_R =0 ~, \cr
 (f)\ \Rightarrow\ &
\mu_1^* \hat{D}_+(c_Q) \check{d}'_L - \hat{D}_+(c_{D+})\check{D}^+_L 
 =0~, \cr
 (g)\ \Rightarrow\ &
\hat{D}_-(c_{D-}) \check{D}^-_R + \tilde{m}_D^* \mu_1^* \check{d}'_R =0 ~,
\label{Eq:BCs-down-type-quark-2}
\end{align}
where the equations of motion $(e)$ and $(d)$ at $y=+\epsilon$ have been made use of.
Relations (\ref{Eq:BCs-down-type-quark-1}) and (\ref{Eq:BCs-down-type-quark-2}) specify
the boundary conditions at $z= 1^+$.
We examine the spectrum in two cases, $c_{D^+}=c_{D^-}$ and $c_{D^+}= - c_{D^-}$ below.

\subsubsection*{Case I: $c_{D^+}=c_{D^-}= c_D$}

The BCs at $z=z_L$ are given by 
\begin{align}
 \left\{
 \begin{array}{l}
  d_{R}=0,\\
  D_{+}(c_Q)d_{L}=0,\\
  D_{-}(c_Q)d_{R}'=0,\\
  d_{L}'=0,\\
 \end{array}
 \right.\ \ \
 \left\{
 \begin{array}{l}
  D_{R}^{+}=0,\\
  D_{+}(c_D)D_{L}^{+}=0,\\
  D_{-}(c_D)D_{R}^{-}=0,\\
  D_{L}^{-}=0.\\
 \end{array}
 \right.
\label{Eq:down-type-quark-BC-z=zL}
\end{align}
In the twisted gauge, the BCs in  (\ref{Eq:down-type-quark-BC-z=zL}) are satisfied by 
mode functions  in (\ref{fermionF2}) and (\ref{MfermionWave2}) so that one can write as
\begin{align}
\begin{pmatrix}  \widetilde{\check{d}}_{R} \cr   \widetilde{\check{d'}}_{R} \cr 
\widetilde{\check D}{}_{R}^{+} \cr   \widetilde{\check D}{}_{R}^{-} \end{pmatrix}
&=\begin{pmatrix}   \alpha_{d} S_R(z;\lambda, c_Q) \cr   
\mynoalign
\alpha_{d'}C_R(z;\lambda, c_Q) \cr
\mynoalign
a_{d}{\cal S}_{R2}(z;\lambda, c_D, \tilde m_D)+b_{d}{\cal S}_{R1}(z;\lambda, c_D, \tilde m_D) \cr 
\mynoalign
a_{d}{\cal C}_{R1}(z;\lambda, c_D, \tilde m_D)+b_{d}{\cal C}_{R2}(z;\lambda, c_D, \tilde m_D) \cr 
\end{pmatrix} , \cr
\noalign{\kern 10pt}
\begin{pmatrix}  \widetilde{\check{d}}_{L} \cr \widetilde{\check{d'}}_{L} \cr
\widetilde{\check D}{}_{L}^{+} \cr  \widetilde{\check D}{}_{L}^{-} \end{pmatrix}
&=\begin{pmatrix}  \alpha_{d} C_L(z; \lambda, c_Q)\cr
\mynoalign
\alpha_{d'}S_L(z; \lambda, c_Q)\cr
\mynoalign
a_{d}{\cal C}_{L2}(z;\lambda, c_D, \tilde m_D)+b_{d}{\cal C}_{L1}(z;\lambda, c_D, \tilde m_D) \cr
\mynoalign
a_{d}{\cal S}_{L1}(z;\lambda, c_D, \tilde m_D)+b_{d}{\cal S}_{L2}(z;\lambda, c_D, \tilde m_D) \cr
\end{pmatrix} ,
\label{Eq:solution-down-type-quark} 
\end{align}
where $\alpha_{d}$, $\alpha_{d'}$, $a_{d}$, $b_{d}$ are parameters. 


Boundary conditions at $z=1^+$ for the left-handed fields 
$\check{d}_L, \check{d}'_L, \check{D}^+_L, \check{D}^-_{L}$ are found from 
Eqs.~(\ref{Eq:BCs-down-type-quark-1}) and (\ref{Eq:BCs-down-type-quark-2}) to be
\begin{align}
 (c): ~&  \lambda \Big( \cos\frac{\theta_H}{2} \alpha_{d} S_R^Q - i \sin\frac{\theta_H}{2}
\alpha_{d'} C_R^Q \Big) = 0 ~ , \cr
\noalign{\kern 5pt}
 (d): ~&  -i \sin\frac{\theta}{2} \alpha_{d} C_L^Q
 +\cos\frac{\theta}{2} \alpha_{d'} S_L^Q
 +\mu_1\Big( a_{d}{\cal C}_{L2}^{D}+b_{d}{\cal C}_{L1}^{D} \Big) = 0 ~, \cr
 \noalign{\kern 5pt}
 (f):~& \lambda \mu_1^* \Big(-i\sin\frac{\theta_H}{2} \alpha_{d} S_R^Q
 + \cos\frac{\theta_H}{2} \alpha_{d'} C_R^Q \Big) \cr
 \noalign{\kern 5pt}
 &\hskip 2.cm
 -\lambda\left(a_{d}{\cal S}_{R2}^{D}+b_{d}{\cal S}_{R1}^{D}\right)
 +\tilde{m}_D\left(a_{d}{\cal S}_{L1}^{D}+b_{d}{\cal S}_{L2}^{D}\right) = 0 ~, \cr
\noalign{\kern 5pt} 
 (h):  ~ &a_{d}{\cal S}_{L1}^{D}+b_{d}{\cal S}_{L2}^{D} = 0 ~,
\label{Eq:BCs-down-type-quark-dhcf-I'}
\end{align}
where $S_{L/R}^Q:=S_{L/R}(z=1;\lambda, c_Q)$,
${\cal S}_{L/Rj}^D:={\cal S}_{L/Rj}(z=1;\lambda, c_D, \tilde m_D)$ etc..
Conditions in (\ref{Eq:BCs-down-type-quark-dhcf-I'}) are summarized as 
\begin{align}
{\cal M}_L^D V^D = 
\begin{pmatrix}
\cos\frac{\theta_H}{2} S_R^Q & -i \sin\frac{\theta_H}{2} C_R^Q & 0 & 0 \\
 -i\sin\frac{\theta_H}{2} C_L^Q & \cos\frac{\theta_H}{2} S_L^Q
 & \mu_1 {\cal C}_{L2}^{D} & \mu_1 {\cal C}_{L1}^{D}\\
-i\mu_1^* \sin\frac{\theta_H}{2} S_R^Q &
\mu_1^* \cos\frac{\theta_H}{2} C_R^Q & 
-{\cal S}_{R2}^{D} & -{\cal S}_{R1}^{D} \\
0 & 0 & {\cal S}_{L1}^{D} & {\cal S}_{L2}^{D}\\
\end{pmatrix}
\begin{pmatrix}
 \alpha_d\\
 \alpha_{d'}\\
 a_{d}\\
 b_{d}\\
\end{pmatrix}
= 0 ~.
\label{downBC1}
\end{align}
The mass spectrum is determined by 
\begin{align}
&\det{\cal M}_L^D=\Big( S_L^Q S_R^Q +\sin^2\frac{\theta_H}{2} \Big)
 \big({\cal S}_{L1}^{D}{\cal S}_{R1}^{D}
 -{\cal S}_{L2}^{D}{\cal S}_{R2}^{D}\big) \cr
 \noalign{\kern 5pt}
 &\hskip 3.cm
+|\mu_1|^2 C_R^Q S_R^Q
 \left({\cal S}_{L1}^{D}{\cal C}_{L1}^{D} -{\cal S}_{L2}^{D}{\cal C}_{L2}^{D}\right)=0 ~.
\label{Eq:Down-type-quark-mass-spectra}
\end{align}
Note the relations (\ref{MfermionSpectrum2}). 

To lift the degeneracy between the up-type and down-type quark masses, 
the $\mu_1$ term in (\ref{Eq:Down-type-quark-mass-spectra}) is necessary.  
Its coefficient contains the factor $C_R^Q = C_R(1; \lambda, c_Q)$.
For the first and second generations $|c_Q| = |c_u|, |c_c| > \onehalf$.
For $\lambda z_L \ll 1$, $C_R(1; \lambda, c) \sim z_L^{-c} \ll 1$ for $c>\onehalf$ and
$C_R(1; \lambda, c)   \gg 1$ for $c< - \onehalf$.  The detailed study shows that
with $c > \onehalf$ Eq.\ (\ref{Eq:Down-type-quark-mass-spectra}) necesarrily yields the first KK mode
with a mass much less than $m_\KK$, 
which contradicts with observation.  One needs to take $c_u, c_c <0$.  
For the third generation $|c_t| < \onehalf$, and this problem does not show up.

Consider  Case I, $c_{D^+}=c_{D^-}=c_D>0$,  with $\tilde{m}_D>1/2$ and $c_D -\tilde{m}_D>1/2$.
The up-type quark mass $m_u$ for $|c_Q| > \onehalf$  is approximately given by 
\begin{align}
 m_u=\lambda_u z_L\simeq \sqrt{4c_Q^2-1}z_L^{-|c_Q| +\frac{1}{2}}
 \sin\frac{\theta_H}{2}
\end{align}
from Eq.~(\ref{Eq:Up-type-quark-mass-spectra}). Substituting
\begin{align}
 S_L^QS_R^Q+\sin^2\frac{\theta_H}{2} &\simeq
 -\frac{\lambda^2 z_L^{2|c_Q|+1}}{4c_Q^2-1}+\sin^2\frac{\theta_H}{2}
 \simeq-\frac{(\lambda^2-\lambda_u^2)z_L^{2|c_Q|+1}}{4c_Q^2-1},
 \nonumber\\
 \noalign{\kern 5pt}
 {\cal S}_{L1}^Q{\cal S}_{R1}^Q-{\cal S}_{L2}^Q{\cal S}_{R2}^Q
 &\simeq  z_L^{2\tilde{m}_D}-\lambda^2z_L^{2c_D+1}
 \left(\frac{1}{4c_D^2-(2\tilde{m}_D+1)^2}
 +\frac{1}{4c_D^2-(2\tilde{m}_D-1)^2}\right),
 \nonumber\\
\noalign{\kern 5pt}
 C_R^QS_R^Q
 &\simeq 
 \left\{
 \begin{array}{cc}
\myfrac{\lambda}{2c_Q-1}&\ \mbox{for}\ c_Q>0,\\
\noalign{\kern 5pt}
\myfrac{\lambda z_L^{2|c_Q|+1}}{2|c_Q|+1}&\ \mbox{for}\ c_Q<0,\\
  \end{array}
 \right.
 \nonumber\\
\noalign{\kern 5pt}
 {\cal S}_{L1}^Q{\cal C}_{L1}^Q-{\cal S}_{L2}^Q{\cal C}_{L2}^Q
 &\simeq  -2\lambda z_L^{2c_D+1}
 \left(\frac{1}{2(c_D+\tilde{m}_D)+1}
 +\frac{1}{2(c_D-\tilde{m}_D)+1}\right),
\end{align}
into ${\cal M}_L^D$ in Eq.~(\ref{Eq:Down-type-quark-mass-spectra}),
we find
\begin{align}
 \det{\cal M}_L^D=&
 -\frac{(\lambda^2-\lambda_u^2)z_L^{2|c_Q|+1}}{4c_Q^2-1}
 \left(z_L^{2\tilde{m}_D}-\lambda^2z_L^{2c_D+1} A
 \right)
 \nonumber\\
 &+|\mu_1|^2\left\{
 \begin{array}{c}
  \displaystyle \frac{1}{2c_Q-1}\\
   \noalign{\kern 5pt}
  \displaystyle \frac{z_L^{2|c_Q|+1}}{2|c_Q|+1}\\
  \end{array}
 \right\}
 \left(-2\lambda^2 z_L^{2c_D+1}\right)B=0
 \ \ \mbox{for}\ \
 \left\{
 \begin{array}{cc}
  c_Q>0\\
   \noalign{\kern 10pt}
  c_Q<0\\
 \end{array}
 \right. ,
\end{align}
\label{downSpectrum1}
where 
\begin{align}
 A&=\frac{1}{4c_D^2-(2\tilde{m}_D+1)^2}
 +\frac{1}{4c_D^2-(2\tilde{m}_D-1)^2}>0 ~,\\
 \noalign{\kern 5pt}
 B&=\frac{1}{2(c_D+\tilde{m}_D)+1}
 +\frac{1}{2(c_D-\tilde{m}_D)+1}>0 ~. 
 \label{downSpectrum2}
\end{align}
Both $A$ and $B$ are $O(1)$.  If $z_L^{2\tilde{m}_D}\gg \lambda^2z_L^{2c_D+1}$, then
it follows from (5.22) that
\begin{align}
\lambda^2\simeq
\left\{
\begin{array}{lll}
\displaystyle \frac{\lambda_u^2}{1+2|\mu_1|^2(2c_Q+1)z_L^{-2c_Q+2c_D-2\tilde{m}_D}B}
& < \lambda_u^2 & \mbox{for}\ c_Q> \onehalf ~,\\
\noalign{\kern 10pt}
\displaystyle \frac{\lambda_u^2}{1+2|\mu_1|^2(2|c_Q|-1)z_L^{2c_D-2\tilde{m}_D+1}B}
   & < \lambda_u^2 & \mbox{for}\ c_Q< - \onehalf  ~.\\
 \end{array}
 \right.
\label{downSpectrum2b}
\end{align}
In other words the spectrum for the second generation $m_s < m_c$ can be
reproduced with appropriate $\mu_1$, $c_Q$ and $\tilde m_D$.

Indeed, one can show that the smallest value of $\lambda^2$ determined from 
Eq.\ (\ref{Eq:Down-type-quark-mass-spectra}) necessarily becomes smaller than $\lambda_u^2$
with general $\mu_1 \not= 0$, $c_Q$ and $\tilde m_D$.
For $\lambda z_L \ll 1$, Eq.\ (\ref{Eq:Down-type-quark-mass-spectra}) reduces to the form
$(\lambda^2 - \lambda_u^2) (\lambda^2 - a) - b |\mu_1|^2 \lambda^2 =0$ where
$a \gg  \lambda_u^2$ and $b >0$. Consequently the two roots $\lambda^2 = \lambda^2_\pm$
satisfy $\lambda^2_- < \lambda_u^2$ and $\lambda^2_+ > a$.
This implies that the spectrum $m_d > m_u$ cannot be realized at the tree level 
in the current scheme.   It is left for future investigation to find a solution to this problem.

\ignore{
In the opposite case, $z_L^{2\tilde{m}_D}\ll \lambda^2z_L^{2c_D+1}$,  one has
\begin{align}
 \lambda^2&=\lambda_u^2+2|\mu_1|^2\frac{B}{A}
 \left\{
 \begin{array}{lll}
  (2c_Q+1)z_L^{-2c_Q-1}   & >\lambda_u^2 & \mbox{for}\ c_Q>\onehalf ,  \\
  \noalign{\kern 10pt}
  2|c_Q|-1   &  >\lambda_u^2 & \mbox{for}\ c_Q<- \onehalf .\\
 \end{array}
 \right.
 \label{downSpectrum3}
 \end{align}
Note that in this case
\begin{align}
\left(\frac{m}{m_u}\right)^2 \sin^2\frac{\theta_H}{2}
\gg z_L^{2|c_Q|-2c_D+2\tilde{m}_D}.
\end{align}
Hence, with $c_D-\tilde{m}_D>|c_Q|$ one can reproduce the spectrum
for the first generation, $m_d > m_u$.
}

\begin{figure}[tbh]
\begin{center}
\includegraphics[bb=26 4 352 224, height=6cm]{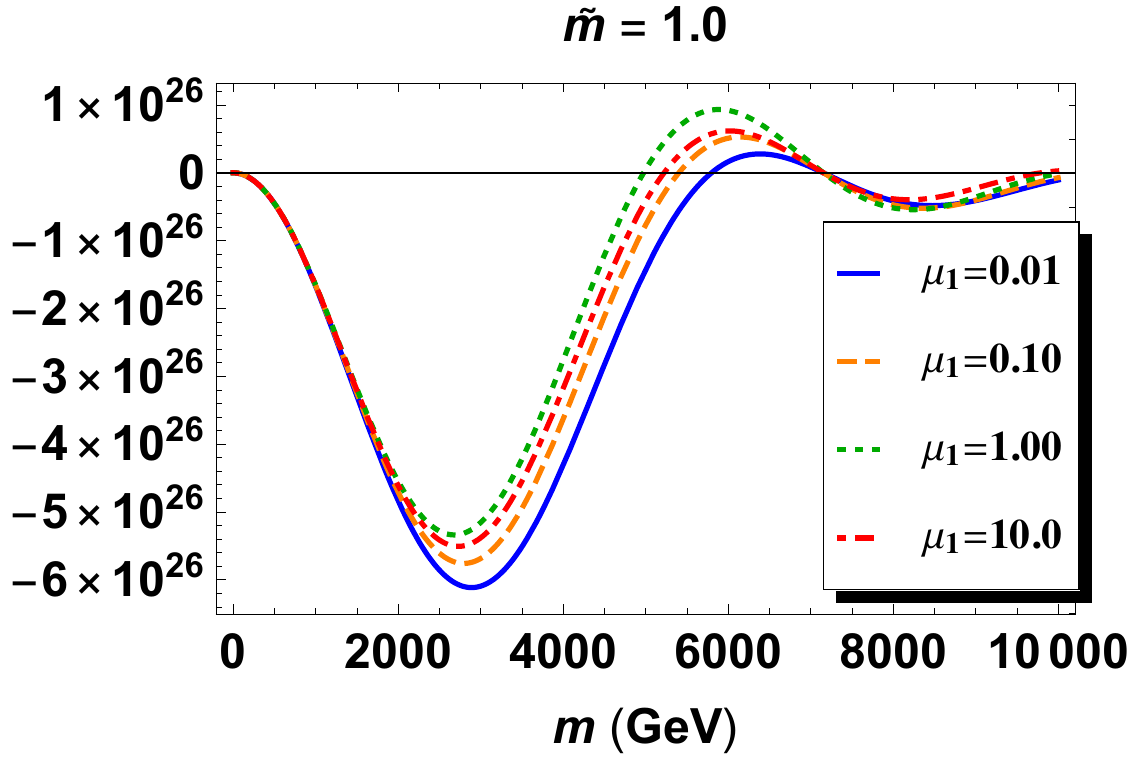}
\end{center}
\vskip -10pt
\caption{Spectrum of strange quark tower.  $\det {\cal M}_L^D$ in 
Eq.\  (\ref{Eq:Down-type-quark-mass-spectra}) is plotted as a function of $m = k\lambda$ 
for $\tilde m_D = 1.0$ and various values of $\mu_1$.
The mass spectrum $\{ m_n = k \lambda_n \}$ is determined by  roots of
$\det {\cal M}_L^D = 0$.
$ m_\KK = 8062\, $GeV.
}
\label{Figure:strange1}
\end{figure}

Typical values of the parameters reproducing the quark mass spectrum (except for $m_d$)
are tabulated in Table~\ref{Tab:quarkSpectrum}.
$\det {\cal M}_L^D$  in  Eq.\  (\ref{Eq:Down-type-quark-mass-spectra}) for the second generation
is plotted as a function of $\lambda$ 
for $\tilde m_D = 1.0$ and various values of $\mu_1$ in Figure~\ref{Figure:strange1}.

\begin{table}[tbh]
\renewcommand{\arraystretch}{1.2}
\begin{center}
\caption{Parameters which reproduce the spectrum of quarks
for $\theta_H=0.15$, $z_L=10^{10}$.  $m_\KK = 8.062 \,$TeV.
The masses of the 1st KK modes of up-type and down-type quarks  are also shown.
$c_u, c_c < 0$ for the reason described below Eq.\  (\ref{Eq:Down-type-quark-mass-spectra}).
The values $m_u = 1.27\,$MeV, $m_s = 55\,$MeV, $m_c = 619\,$MeV, $m_b = 2.89\,$GeV,
and $m_t=171.17\,$GeV have been used.  $m_d = 0.9\, m_u$ has been used for the first generation.
}
\vskip 10pt
\begin{tabular}{|c|c|c|c|c|c|c|}
\hline
Quarks & $c_Q$ &$\mu_1$ & $c_D$ & $\tilde m_D$ &$m_{d^{(1)}}$ &$m_{u^{(1)}}$\\
& &  &  &  &(TeV) &(TeV)\\
\hline
$(u,d)$ &$-1.044$ &$0.01$ &$0.6194 $ &$1.0$ &4.59 &8.23\\
\cline{3-6}
&&$0.1$ &$0.4612$ &$1.0$ &4.80&\\
\hline
$(c,s)$ &$-0.7546 $ &$0.1$ &$0.6808$ &$1.0$ &$5.40$ &7.16\\
\cline{3-6}
&&$10.$ &$0.0949$ &$1.0$ &$5.22$&\\
\hline
$(t,b)$ &$+0.2287 $ &$0.1$ &$0.5838$ &$0.1$ &2.84 &7.20\\
\cline{3-6}
&&$10.$ &$0.3791$ &$0.1$ &2.84&\\
\cline{2-6}
&$-0.2287 $ &$0.1$ &$1.044$ &$1.0$ &5.06 &\\
\cline{3-6}
&&$10.$ &$0.8352$ &$1.0$ &5.06&\\
\hline
\end{tabular}
\label{Tab:quarkSpectrum}
\end{center}
\end{table}

\subsubsection*{Case II:  $c_{D^+}=-c_{D^-}=c_D$}

The BCs at $z=z_L$ are given by 
\begin{align}
 \left\{
 \begin{array}{l}
  d_{R}=0,\\
  D_{+}(c_Q)d_{L}=0,\\
  D_{-}(c_Q)d_{R}'=0,\\
  d_{L}'=0,\\
 \end{array}
 \right.\ \ \
 \left\{
 \begin{array}{l}
  D_{R}^{+}=0,\\
  D_{+}(c_D)D_{L}^{+}=0,\\
  D_{+}(c_D)D_{R}^{-}=0,\\
  D_{L}^{-}=0.\\
 \end{array}
 \right.
\label{Eq:down-type-quark-BC-z=zL-Case-II}
\end{align}
In the twisted gauge, the BCs in
Eq.~(\ref{Eq:down-type-quark-BC-z=zL-Case-II}) are satisfied by mode functions
in (\ref{fermionF2}) and (\ref{M2fermionWave2}) so that one can write as
\begin{align}
\begin{pmatrix}  \widetilde{\check{d}}_{R} \cr   \widetilde{\check{d'}}_{R} \cr 
\widetilde{\check D}{}_{R}^{+} \cr    \widetilde{\check D}{}_{R}^{-} \end{pmatrix}
&=\begin{pmatrix}   \alpha_{d} S_R(z;\lambda, c_Q) \cr  
\mynoalign
\alpha_{d'}C_R(z;\lambda, c_Q) \cr
\mynoalign
a_{d} \hat{\cal S}_{R2}(z;\lambda, c_D, \tilde m_D)+b_{d} \hat{\cal S}_{R1}(z;\lambda, c_D, \tilde m_D) \cr 
\mynoalign
a_{d} \hat{\cal C}_{L1}(z;\lambda, c_D, \tilde m_D)+b_{d} \hat{\cal C}_{L2}(z;\lambda, c_D, \tilde m_D) \cr 
\end{pmatrix} , \cr
 \noalign{\kern 10pt}
\begin{pmatrix}  \widetilde{\check{d}}_{L} \cr \widetilde{\check{d'}}_{L} \cr
\widetilde{\check D}{}_{L}^{+} \cr  \widetilde{\check D}{}_{L}^{-} \end{pmatrix}
&=\begin{pmatrix}  \alpha_{d} C_L(z; \lambda, c_Q)\cr   
\mynoalign
\alpha_{d'}S_L(z; \lambda, c_Q)\cr
\mynoalign
a_{d} \hat{\cal C}_{L2}(z;\lambda, c_D, \tilde m_D)+b_{d} \hat{\cal C}_{L1}(z;\lambda, c_D, \tilde m_D) \cr
\mynoalign
- a_{d} \hat{\cal S}_{R1}(z;\lambda, c_D, \tilde m_D) -b_{d} \hat{\cal S}_{R2}(z;\lambda, c_D, \tilde m_D) \cr
\end{pmatrix} ,
\label{Eq:solution-down-type-quark-Case-II} 
\end{align}
where $\alpha_{d}$, $\alpha_{d'}$, $a_{d}$, $b_{d}$ are parameters.

From Eqs.~(\ref{Eq:BCs-down-type-quark-1}) and (\ref{Eq:BCs-down-type-quark-2}),
we find the boundary conditions at $z=1$ for the left-handed fields.
The manipulation is similar to that in Case I.   The difference appears only for terms involving
$D_{L/R}^-$.  It is straightforward to see

\begin{align}
{\cal M}_L^D \, V^D = 
\begin{pmatrix}
\cos\frac{\theta_H}{2} S_R^Q & -i \sin\frac{\theta_H}{2} C_R^Q & 0 & 0 \\
 -i\sin\frac{\theta_H}{2} C_L^Q & \cos\frac{\theta_H}{2} S_L^Q
 & \mu_1 \hat{\cal C}_{L2}^{D} & \mu_1 \hat{\cal C}_{L1}^{D}\\
-i\mu_1^* \sin\frac{\theta_H}{2} S_R^Q &
\mu_1^* \cos\frac{\theta_H}{2} C_R^Q & 
-\hat{\cal S}_{R2}^{D} & - \hat{\cal S}_{R1}^{D} \\
0 & 0 & \hat{\cal S}_{R1}^{D} & \hat{\cal S}_{R2}^{D}\\
\end{pmatrix}
\begin{pmatrix}
 \alpha_d\\
 \alpha_{d'}\\
 a_{d}\\
 b_{d}\\
\end{pmatrix}
= 0 
\label{down2BC1}
\end{align}
where $S_{L/R}^Q:=S_{L/R}(z=1;\lambda, c_Q)$,
$\hat{\cal S}_{L/Rj}^D =\hat{\cal S}_{L/Rj}(z=1;\lambda, c_D, \tilde m_D)$ etc..
The spectrum is determined by
\begin{align}
&\det{\cal M}_L^D=
\Big(S_L^Q S_R^Q +\sin^2\frac{\theta_H}{2} \Big)
 \Big\{(\hat{\cal S}_{R1}^{D})^2-(\hat{\cal S}_{R2}^{D})^2\Big\} \cr
\noalign{\kern 10pt}
&\hskip 3.cm
 +|\mu_1|^2 C_R^Q S_R^Q
(\hat{\cal S}_{R1}^{D} \hat{\cal C}_{L1}^{D}
 - \hat{\cal S}_{R2}^{D} \hat{\cal C}_{L2}^{D} )=0 ~.
\label{Eq:Down-type-quark-mass-spectra-II}
\end{align}
Note the relation (\ref{M2fermionIdentity2}).

For $|c_Q |, \hat c > \onehalf$ , $c_D  >0$ and $\lambda z_L \ll 1$, 
we have
\begin{align}
& S_L^QS_R^Q+\sin^2\frac{\theta_H}{2}
 \simeq-\frac{(\lambda^2-\lambda_u^2)z_L^{2|c_Q|+1}}{4c_Q^2-1} ~, \cr
 \noalign{\kern 5pt}
&(\hat{\cal S}_{R1}^D)^2-(\hat{\cal S}_{L2}^D)^2
 \sim-\alpha_+^2z_L^{2\hat{c}} ~, \cr
\noalign{\kern 5pt}
&\hat{\cal S}_{R1}^D {\cal C}_{L1}^D - \hat{\cal S}_{R2}^D {\cal C}_{L2}^D
\sim (1+\alpha_+^2) \, \frac{\lambda z_L^{2\hat{c}}}{2\hat{c} - 1} ~,
\label{down2Approx1}
\end{align}
so that
\begin{align}
&\det{\cal M}_L^D \simeq
 -\frac{(\lambda^2-\lambda_u^2)z_L^{2|c_Q|+1}}{4c_Q^2-1}
 \cdot \left(-\alpha_+^2 z_L^{2\hat{c}}\right) \cr
 \noalign{\kern 5pt}
 &\hskip 1.cm
 +|\mu_1|^2\left\{
 \begin{array}{c}
  \displaystyle \frac{1}{2c_Q-1}\\
  \noalign{\kern 5pt}
  \displaystyle \frac{z_L^{2|c_Q|+1}}{2|c_Q|+1}\\
  \end{array}
 \right\}
 (1+\alpha_+^2)\frac{\lambda^2 z_L^{2\hat{c}}}{2\hat{c}+1}=0
 \ \ \mbox{for}\ \
 \left\{
 \begin{array}{l}
  c_Q> \onehalf\\
  \noalign{\kern 10pt}
  c_Q< - \onehalf\\
 \end{array}
 \right..
 \label{down2Spectrum1}
\end{align}
Thus we find
\begin{align}
 \lambda^2 \left[1+\frac{|\mu_1|^2}{2\hat{c}-1}
 \left\{
 \begin{array}{c}
  \displaystyle (2c_Q+1)z_L^{-2|c_Q|-1}\\
  \noalign{\kern 5pt}
  \displaystyle 2|c_Q|-1\\
  \end{array}
 \right\}
 \frac{1+\alpha_+^2}{\alpha_+^2}
 \right]=\lambda_u^2 
 \ \ \mbox{for}\ \
 \left\{
 \begin{array}{l}
  c_Q> \onehalf \\
  \noalign{\kern 5pt}
  c_Q< - \onehalf \\
 \end{array}
 \right..
\end{align}
We observe that $\lambda^2<\lambda_u^2$ so that $m_d > m_u$ cannot be realized
with this parametrization, as in Case I.

\subsection{Charged lepton}
\label{Sec:Charged-lepton}

\underline{$Q_{\rm EM}=-1$: $e,e'$ $(\Psi_{({\bf 1,4})})$}

In general $\Psi_{({\bf 1,4})}$  may couple with $\Psi_{({\bf 1,5})}^\pm$
through the brane interaction ${\cal L}_2^m$ in (\ref{braneFmass1}).
We suppose that $\tilde \mu_2$ there is sufficiently small so that
the effect of ${\cal L}_2^m$  can be ignored.
In this case the equations and boundary conditions for $e,e'$ take the same form
as those for $u, u'$.   Mode functions and boundary conditions are summarized as
\begin{align}
&\begin{pmatrix}  \tilde{\check e}{}_{R} \cr  \tilde{\check e}{}_{R}' \end{pmatrix} 
= \begin{pmatrix}  \alpha_e S_R (z,\lambda,c_L)\cr \alpha_{e'} C_R (z,\lambda,c_L)\end{pmatrix} , \cr
\noalign{\kern 5pt}
&\begin{pmatrix}  \tilde{\check e}{}_{L} \cr  \tilde{\check e}{}_{L}' \end{pmatrix}
= \begin{pmatrix}  \alpha_e C_L (z,\lambda,c_L)\cr \alpha_{e'} S_L(z,\lambda,c_L) \end{pmatrix} , \cr
\noalign{\kern 5pt}
&\begin{pmatrix} \cos \onehalf \theta_H S_R^{L} & -i \sin \onehalf \theta_H  C_R^{L} \cr
 -i \sin \onehalf \theta_H  C_L^{L} & \cos \onehalf \theta_H S_L^{L} \end{pmatrix} 
 \begin{pmatrix} \alpha_e \cr  \alpha_{e'} \end{pmatrix}  = 0 ~,
\label{electronWave1}
\end{align}
where $S_{L/R}^{L} = S_{L/R}(1,\lambda,c_L)$ etc. in the last equation.
The mass spectrum is determined by
\begin{align}
S_L^{L}S_R^{L}+\sin^2\frac{\theta_H}{2}=0 ~.
\label{electronMass1}
\end{align}
The mass of the lowest mode (charged lepton)  $m=k\lambda$ is given by
\begin{align}
 &m_e=
\pi^{-1}\sqrt{4c_{L}^2-1} \, z_L^{-|c_{L}|+0.5} \sin \onehalf \theta_H  \, m_{\rm KK} ~.
\label{electronMass2}
\end{align}
Note $|c_L| > \onehalf$.

\subsection{Neutrino}
\label{Sec:Neutrino}

\underline{$Q_{\rm EM}=0$: $\nu,\nu',  \chi$ $(\Psi_{({\bf 1,4})(-3)}, \chi)$}

As mentioned above, 
we assume that ${\cal L}_2^m$  can be ignored.
The brane interaction ${\cal L}_3$ in (\ref{Eq:Action-brane-fermion}) yields the coupling  between $\nu'$ 
and $\chi$,  ${\cal L}_3^m$ in  (\ref{braneFmass1}).    It leads to the gauge-Higgs seesaw
mechanism.\cite{HosotaniYamatsu2017}
In the present paper we treat the case in which all brane interactions are diagonal in generations.  
In particular we set $M^{\alpha \beta} = - M_\alpha \delta^{\alpha \beta}$ in (\ref{braneFmass1}).

Equations of motion are given by
\begin{align}
\begin{matrix}  (a) \cr  (b) \end{matrix}:\quad
-i\delta  \begin{pmatrix}  \nu_{L}^{\dag} \cr \nu_{L}^{\prime\dag} \end{pmatrix} :\
 &  -k\hat{D}_-(c_{L}) \begin{pmatrix}  \check{\nu}_{R} \cr \check{\nu}_{R}^{\prime} \end{pmatrix}
 +\sigma^\mu\partial_\mu \begin{pmatrix}  \check{\nu}_{L} \cr \check{\nu}_{L}^{\prime} \end{pmatrix}
 =0 ~, \cr
\noalign{\kern 5pt}
 \begin{matrix}  (c) \cr  (d) \end{matrix}:\quad  \hspace{0.7em}
 i\delta  \begin{pmatrix}  \nu_{R}^{\dag} \cr \nu_{R}^{\prime\dag} \end{pmatrix} :\
 & \overline{\sigma}^\mu\partial_\mu 
 \begin{pmatrix}  \check{\nu}_{R} \cr \check{\nu}_{R}^{\prime} \end{pmatrix}
 -k\hat{D}_+(c_{L}) \begin{pmatrix}  \check{\nu}_{L} \cr \check{\nu}_{L}^{\prime} \end{pmatrix}
 = \frac{2 m_B}{\sqrt{k}} \, \delta(y) \begin{pmatrix} 0 \cr \eta \end{pmatrix} , \cr
\noalign{\kern 5pt}
(e):\quad \hspace{1.9em}
 i\delta\eta^\dag  ~~
:\ &
\Big\{ {\sigma}^\mu\partial_\mu\eta
- \frac{m_B}{\sqrt{k}} \nu_{R}' + M\eta^c \Big\} \, \delta(y)=0 ~.
\label{neutrinoEq1}
\end{align}
$\nu_{R}$ and $\nu_{L}'$ are parity-odd at $y=0$, whereas $\nu_{L}$ and $\nu_{R}'$
are parity-even.  We integrate the equations
$(a)$, $(d)$  in the vicinity of $y=0$ 
and evaluate the equations $(b)$, $(c)$ at $y= + \ep$ to find 
boundary conditions at $y=+ \ep$ as
\begin{align}
(a)\ \Rightarrow\ & \check{\nu}_{R}(x,\epsilon)  =0 ~, \cr
\noalign{\kern 5pt}
(d)\ \Rightarrow\ &  -\check{\nu}_{L}'(x,\epsilon)=  + \frac{m_B}{ \sqrt{k}} \, \eta(x) ~ ,\cr
\noalign{\kern 5pt}
(b)\ \Rightarrow\ &
-\hat{D}_-(c_{L})\check{\nu}_{R}' - \frac{m_B^2}{k^2}\check{\nu}_{R}'
+ \frac{m_B M}{k^{3/2}}\eta^c = 0 ~, \cr
\noalign{\kern 5pt}
(c)\ \Rightarrow\ &
\hat{D}_+(c_{L})\check{\nu}_{L} =0 ~.
\label{neutrinoBC1}
\end{align}
Boundary conditions at $z=z_L$ are given by
${D}_+(c_{L})\check{\nu}_{L} = \check{\nu}_{R} =0$ and 
$\check{\nu}_{L}' = {D}_-(c_{L})\check{\nu}_{R}' =0$.

Mode functions of these fields in the twisted gauge can be written as
\begin{align}
&\begin{pmatrix} \tilde{\check{\nu}}_R \cr \tilde{\check{\nu}}_R' \cr \eta^c \end{pmatrix}
= \begin{pmatrix} \alpha_\nu S_R^L \cr  i\alpha_{\nu'} C_R^L \cr   \mp i\alpha_\eta^*/\sqrt{k} \end{pmatrix}
 f_{\pm R}(x) ~, ~
\begin{pmatrix} \tilde{\check{\nu}}_L \cr \tilde{\check{\nu}}_L' \cr  \eta \end{pmatrix}
 = \begin{pmatrix} \alpha_\nu C_L^L \cr  i\alpha_{\nu'} S_L^L \cr  i\alpha_\eta/\sqrt{k} \end{pmatrix}
f_{\pm L}(x) ~, \cr
\noalign{\kern 10pt}
&\bar{\sigma}^\mu\partial_\mu f_{\pm R}(x)=k\lambda f_{\pm L}(x) ~,~~ 
{\sigma}^\mu\partial_\mu f_{\pm L}(x) =k\lambda f_{\pm R}(x) ~,\cr
\noalign{\kern 5pt}
&f_{\pm L}(x)^c = e^{i\delta_C}\sigma^2 f_{\pm L}(x)^* = \pm f_{\pm R}(x) 
\label{neutrinoWave1}
\end{align}
where
$S_{L/R}^L=S_{L/R}(z;\lambda,c_{L})$ and  $C_{L/R}^L=C_{L/R}(z;\lambda,c_{L})$,
and $\delta_C$ is defined in Eq.~(\ref{Majorana1}).
Explicit forms of $f_{\pm L/R}$ are given in Appendix C. 
One can take $\alpha_\nu , \alpha_{\nu'}, \alpha_\eta$ to be real.
In this case  $\sigma^\mu \partial_\mu \eta  = \mp k\lambda \eta^c$ is
satisfied so that the equation $(e)$  in Eq.~(\ref{neutrinoEq1}) implies that 
\begin{align}
\frac{m_B}{\sqrt{k}} \, \check{\nu}_{R}'\Big|_{y=0} - (M \mp k\lambda )\eta^c=0 ~.
\label{neutrinoBC2}
\end{align}
With this identity the third relation in Eq.~(\ref{neutrinoBC1}) can be 
rewritten as 
\begin{align}
&\hat{D}_-(c_{L})\check{\nu}_{R}^{\prime}
\mp \frac{m_B\lambda}{\sqrt{k}} \, \eta^c =0 ~.
\label{neutrinoBC3}
\end{align}

Substituting (\ref{neutrinoWave1}) into (\ref{neutrinoBC1}), one finds
\begin{align}
 K_\nu \begin{pmatrix}   \alpha_\nu \cr  \alpha_{\nu'} \cr \alpha_\eta \end{pmatrix} 
=\begin{pmatrix} 
\cos \frac{\theta_H}{2}S_R^L &\sin \frac{\theta_H}{2}C_R^L &0 \cr
\noalign{\kern 5pt}
-\sin \frac{\theta_H}{2}C_L^L &\cos \frac{\theta_H}{2}S_L^L& \mfrac{m_B}{k} \cr
\noalign{\kern 5pt}  
m_B \sin \frac{\theta_H}{2}S_R^L &- m_B \cos\frac{\theta_H}{2}C_R^L 
& k\lambda \mp M \end{pmatrix}
\begin{pmatrix}   \alpha_\nu \cr  \alpha_{\nu'} \cr \alpha_\eta \end{pmatrix} =0 
\label{neutrinoBC4}
\end{align}
where $S_{L/R}^L=S_{L/R}(1;\lambda,c_{L})$ etc..
From $\det K_\nu =0$, we find the mass spectrum formula for the  neutrino
sector:\footnote{There was an error of a factor 2 in the right side of Eq.\ $(d)$ in (\ref{neutrinoEq1})
in the previous papers \cite{HosotaniYamatsu2017, HosotaniYamatsu2018}.   
The formulas (\ref{neutrinoSpectrum1}),  (\ref{neutrinoSpectrum2}) reflect this correction.}
\begin{align}
\det K_\nu = 
(k\lambda \pm M )
\Big\{ S_L^L S_R^L +\sin^2\frac{\theta_H}{2} \Big\}
+ \frac{m_B^2}{k} S_R^L C_R^L=0 ~.
\label{neutrinoSpectrum1}
\end{align}
One of the solutions with $f_{+R/L} (x)$ or $f_{-R/L} (x)$ allows a small mass 
eigenvalue $m_\nu = k \lambda_\nu > 0$.
For $M > 0$, the neutrino mode is obtained with $f_{+R/L} (x)$.
Noting $\lambda z_L\ll 1$ and $k\lambda\ll M$, one finds the neutrino mass  given
by
\begin{align}
&m_\nu \simeq 
\begin{cases}
\myfrac{m_e^2 M  z_L^{2 c_L +1}}{(2c_L +1)m_B^2}  &{\rm for~} c_L > \onehalf ~,\cr
\noalign{\kern 5pt}
\myfrac{m_e^2 M}{(2|c_L| -1)m_B^2}  &{\rm for~} c_L < - \onehalf ~.
\end{cases}
\label{neutrinoSpectrum2}
\end{align}
The gauge-Higgs seesaw mechanism\cite{HosotaniYamatsu2017, Minkowski1977, Mohapatra1986}
is characterized by
a $3 \times 3$ mass matrix
\begin{align}
\frac{i}{2} (\nu_{0L}^{c\dagger}, \nu_{0R}^{\prime\dagger}, \eta^{c \dagger})
\begin{pmatrix} 0 &m_e&0 \cr  m_e&0&\tilde{m}_B \cr  0 &\tilde{m}_B&M \end{pmatrix}
\begin{pmatrix} \nu_{0L} \cr \nu_{0R}^{\prime c} \cr \eta \end{pmatrix} 
+ {h.c.} ~, 
\label{neutrinoSpectrum3}
\end{align}
where $m_e$ is its corresponding charged lepton mass.
The structure takes the same form as the inverse seesaw mechanism in Ref.\ \cite{Mohapatra1986}, 
and yields very light neutrino mass $m_\nu \sim m_e^2 M/\tilde m_B^2$.
The Majorana mass $M$ may take a moderate value.
In particular, for $c_L < - \onehalf$, $m_\nu \sim 1\,$meV is obtained 
with $m_B \sim 1\,$TeV and $M \sim 50\,$GeV.
For $c_L >\onehalf$, $m_B$ has to take a rather large value, larger than the Planck mass.

\begin{table}[bth]
\renewcommand{\arraystretch}{1.2}
\begin{center}
\caption{Parameters which reproduce the spectrum of leptons
for $\theta_H=0.15$, $z_L=10^{10}$.  $m_\KK = 8.062 \,$TeV.
The masses of the 1st KK modes leptons   are also shown in the unit of TeV.
For $c_L >0$, there appear light neutrino excitation modes, $\nu_s$.
The values $m_e = 0.511\,$MeV, $m_\mu = 105.7\,$MeV, $m_\tau = 1.776\,$GeV, and
$m_\nu = 1\,$meV  have been used.  
}
\vskip 10pt
\begin{tabular}{|c|c|c|c|c|c|c|}
\hline
Leptons & $c_L$ &$M$ & $m_B$  &$m_{\nu_s}$ &$m_{\nu^{(1)}}$ & $m_{e^{(1)}}$\\
& &(GeV) & (GeV) &  &(TeV) &(TeV)\\
\hline
$(\nu_e, e)$ &$1.086$ &$10^3$ &$6.6\times 10^{19} $ &$6.8\,$MeV &$8.38$ &$8.38$ \\
\cline{3-7}
&&$1$ &$2.1\times 10^{18} $ &$6.8\,$MeV &$8.38$ &$8.38$ \\
\cline{2-7}
&$-1.086$&$10^3$ &$1.5 \times 10^4$  & -- &$8.38$ &$8.38$ \\
\cline{3-7}
& &$1 $ &$4.7 \times 10^2$  & -- &$0.51$ &$8.38$ \\
\hline
$(\nu_\mu, \mu)$ &$0.839 $ &$10^3$ &$5.0 \times 10^{19}$  &$1.4 \,$GeV &7.47 &$7.47$\\
\cline{2-7}
&$-0.839 $&$10^3$ &$1.2 \times 10^7$ &-- &$7.47$  &$7.47$\\
\hline
$(\nu_\tau, \tau)$ &$0.703 $ &$10^3$ &$3.9 \times 10^{19}$  &$24. \,$GeV &6.96 &$6.96$\\
\cline{2-7}
&$-0.703 $&$10^3$ &$8.8 \times 10^8$ &-- &6.96 &$6.96$ \\
\hline
\end{tabular}
\label{Tab:leptonSpectrum}
\end{center}
\end{table}

\begin{figure}[tbh]
\begin{center}
\includegraphics[bb=24 5 321 224, height=5.5cm]{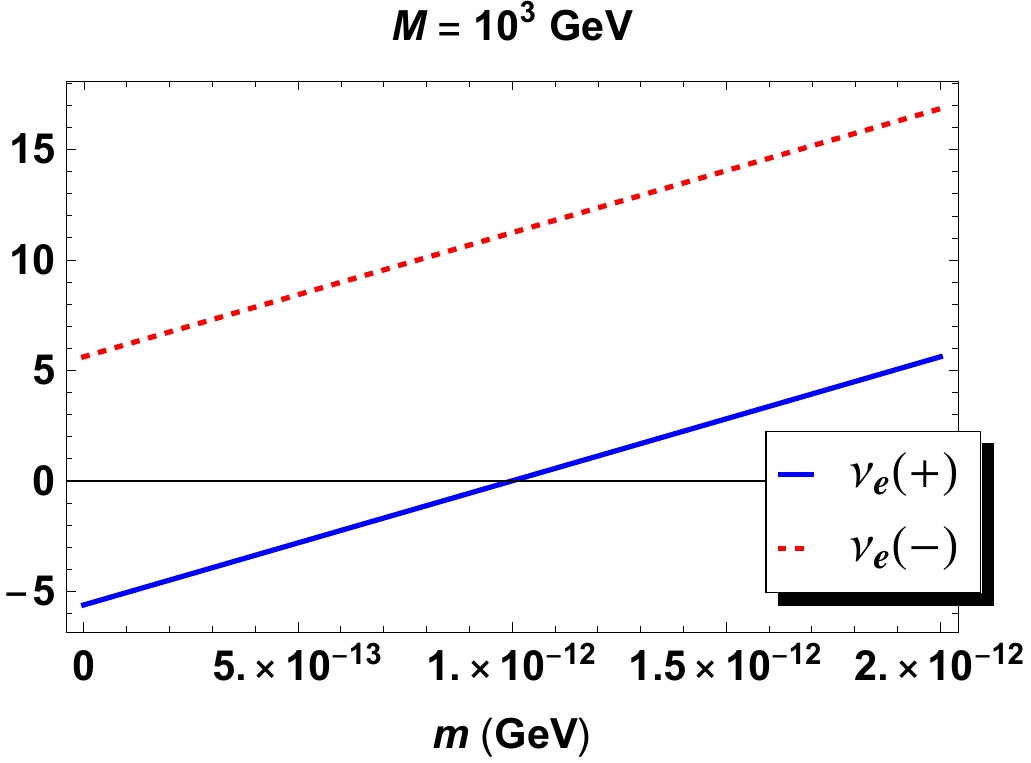}
\hskip 0pt
\includegraphics[bb=23 4 356 225, height=5.5cm]{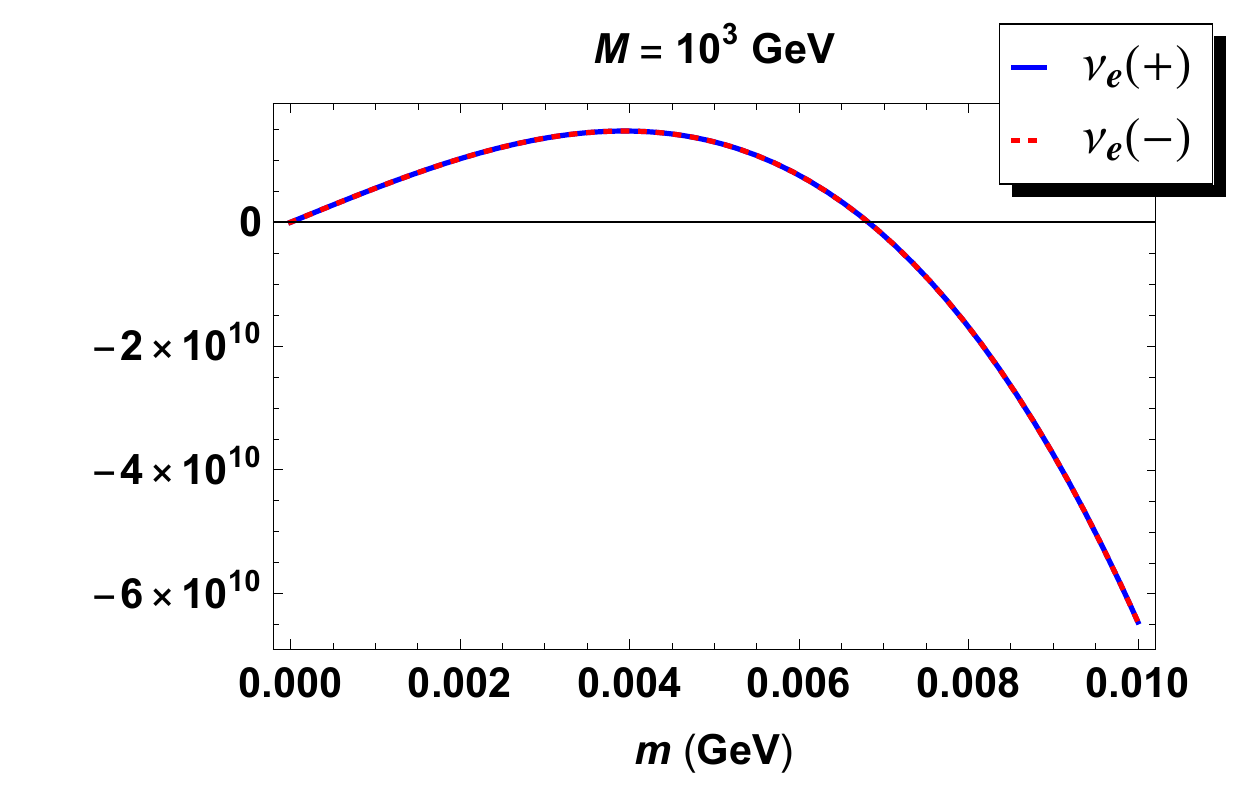}\\
(a) \hskip 7.cm (b)\\
\vskip 5pt
\includegraphics[bb=23 5 355 224, height=5.5cm]{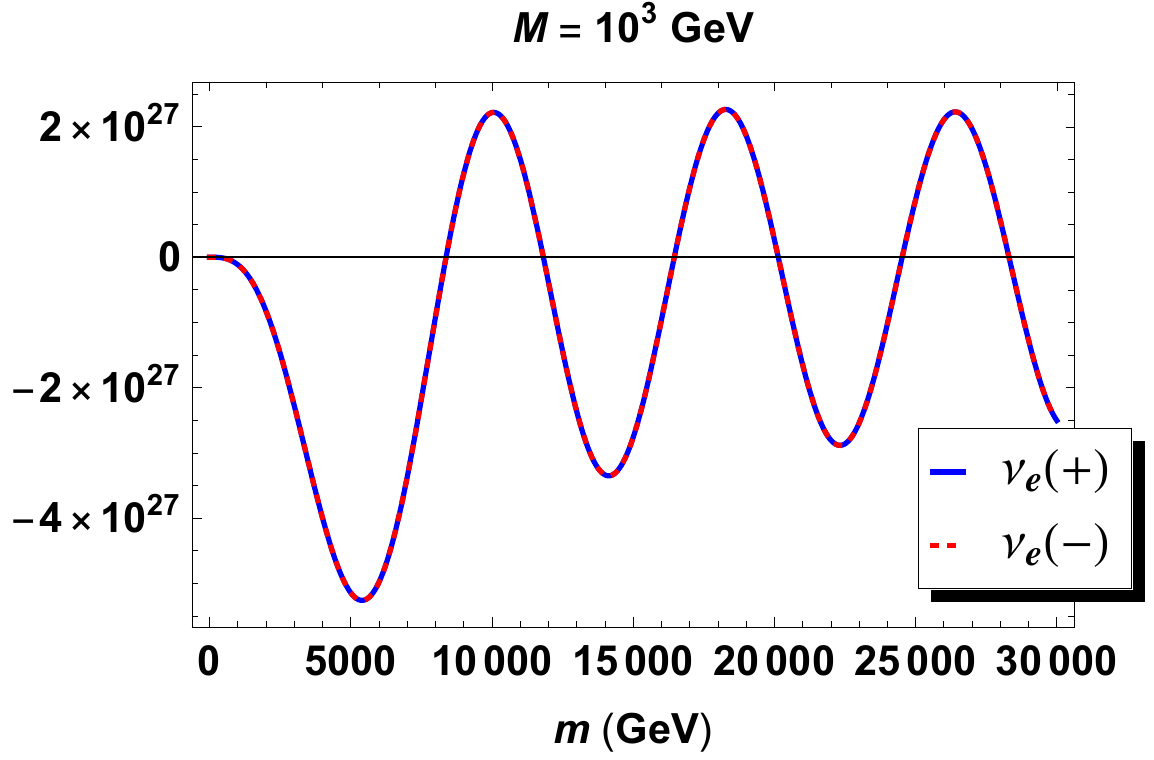}\\
(c)
\end{center}
\vskip -10pt
\caption{Spectrum of electron neutrino tower for $c_e >0$.  
$\det K_\nu $ in (\ref{neutrinoSpectrum1})
is plotted as a function of $m= k \lambda$
in various mass ranges for $\theta_H=0.15$, $z_L=10^{10}$, 
$m_\KK = 8.062 \,$TeV and $M= 1\,$TeV.
The mass spectrum $\{ m_n = k \lambda_n \}$ is determined by  roots of $\det K_\nu  = 0$.
$\nu_e (\pm)$ indicates the case of $f_{\pm L/R} (x)$ in (\ref{neutrinoWave1}).
Only $\nu_e (+)$ has a solution corresponding to $\nu_e$ with $m_{\nu_e} = 1\,$meV. 
In (b) and (c) the curves for $\nu_e (+)$ and $\nu_e (-)$ almost overlap with each other 
at this scale.
}
\label{Figure:neutrino1}
\end{figure}

\begin{figure}[tbh]
\begin{center}
\includegraphics[bb=23 4 322 225, height=5.5cm]{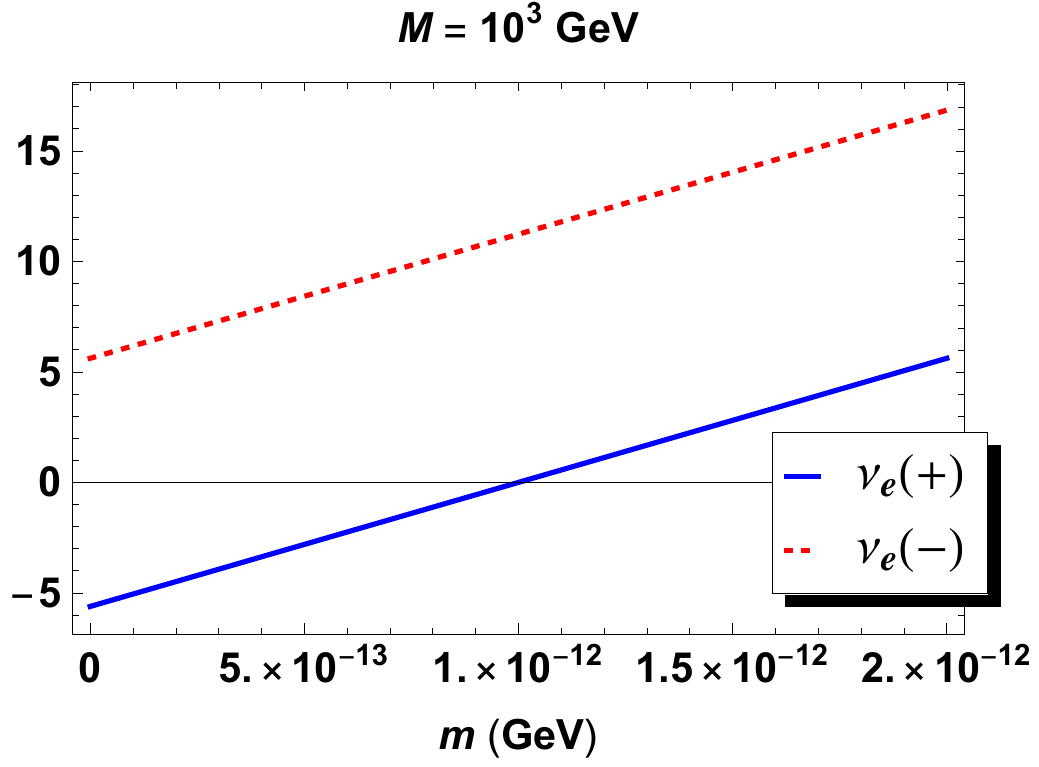}
\hskip 0pt
\includegraphics[bb=25 3 354 224, height=5.5cm]{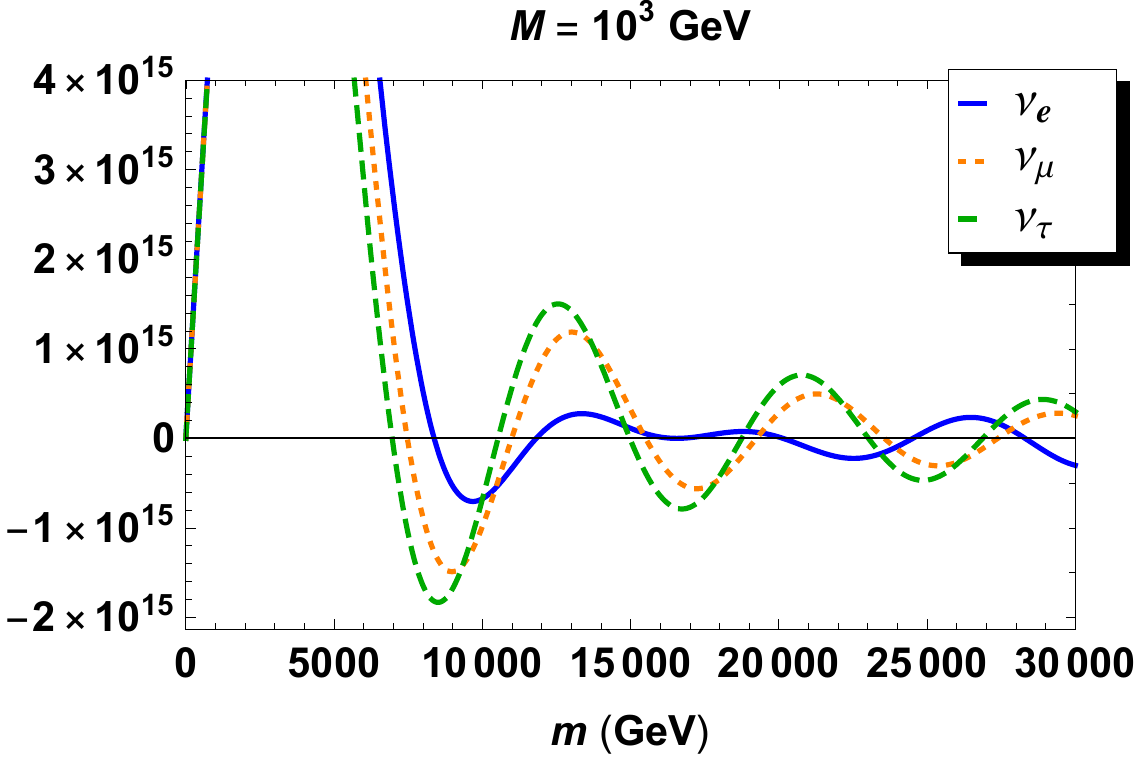}\\
(a) \hskip 7.cm (b)\\
\end{center}
\vskip -10pt
\caption{Spectrum of  neutrino towers for $c_L <0$.  
As in Fig.\ \ref{Figure:neutrino1},  $\det K_\nu $ in  (\ref{neutrinoSpectrum1})
is plotted as a function of $m= k \lambda$
in two mass ranges for $\theta_H=0.15$, $z_L=10^{10}$, 
$m_\KK = 8.062 \,$TeV and $M= 1\,$TeV.
The mass spectrum $\{ m_n = k \lambda_n \}$ is determined by  roots of $\det K_\nu  = 0$.
(a) Only $\nu_e (+)$ has a solution corresponding to $\nu_e$ with $m_{\nu_e} = 1\,$meV. 
(b) The spectrum of $\nu_e, \nu_\mu, \nu_\tau$  towers are shown. 
$\nu(+)$ and $\nu(-)$ towers almost overlap in this figure.
For the $\nu_e$ tower the masses of the 3rd and 4th KK modes are 16.46 TeV and 16.67 TeV,
respectively.
}
\label{Figure:neutrino2}
\end{figure}

Typical parameters in the lepton sector are summarized in Table \ref{Tab:leptonSpectrum}.
$|c_L|$ and $m_{e^{(1)}}$ are fixed by $m_e$.  
The value of $M$ can be varied.  The spectrum does not depend on $M$  very much.
As is seen in the table, very light neutrino excited mode $\nu_s$ appears for positive $c_L$.
This does not necessarily mean the inconsistency with the observation.
The $\nu_s$ mode may become a candidate for warm dark matter,\cite{WarmDM2012} 
though more detailed investigation of gauge couplings is necessary to see the feasibility.
For negative $c_L$ very light neutrino excited mode appears only when $M$ becomes very small.
The spectrum of the neutrino towers are shown in fig. \ref{Figure:neutrino1} for $c_L >0$ and
in fig. \ref{Figure:neutrino2} for $c_L <0$.

\subsection{$W$ couplings of quarks and leptons}

As have been shown above, the quark and lepton mass spectrum can be reproduced except that
the down quark mass turns out  lighter than the up quark mass.
At this stage one might worry about the $W$ couplings of quarks and leptons in the current scheme.
In the gauge-Higgs unification the $W$ boson at $\theta_H \not= 0$ necessarily contains 
the original $SU(2)_R$ component as seen in Section 4.1.     If quarks and leptons originated from
only spinor representation multiplets in $SO(5)$, right-handed components of quarks and
leptons also would have had non-vanishing couplings to $W$, which contradicts with the observation.

The left-handed quark and lepton doublets are mainly in the spinor representation of $SO(5)$, which
have nominal $W$ couplings.  The mechanism in the current model for making right-handed quarks and 
leptons having almost vanishing $W$ couplings is the following.  
The up-type quarks are contained solely in the spinor multiplets.
The down-type quarks are contained in both of the  spinor and singlet representations of $SO(5)$.
Left-handed down-type quarks are mostly in the spinor representation multiplets, whereas  right-handed 
down-type quarks are mostly in the singlet representation multiplets so that right-handed up-type quarks
have almost vanishing $W$ couplings to right-handed down-type quarks.

The mechanism in the lepton sector is different.  With the presence of brane fermions $\chi$, 
the gauge-Higgs seesaw mechanism functions in the neutrino sector.  Right-handed neutrinos
become heavy, acquiring $O(m_\KK)$ masses, and decouple from right-handed charged leptons.

Indeed, one can evaluate the $W$ couplings of quarks and leptons by determining wave functions
of quarks and leptons from the mass-determining  matrices explained above and inserting them
to the original action.  The result is shown in Table \ref{Tab:Wcoupling}.
It is seen that the $\mu$-$e$ universality in the charged current interactions holds
to high accuracy, provided the same sign of $c_L$ is adopted.  
It is also confirmed that the $W$ couplings of right-handed quarks and leptons are
strongly suppressed.
More detailed study of gauge couplings, including $Z$ and $Z'$ couplings, will be given
separately.

\begin{table}[tbh]
\renewcommand{\arraystretch}{1.2}
\begin{center}
\caption{$W$ couplings of quarks and leptons
for $\theta_H=0.15$, $z_L=10^{10}$.  $m_\KK = 8.062 \,$TeV.
The couplings are defined by 
${\cal L} = W_\mu \big(  g_L^W \bar u_L \gamma^\mu d_L  +  g_R^W \bar u_R \gamma^\mu d_R)$
for $(u,d)$ doublet.  In the SM $g_L^W= g_w/\sqrt{2}$ and $g_R^W  = 0$.
For $(u,d)$ doublet, we set $m_d = 0.9 \, m_u$.
}
\vskip 10pt
\begin{tabular}{|c|c|c|c|c|}
\hline
Leptons & $c_L$ &$M$&$\myfrac{g_L^W}{g_w/\sqrt{2}} - 1$ & $\myfrac{g_R^W}{g_w/\sqrt{2}}$  \\
\hline
$(\nu_e, e)$ &$1.086$ &$1\,$TeV  &$-2.64 \times 10^{-3}$ &$O(10^{-11} )$  \\
\cline{2-5}
&$-1.086$ &$1\,$TeV  &$-5.24 \times 10^{-3}$ &$O(10^{-23} )$  \\
\hline
$(\nu_\mu, \mu)$ &$0.839 $ &$1\,$TeV  &$-2.64 \times 10^{-3}$ &$O(10^{-14} )$ \\
\cline{2-5}
&$-0.839 $  &$1\,$TeV &$-5.25 \times 10^{-3}$ &$O(10^{-21} )$  \\
\hline
$(\nu_\tau, \tau)$ &$0.703 $  &$1\,$TeV &$-2.64 \times 10^{-3}$ &$O(10^{-15} )$   \\
\cline{2-5}
&$-0.703 $  &$1\,$TeV &$-5.25 \times 10^{-3}$ &$O(10^{-19} )$   \\
\hline
\end{tabular}
\vskip 10pt
\begin{tabular}{|c|c|c|c|c|c|}
\hline
Quarks & $c_Q$ &$\mu_1$ &$\tilde m_D$ &$\myfrac{g_L^W}{g_w/\sqrt{2}} - 1$ 
& $\myfrac{g_R^W}{g_w/\sqrt{2}}$  \\
\hline
$(u,d)$ &$-1.044$ &$0.1$ &$1.0$ &$-5.24 \times 10^{-3}$ &$O(10^{-14} )$  \\
\hline
$(c, s)$ &$- 0.7546 $ &$0.1$ &$1.0$ &$-5.25 \times 10^{-3}$ &$O(10^{-9} )$ \\
\hline
$(t, b)$ &$0.2287$ &$0.1$ &$0.1$ &$-3.43 \times 10^{-3}$ &$O(10^{-4} )$   \\
\cline{2-6}
&$-0.2287$  &$0.1$ &$1.0$ &$-4.41 \times 10^{-3}$ &$O(10^{-5} )$   \\
\hline
\end{tabular}
\label{Tab:Wcoupling}
\end{center}
\end{table}

\section{Summary and discussions}

In this paper we have presented a new model of the $SO(5) \times U(1) \times SU(3)$ gauge-Higgs
unification in which quark and lepton multiplets are introduced in the spinor, vector, and singlet
representations of $SO(5)$ such that they can be implemented in the $SO(11)$ gauge-Higgs
grand unification scheme.  This should be contrasted to the previous model in which 
all quark and lepton multiplets are introduced in the vector representation of $SO(5)$.
The up-type quarks are contained solely in the spinor representation.
The right-handed down-type quarks are mainly contained in the singlet representation of $SO(5)$.
$SO(5) \times U(1) \times SU(3)$ singlet brane Majorana fermions are introduced on the UV brane.
The coupling of these brane fermions to bulk fermion multiplets induces the gauge-Higgs 
seesaw mechanism in the neutrino sector, which takes the same form as the inverse seesaw
mechanism in four-dimensional GUT theories.

With $SO(5) \times U(1) \times SU(3)$ gauge-invariant brane interactions taken into account 
the quark-lepton mass spectrum has been reproduced with the exception that down quark mass
($m_d$) becomes lighter than up quark mass ($m_u$).  A solution to this problem is yet
to be found.    The compatibility with grand unification
severely restricts matter content and interactions in the gauge-Higgs unification.
Nevertheless it is very encouraging that the model yields almost the same $W$ couplings 
of quarks and leptons.

The present model serves as a viable alternative to the standard model.    
If it is the case, phenomenological consequences of the model need to be clarified.
As in the previous model   $Z'$ bosons (the first KK modes of $\gamma$, $Z$ and $Z_R$) 
are predicted around $7\,$TeV to $10\,$TeV range.
We have seen in Section 5 that the bulk mass parameters ($c_u, c_c$) of quark multiplets
$\Psi_{({\bf 3,4})}$ in the first and second generations must be negative
to avoid exotic light excitation modes of down-quark-type.  
The bulk mass parameters $c_L$ of lepton multiplets can be either positive or negative.
The sign of the bulk mass parameters is critically important to determine the behavior of
wave functions.  For $c> + \onehalf$ ($c < - \onehalf)$ left-handed quarks/leptons are 
localized near the UV (IR) brane, whereas right-handed ones near the IR (UV) brane.
As  $Z'$ bosons are localized near the IR brane, right-handed (left-handed) quarks/leptons 
have larger couplings to $Z'$ bosons for $c> + \onehalf$ ($c < - \onehalf)$.
The effect of the large parity violation can be seen in the $e^+ e^-$ collisions
through interference terms.  In particular, cross sections of various fermion-pair production processes
should reveal distinct dependence on the $e^-$ polarization.\cite{FHHO2017ILC}

With the mass spectra of all fields having been determined, one can investigate the effective
potential $V_\eff (\theta_H)$ to show that EW symmetry is dynamically broken.
The flavor mixing in the quark and lepton sectors and the dark matter are also among the problems
to be solved in the gauge-Higgs unification scenario.  We shall come back to these issues
in the near future.

\section*{Acknowledgements}

This work was supported in part 
by European Regional Development Fund-Project Engineering Applications of 
Microworld Physics (No.\ CZ.02.1.01/0.0/0.0/16-019/0000766) (Y.O.), 
by the National Natural Science Foundation of China (Grant Nos.~11775092, 
11675061, 11521064 and 11435003) (S.F.), 
by the International Postdoctoral Exchange Fellowship Program (IPEFP) (S.F.), 
and by Japan Society for the Promotion of Science, 
 Grants-in-Aid  for Scientific Research,  No.\ 15K05052 (Y.H.) and No.\  18H05543 (N.Y.).

\appendix

\section{$SO(5)$}
\label{Sec:SO5}

The generators of $SO(5)$,
$T_{jk}=-T_{kj}=T_{jk}^\dag $ ($j,k=1,2,3,4,5$), satisfy
the algebra
\begin{align}
 [T_{ij},T_{kl}]=
 i(\delta_{ik} T_{jl}-\delta_{il}T_{jk}
 +\delta_{jl}T_{ik}-\delta_{jk}T_{il}).
\label{SO5-algebra}
\end{align}
In the adjoint representation, 
\begin{align}
 &(T_{ij})_{pq}=
 -i (\delta_{ip} \delta_{jq} - \delta_{iq} \delta_{jp}),\nonumber\\ 
 &\mbox{tr}(T_{jk}T_{lm})=
 2(\delta_{jl}\delta_{km}-\delta_{jm}\delta_{kl}),
 \hspace{1em}
 \mbox{tr}(T_{jk})^2=2.
\end{align}
We take the following basis of $SO(5)$ Clifford algebra:
\begin{align}
&\{\Gamma_j,\Gamma_k\}=2\delta_{jk}I_{4} ~,\cr
&\Gamma_a=\sigma^a\otimes\sigma^1 \quad (a=1,2,3) ~,\cr
&\Gamma_4=\sigma^0\otimes\sigma^2 ~,~~
\Gamma_5=\sigma^0\otimes\sigma^3=-\Gamma_1\Gamma_2\Gamma_3\Gamma_4 ~,
\label{SO5-Gamma-matrix}
\end{align}
where $\sigma^0=I_2$ and $\{\sigma^a \}$ are Pauli matrices.
In terms of $\Gamma_j$ the $SO(5)$ generators in the spinor
representation are given by
\begin{align}
&T_{jk}=-\frac{i}{4}[\Gamma_j,\Gamma_k] \hskip .3cm 
 \Big(= -\frac{i}{2} \Gamma_j\Gamma_k \quad  \mbox{for}\ j\not= k \Big), 
 \nonumber\\
&(T_{jk})^2=\frac{1}{4}I_{4} ~,~~  \mbox{tr}(T_{jk})^2=1 ~.
\end{align}

The orbifold boundary conditions $P_0, P_1$ in Eqs~(\ref{Eq:SO5-BCs})
break $SO(5)$ to $SO(4)\simeq SU(2)_L\times SU(2)_R$.
The generators of the corresponding
$SO(4)\simeq SU(2)_L\times SU(2)_R$ in the spinor representation are
given by
\begin{align}
\vec{T}_L&= \frac{1}{2} 
\begin{pmatrix}  T_{23} + T_{14} \cr  T_{31} + T_{24} \cr T_{12} + T_{34} \end{pmatrix}
= \frac{1}{2}\vec{\sigma}\otimes \begin{pmatrix} 1 & 0 \cr 0 & 0 \end{pmatrix} , \cr
\noalign{\kern 5pt}
\vec{T}_R&= \frac{1}{2}
\begin{pmatrix}  T_{23} - T_{14} \cr  T_{31} - T_{24} \cr T_{12} - T_{34} \end{pmatrix}
= \frac{1}{2}\vec{\sigma}\otimes \begin{pmatrix} 0 & 0 \cr 0 & 1 \end{pmatrix}.
\label{su2LR} 
\end{align}
These generators become block-diagonal so that
an $SO(5)$ spinor representation {\bf 4} can be decomposed into
$({\bf 2,1})\oplus({\bf 1,2})$ of $SO(4)\simeq SU(2)_L\times SU(2)_R$ :
\begin{align}
\Psi_{\bf 4}= \begin{pmatrix}   \Psi_{({\bf 2,1})} \cr  \Psi_{({\bf 1,2})} \end{pmatrix} .
\end{align}

In the representation (\ref{SO5-Gamma-matrix})  one finds that
\begin{align}
&\Gamma_j^* = (-1)^{j+1}\Gamma_j,\nonumber\\
&R:=-i\Gamma_2\Gamma_4=R^\dag=R^{-1}
= \sigma^2\otimes\sigma^3,\nonumber\\
&R\Gamma_jR=(-1)^{j+1}\Gamma_j,\ \ \ R\Gamma_j^*R=\Gamma_j,\nonumber\\
&RT_{jk}^*R=-T_{jk}.
\end{align}
It follows that for an $SO(5)$ spinor $\Psi_{\bf 4}$, the
$R$-transformed one also transforms as {\bf 4}.
\begin{align}
&\tilde{\Psi}_{\bf 4}:=iR\Psi_{\bf 4}^* ~, \cr
\noalign{\kern 5pt}
&\Psi_{\bf 4}'=\bigg(1+\frac{i}{2}\epsilon_{jk}T_{jk}\bigg)\Psi_{\bf 4}
\ \ \ \Rightarrow \ \ \
 \tilde{\Psi}_{\bf 4}'=\bigg(1+\frac{i}{2}\epsilon_{jk}T_{jk}\bigg)
 \tilde{\Psi}_{\bf 4}.
 \label{so5spinor1}
\end{align}
Its $SO(5)$ content is given by
\begin{align}
\tilde{\Psi}_{\bf 4}= 
\begin{pmatrix}  \tilde{\Psi}_{({\bf 2,1})} \cr \tilde{\Psi}_{({\bf 1,2})} \end{pmatrix}
  =\begin{pmatrix}   i\sigma^2{\Psi}_{({\bf 2,1})}^* \cr  
   -i\sigma^2{\Psi}_{({\bf 1,2})}^* \end{pmatrix}.
\label{so5spinor2}
\end{align}

\section{Basis functions}
\label{Sec:basis-functions}

We summarize basis functions in the RS space.

\subsection{Gauge fields}

We define
\begin{align}
&F_{\alpha,\beta}(u,v) \equiv J_\alpha(u) Y_\beta(v) - Y_\alpha(u)J_\beta(v)
\end{align}
where $J_\alpha(x)$ and $Y_\alpha(x)$
are Bessel functions of the 1st and 2nd kind, respectively.
For gauge bosons $C = C(z;\lambda)$ and $S = S(z;\lambda)$ are defined as solutions of
\begin{align}
&- {\cal P}_4 \begin{pmatrix}C \\ S \end{pmatrix}  = 
\left( - \frac{d^2}{dz^2} + \frac{1}{z}\frac{d}{dz} \right) 
\begin{pmatrix}C \\ S \end{pmatrix}  
=  \lambda^2 \begin{pmatrix}C \\ S \end{pmatrix} ,  
\end{align}
with boundary conditions $C = z_L$,  $S = 0$,  $C' = 0$,   and $S' = \lambda$ at $z=z_L$.
They are given by
\begin{align}
C(z;\lambda) &= + \frac{\pi}{2} \lambda z z_L F_{1,0}(\lambda z, \lambda z_L), \cr
C'(z;\lambda) &= + \frac{\pi}{2} \lambda^2 z z_L F_{0,0}(\lambda z, \lambda z_L), \cr
S(z;\lambda) &= - \frac{\pi}{2} \lambda z F_{1,1}(\lambda z, \lambda z_L), \cr
S'(z;\lambda) &= - \frac{\pi}{2} \lambda^2 z F_{0,1}(\lambda z, \lambda z_L). 
\label{gaugeF1}
\end{align}
We note that
\begin{align}
- {\cal P}_z \begin{pmatrix}C' \\ S' \end{pmatrix}  
&= \lambda^2 \begin{pmatrix}C' \\ S' \end{pmatrix},\cr
\noalign{\kern 5pt}
C S' - SC' &= \lambda z~.
\label{gaugeF2}
\end{align}

\subsection{Massless fermion fields}

For  massless fermions in five dimensions we define
\begin{align}
\begin{pmatrix} C_L \cr S_L \end{pmatrix} (z;\lambda,c)   
&= \pm\frac{\pi}{2} \lambda \sqrt{z z_L} \, F_{c+\frac{1}{2},c \mp \frac{1}{2}}(\lambda z,\,\lambda z_L) ~, \cr
\noalign{\kern 5pt}
\begin{pmatrix}C_R \cr S_R \end{pmatrix} (z;\lambda,c) 
&= \mp \frac{\pi}{2} \lambda \sqrt{z z_L} \, F_{c-\frac{1}{2},c \pm \frac{1}{2}}(\lambda z,\,\lambda z_L) ~,
\label{fermionF1}
\end{align}
which satisfy
\begin{align}
&
D_+ \begin{pmatrix} C_L \\ S_L \end{pmatrix} = \lambda \begin{pmatrix} S_R \\ C_R \end{pmatrix},
~~
D_- \begin{pmatrix} C_R \\ S_R \end{pmatrix} = \lambda \begin{pmatrix} S_L \\ C_L \end{pmatrix}, \cr
\noalign{\kern 5pt}
& C_L C_R - S_L S_R = 1~, \cr
\noalign{\kern 5pt}
& C_R = C_L = 1 ~, ~~ S_R = S_L = 0 ~, ~~{\rm at~} z= z_L.
\label{fermionF2}
\end{align}
They also satisfy
\begin{align}
C_L(z;\lambda, -c) = C_R(z; \lambda, c) ~,~~
S_L(z;\lambda, -c) = - S_R(z; \lambda, c) ~.
\label{fermionF3}
\end{align}

\subsection{Massive fermion fields}

As seen in (\ref{fermionAction2}), $\check \Psi_{({\bf 3},{\bf 1})}^{\pm \alpha}$ and
$\check \Psi_{({\bf 1},{\bf 5})}^{\pm \beta}$ have additional pseudo-Dirac
bulk mass terms in the action. To find basis functions for these massive fermions,
we consider the action for $N^\pm$ fields given by
\begin{align}
&\int d^4x \int_1^{z_L} \frac{dz}{k} \Big\{ 
\overline{\check N}{}^+ {\cal D}_0 (c_+) \check N^+
+ \overline{\check N}{}^- {\cal D}_0 (c_-) \check N^-
- \frac{k \tilde m}{z} \big( \overline{\check N}{}^+ \check N^- 
+ \overline{\check N}{}^- \check N^+ \big) \Big\}  \cr
\noalign{\kern 5pt}
&\hskip 2.cm
\hbox{where~}
{\cal D}_0(c)= \begin{pmatrix}   -kD_-(c) &\sigma^\mu\partial_\mu \cr
 \bar{\sigma}^\mu\partial_\mu &-kD_+(c) \end{pmatrix} .
\label{Mfermion1}
\end{align}
$\tilde m$ is dimensionless, and $k \tilde m$ corresponds to 
$m_D^\alpha$ and $m_V^\beta$ in (\ref{fermionAction2}).

To find eigenmodes  with four-dimensional mass $k \lambda$, we write
$\check N_R^\pm (x,z)= N_{\pm R} (z) f_R(x)$ and 
$\check N_L^\pm (x,z) = N_{\pm L} (z) f_L(x)$ as described below Eq.\ (\ref{upWave1}).
Then $N_{\pm R} (z)$  and $N_{\pm L} (z)$ must satisfy
\begin{align}
&D_{-}(c_\pm)N_{\pm R} - \lambda N_{\pm L} + \frac{\tilde{m}}{z}\, N_{\mp R} =0 ~, \cr
\noalign{\kern 5pt}
&D_{+}(c_\pm)N_{\pm L}- \lambda N_{\pm R} + \frac{\tilde{m}}{z}\, N_{\mp L} =0 ~.
\label{Mfermion2}
\end{align}
Note
\begin{align}
D_{\pm}(c)D_{\mp}(c)= - \frac{d^2}{dz^2}+\frac{c(c \mp 1)}{z^2} ~.
\label{Mfermion3}
\end{align}
We consider two cases; $c_+ = c_-$ and $c_+ = - c_-$.


\subsubsection{Case I. $c_+ = c_- = c$}

It follows immediately from (\ref{Mfermion2}) that
\begin{align}
 D_- (c \pm \tilde{m})(N_{+R} \pm N_{-R})&=  \lambda\left(N_{+L} \pm  N_{-L}\right) ~, \cr
 D_+ (c \pm \tilde{m}) (N_{+L} \pm N_{-L})&=  \lambda\left(N_{+R} \pm N_{-R}\right) ~.
\label{Mfermion4}
\end{align}
General solutions are given by
\begin{align}
&\begin{pmatrix} N_{\pm R} \cr N_{\pm L} \end{pmatrix}
= a \begin{pmatrix}  C_R^{c+\tilde{m}} \cr  S_L^{c+\tilde{m}} \end{pmatrix}
+ b  \begin{pmatrix}  S_R^{c+\tilde{m}} \cr  C_L^{c+\tilde{m}} \end{pmatrix}
\pm a' \begin{pmatrix}  C_R^{c- \tilde{m}} \cr  S_L^{c - \tilde{m}} \end{pmatrix}
\pm b' \begin{pmatrix}  S_R^{c- \tilde{m}} \cr  C_L^{c - \tilde{m}} \end{pmatrix} .
\label{MfermionSol1}
\end{align}
Here $C_{L/R}^{c \pm \tilde{m}} = C_{L/R}(z;\lambda, {c\pm\tilde{m}})$ and 
$S_{L/R}^{c \pm \tilde{m}} =S_{L/R} (z; \lambda, {c\pm\tilde{m}})$.

At this stage we define basis functions by
\begin{align}
{\cal C}_{R1}(z; \lambda, c, \tilde m) &= C_R(z; \lambda, c+\tilde{m})+C_R(z; \lambda, c-\tilde{m}) ~, \cr
{\cal C}_{R2}(z; \lambda, c, \tilde m) &= S_R(z; \lambda, c+\tilde{m})-S_R(z; \lambda,c-\tilde{m}) ~, \cr
{\cal S}_{L1}(z; \lambda, c, \tilde m) &= S_L(z; \lambda, c+\tilde{m})+S_L(z; \lambda,c-\tilde{m}) ~, \cr
{\cal S}_{L2}(z; \lambda, c, \tilde m) &= C_L(z; \lambda, c+\tilde{m})-C_L(z; \lambda, c-\tilde{m}) ~, \cr
{\cal C}_{L1}(z; \lambda, c, \tilde m) &= C_L(z; \lambda, c+\tilde{m})+C_L(z; \lambda, c-\tilde{m}) ~, \cr
{\cal C}_{L2}(z; \lambda, c, \tilde m) &= S_L(z; \lambda, c+\tilde{m})-S_L(z; \lambda, c-\tilde{m}) ~, \cr
{\cal S}_{R1}(z; \lambda, c, \tilde m) &= S_R(z; \lambda, c+\tilde{m})+S_R(z; \lambda, c-\tilde{m}) ~, \cr
{\cal S}_{R2}(z; \lambda, c, \tilde m) &= C_R(z; \lambda, c+\tilde{m})-C_R(z; \lambda, c-\tilde{m}) ~,
\label{MfermionBasis1}
\end{align}
which satisfy the equations and boundary conditions
\begin{align}
&D_- (c) \begin{pmatrix} {\cal C}_{R1} \cr {\cal C}_{R2} \end{pmatrix}
= \lambda \begin{pmatrix} {\cal S}_{L1} \cr {\cal S}_{L2} \end{pmatrix}
- \frac{\tilde m}{z} \begin{pmatrix} {\cal S}_{R2} \cr {\cal S}_{R1} \end{pmatrix} , \cr
\noalign{\kern 5pt}
&D_- (c) \begin{pmatrix} {\cal S}_{R1} \cr {\cal S}_{R2} \end{pmatrix}
= \lambda \begin{pmatrix} {\cal C}_{L1} \cr {\cal C}_{L2} \end{pmatrix}
- \frac{\tilde m}{z} \begin{pmatrix} {\cal C}_{R2} \cr {\cal C}_{R1} \end{pmatrix} , \cr
\noalign{\kern 5pt}
&D_+ (c) \begin{pmatrix} {\cal C}_{L1} \cr {\cal C}_{L2} \end{pmatrix}
= \lambda \begin{pmatrix} {\cal S}_{R1} \cr {\cal S}_{R2} \end{pmatrix}
- \frac{\tilde m}{z} \begin{pmatrix} {\cal S}_{L2} \cr {\cal S}_{L1} \end{pmatrix} , \cr
\noalign{\kern 5pt}
&D_+ (c) \begin{pmatrix} {\cal S}_{L1} \cr {\cal S}_{L2} \end{pmatrix}
= \lambda \begin{pmatrix} {\cal C}_{R1} \cr {\cal C}_{R2} \end{pmatrix}
- \frac{\tilde m}{z} \begin{pmatrix} {\cal C}_{L2} \cr {\cal C}_{L1} \end{pmatrix} , \cr
\noalign{\kern 5pt}
&{\cal S}_{Rj} = {\cal S}_{Lj} = D_-(c)\,  {\cal C}_{Rj} = D_+(c) \,  {\cal C}_{Lj} = 0 \quad
{\rm at~} z = z_L ~.
\label{MfermionBasis2}
\end{align}
Note also 
\begin{align}
&{\cal C}_{Rj}  (z; \lambda, -c, \tilde m) =  {\cal C}_{Lj}  (z; \lambda, c, \tilde m) ~,\cr
&{\cal S}_{Rj}  (z; \lambda, -c, \tilde m) =  - {\cal S}_{Lj}  (z; \lambda, c, \tilde m) ~, \cr
\noalign{\kern 5pt}
&{\cal C}_{R/Lj}  (z; \lambda, c, - \tilde m) = (-1)^{j-1} \, {\cal C}_{R/Lj}  (z; \lambda, c, \tilde m) ~,\cr
&{\cal S}_{R/Lj}  (z; \lambda, c, - \tilde m) = (-1)^{j-1} \, {\cal S}_{R/Lj}  (z; \lambda, c, \tilde m) ~.
\label{MfermionBasis3}
\end{align}
In the $\tilde m \go 0$ limit
\begin{align}
&{\cal C}_{R1}\to 2C_{R} ~,~~
{\cal S}_{R1}\to 2S_{R} ~,~~
{\cal C}_{L1}\to 2C_{L} ~,~~
{\cal S}_{L1}\to 2S_{L} ~,\cr
&{\cal C}_{R2} , ~ {\cal S}_{R2} ,~  {\cal C}_{L2} ,~ {\cal S}_{L2} \go 0 ~.
\label{MfermionBasis4}
\end{align}

Two types of boundary conditions appear at $z=z_L$.

\vskip 5pt
\noindent
\underline{Type A: ~$(N_{+R}, N_{-R}, N_{+L}, N_{-L})= (+,-,-,+)$}

When parity assignment  at $y=L$ for $(N_{+R}, N_{-R}, N_{+L}, N_{-L})$ is 
$(+,-,-,+)$, boundary conditions at $z=z_L$ become
\begin{align}
& D_-(c)N_{+R}=0 ~,~~ N_{+L}=0 ~, \cr
&N_{-R}=0 ~, ~~ D_+(c) N_{-L}=0 ~.
\label{MfermionBC1}
\end{align}
In this case $a=a'$ and $b= -b'$ in (\ref{MfermionSol1}) and 
solutions can be written as
\begin{align}
&\begin{pmatrix}   N_{+R} \cr N_{+L} \cr N_{-R} \cr N_{-L} \end{pmatrix} 
= a \begin{pmatrix}  {\cal C}_{R1}(z; \lambda, c, \tilde m) \cr  {\cal S}_{L1}(z; \lambda, c, \tilde m) \cr
 {\cal S}_{R2}(z; \lambda, c, \tilde m) \cr  {\cal C}_{L2}(z; \lambda, c, \tilde m) \end{pmatrix}
+ b \begin{pmatrix}  {\cal C}_{R2}(z; \lambda, c, \tilde m) \cr  {\cal S}_{L2}(z; \lambda, c, \tilde m) \cr
 {\cal S}_{R1}(z; \lambda, c, \tilde m) \cr  {\cal C}_{L1}(z; \lambda, c, \tilde m) \end{pmatrix} ,
 \label{MfermionWave1}
\end{align}
where $a,b$ are arbitrary constants.

If $N$'s have the same parity assignment at $y=0$ as that at $y=L$, then
(\ref{MfermionBC1}) must be satisfied at $z=1$ as well. Substituting (\ref{MfermionWave1})
into (\ref{MfermionBC1}) and evaluating the conditions at $z=1$, one finds
\begin{align}
\begin{pmatrix} {\cal S}_{L1} & {\cal S}_{L2} \cr {\cal S}_{R2} & {\cal S}_{R1} \end{pmatrix}
\begin{pmatrix} a \cr b \end{pmatrix} = 0
\label{MfermionBC2}
\end{align}
where $ {\cal S}_{L1}  =  {\cal S}_{L1} (1; \lambda, c, \tilde m)$ etc..
The mass spectrum is determined by
\begin{align}
{\cal S}_{L1}   {\cal S}_{R1}  -  {\cal S}_{L2}   {\cal S}_{R2} = 0 ~.
\label{MfermionSpectrum1} 
\end{align}
Note 
\begin{align}
& {\cal S}_{L1}{\cal S}_{R1}- {\cal S}_{L2} {\cal S}_{R2} +2 =
{\cal C}_{L1}   {\cal C}_{R1}  -  {\cal C}_{L2}   {\cal C}_{R2}  -2 \cr
&\quad =   S_L^{c+\tilde{m}}S_R^{c-\tilde{m}} +S_L^{c-\tilde{m}}S_R^{c+\tilde{m}}
 +C_L^{c+\tilde{m}}C_R^{c-\tilde{m}} +C_L^{c-\tilde{m}}C_R^{c+\tilde{m}} ~, \cr
 \noalign{\kern 5pt}
&{\cal S}_{L1} {\cal C}_{L1} - {\cal S}_{L2} {\cal C}_{L2}  
= 2\left(  S_L^{c+\tilde{m}} C_L^{c -\tilde{m}} +S_L^{c -\tilde{m}} C_L^{c +\tilde{m}}\right) ~. 
\label{MfermionSpectrum2} 
\end{align}

\vskip 5pt
\noindent
\underline{Type B: ~$(N_{+R}, N_{-R}, N_{+L}, N_{-L})= (-,+,+,-)$}

When parity assignment  at $y=L$ for $(N_{+R}, N_{-R}, N_{+L}, N_{-L})$ is 
$(-,+,+,-)$, boundary conditions at $z=z_L$ become
\begin{align}
&N_{+R}=0 ~, ~~D_+(c) N_{+L}=0 ~, \cr
&D_-(c) N_{-R}=0 ~, ~~N_{-L}=0 ~.
\label{MfermionBC3}
\end{align}
In this case $a=- a'$ and $b= b'$ in (\ref{MfermionSol1}) and 
solutions can be written as
\begin{align}
&\begin{pmatrix}   N_{+R} \cr N_{+L} \cr N_{-R} \cr N_{-L} \end{pmatrix} 
= a \begin{pmatrix}  {\cal S}_{R2}(z; \lambda, c, \tilde m) \cr  {\cal C}_{L2}(z; \lambda, c, \tilde m) \cr
 {\cal C}_{R1}(z; \lambda, c, \tilde m) \cr  {\cal S}_{L1}(z; \lambda, c, \tilde m) \end{pmatrix}
 + b  \begin{pmatrix}  {\cal S}_{R1}(z; \lambda, c, \tilde m) \cr  {\cal C}_{L1}(z; \lambda, c, \tilde m) \cr
 {\cal C}_{R2}(z; \lambda, c, \tilde m) \cr  {\cal S}_{L2}(z; \lambda, c, \tilde m) \end{pmatrix} ,
 \label{MfermionWave2}
\end{align}
where $a,b$ are arbitrary constants.

If $N$'s have the same parity assignment at $y=0$ as that at $y=L$, then
(\ref{MfermionBC3}) must be satisfied at $z=1$ as well. 
Substituting (\ref{MfermionWave2})
into (\ref{MfermionBC3}) and evaluating the conditions at $z=1$, one finds
\begin{align}
\begin{pmatrix} {\cal S}_{R2} & {\cal S}_{R1} \cr {\cal S}_{L1} & {\cal S}_{L2} \end{pmatrix}
\begin{pmatrix} a \cr b \end{pmatrix} = 0 ~. 
\label{MfermionBC4}
\end{align}
The mass spectrum is determined by
\begin{align}
{\cal S}_{L1}   {\cal S}_{R1}  -  {\cal S}_{L2}   {\cal S}_{R2} = 0 ~.
\label{MfermionSpectrum3} 
\end{align}

\subsubsection{Case II. $c_+ = - c_- = c$}

The special case $c_+ = - c_- = c$ naturally emerges in the context of six-dimensional
gauge-Higgs grand unification.\cite{HosotaniYamatsu2017}  The bulk (vector) mass parameter $c$
appears there as a coefficient  in the vector component $\gamma^6$, which becomes 
the bulk mass parameter in the RS space,  $\pm c$, for 6D Weyl ($\gamma^7= \pm$)  components.
In this case Eq.\ (\ref{Mfermion2}) becomes
\begin{align}
&D_{-}(c)N_{+R} - \lambda N_{+L} + \frac{\tilde{m}}{z} N_{-R} =0~, \cr
&D_{+}(c)N_{+L} - \lambda N_{+R} + \frac{\tilde{m}}{z} N_{-L} =0 ~, \cr
-&D_{+}(c) N_{-R} - \lambda N_{-L} + \frac{\tilde{m}}{z} N_{+R} =0 ~,\cr
-&D_{-}(c) N_{-L} - \lambda N_{-R} + \frac{\tilde{m}}{z} N_{+L} =0 ~.
\label{M2fermionEq1}
\end{align}

To find solutions to Eqs.\ (\ref{M2fermionEq1}),  we note that 
\begin{align}
\left\{-\frac{d^2}{dz^2}+\frac{c (c \mp 1)}{z^2}+
\frac{\tilde{m}^2}{z^2}-\lambda^2\right\}N_{\pm R}
-\frac{\tilde{m}}{z^2}N_{\mp R}=0 ~.
\label{M2fermionEq2}
\end{align}
We seek solutions in the form $N_{+R} = f(z)$ and $N_{-R}= \alpha f(z)$.
Solutions exist provided $-c-\alpha\tilde{m}=c-{\tilde{m}}/{\alpha}$ is satisfied, or $\alpha = \alpha_\pm$ where
\begin{align}
&\alpha_{\pm} =\frac{1}{\tilde{m}}  (-c \pm \hat{c} ) ~,~~ \alpha_+\alpha_- = -1 ~, \cr
\noalign{\kern 5pt}
& \hat{c} = \sqrt{c^2+\tilde{m}^2} ~.
\label{alphapm}
\end{align}
With $\alpha = \alpha_\pm$, $f(z)$ satisfies
\begin{align}
&\big\{ D_\pm (\hat{c}) D_\mp (\hat{c}) - \lambda^2 \big\} f(z) = 0 ~.
\end{align}
Hence general solutions are given by
\begin{align}
\begin{pmatrix} N_{+R} \cr N_{-R} \end{pmatrix} =
a \begin{pmatrix}  C_R^{\hat{c}} \cr  \alpha_+ C_R^{\hat{c}} \end{pmatrix}
+ b  \begin{pmatrix}  S_R^{\hat{c}} \cr  \alpha_+ S_R^{\hat{c}} \end{pmatrix}
+ a' \begin{pmatrix}  C_L^{\hat{c}} \cr  \alpha_- C_L^{\hat{c}} \end{pmatrix}
+ b' \begin{pmatrix}  S_L^{\hat{c}} \cr  \alpha_- S_L^{\hat{c}} \end{pmatrix} ,
\label{M2fermionSol1}
\end{align}
where $C_{L/R}^{\hat{c}}=C_{L/R}(z; \lambda, \hat{c})$ and $S_{L/R}^{\hat{c}}=S_{L/R}(z;\lambda, \hat{c})$.

To find the corresponding solutions for $N_{\pm L}$, we make use of the identities
\begin{align}
D_{-}(c) &= + D_{-}(\hat{c})-\frac{\tilde{m}\alpha_+}{z} =  - D_{+}(\hat{c}) -\frac{\tilde{m}\alpha_-}{z} ~, \cr
\noalign{\kern 5pt}
D_{+}(c) &= + D_{+}(\hat{c})-\frac{\tilde{m}\alpha_+}{z} =  - D_{-}(\hat{c}) -\frac{\tilde{m}\alpha_-}{z} 
\label{M2fermionIdentity1}
\end{align}
to find
\begin{align}
\begin{pmatrix} N_{+L} \cr N_{-L} \end{pmatrix} =
a \begin{pmatrix}  S_L^{\hat{c}} \cr  \alpha_+ S_L^{\hat{c}} \end{pmatrix}
+ b  \begin{pmatrix}  C_L^{\hat{c}} \cr  \alpha_+ C_L^{\hat{c}} \end{pmatrix}
- a' \begin{pmatrix}  S_R^{\hat{c}} \cr  \alpha_- S_R^{\hat{c}} \end{pmatrix}
- b' \begin{pmatrix}  C_R^{\hat{c}} \cr  \alpha_- C_R^{\hat{c}} \end{pmatrix} .
\label{M2fermionSol2}
\end{align}

Basis functions for Case II are defined as follows.
\begin{align}
\hat{\cal C}_{R1}(z;\lambda, c, \tilde m) &= C_R(z;\lambda, \hat{c})+\alpha_+^2C_L(z;\lambda, \hat{c}) ~, \cr
\hat{\cal C}_{R2}(z;\lambda, c, \tilde m) &= \alpha_+ \big\{ S_L(z;\lambda, \hat{c})+S_R(z;\lambda, \hat{c}) \big\} ~, \cr
\hat{\cal S}_{L1}(z;\lambda, c, \tilde m) &= S_L(z;\lambda, \hat{c})-\alpha_+^2S_R(z;\lambda, \hat{c}) ~, \cr
\hat{\cal S}_{L2}(z;\lambda, c, \tilde m) &= \alpha_+ \big\{ C_L(z;\lambda, \hat{c})-C_R(z;\lambda, \hat{c}) \big\} ~, \cr
\hat{\cal C}_{L1}(z;\lambda, c, \tilde m) &= C_L(z;\lambda,\hat{c})+\alpha_+^2C_R(z;\lambda,\hat{c}) ~, \cr
\hat{\cal C}_{L2}(z;\lambda, c, \tilde m) &= \alpha_+\big\{ S_R(z;\lambda,\hat{c})+S_L(z;\lambda,\hat{c}) \big\}   ~, \cr
\hat{\cal S}_{R1}(z;\lambda, c, \tilde m) &= S_R(z;\lambda,\hat{c})-\alpha_+^2 S_L(z;\lambda,\hat{c}) ~, \cr
\hat{\cal S}_{R2}(z;\lambda, c, \tilde m) &= \alpha_+\big\{C_R(z;\lambda,\hat{c})-C_L(z;\lambda,\hat{c}) \big\}  ~.
\label{M2fermionBasis1}
\end{align}
We note that $\hat{\cal C}_{L2}(z;\lambda, c, \tilde m) = \hat{\cal C}_{R2}(z;\lambda, c, \tilde m)$ and
$\hat{\cal S}_{L2}(z;\lambda, c, \tilde m) = - \hat{\cal S}_{R2}(z;\lambda, c, \tilde m)$.
With the aid of (\ref{alphapm}) and (\ref{M2fermionIdentity1}), one finds
\begin{align}
&D_- (c) \begin{pmatrix} \hat{\cal C}_{R1} \cr \hat{\cal C}_{R2} \end{pmatrix}
= \lambda \begin{pmatrix} \hat{\cal S}_{L1} \cr \hat{\cal S}_{L2} \end{pmatrix}
+ \frac{\tilde m}{z} \begin{pmatrix} \hat{\cal S}_{L2} \cr \hat{\cal S}_{L1} \end{pmatrix} , \cr
\noalign{\kern 5pt}
&D_+ (c) \begin{pmatrix} \hat{\cal S}_{L1} \cr \hat{\cal S}_{L2} \end{pmatrix}
= \lambda \begin{pmatrix} \hat{\cal C}_{R1} \cr \hat{\cal C}_{R2} \end{pmatrix}
- \frac{\tilde m}{z} \begin{pmatrix} \hat{\cal C}_{R2} \cr \hat{\cal C}_{R1} \end{pmatrix} , \cr
\noalign{\kern 5pt}
&D_- (c) \begin{pmatrix} \hat{\cal S}_{R1} \cr \hat{\cal S}_{R2} \end{pmatrix}
= \lambda \begin{pmatrix} \hat{\cal C}_{L1} \cr \hat{\cal C}_{L2} \end{pmatrix}
- \frac{\tilde m}{z} \begin{pmatrix} \hat{\cal C}_{L2} \cr \hat{\cal C}_{L1} \end{pmatrix} , \cr
\noalign{\kern 5pt}
&D_+ (c) \begin{pmatrix} \hat{\cal C}_{L1} \cr \hat{\cal C}_{L2} \end{pmatrix}
= \lambda \begin{pmatrix} \hat{\cal S}_{R1} \cr \hat{\cal S}_{R2} \end{pmatrix}
+ \frac{\tilde m}{z} \begin{pmatrix} \hat{\cal S}_{R2} \cr \hat{\cal S}_{R1} \end{pmatrix} , \cr
\noalign{\kern 5pt}
&\hat{\cal S}_{Rj} = \hat{\cal S}_{Lj} = D_-(c)\,  \hat{\cal C}_{Rj} = D_+(c) \,  \hat{\cal C}_{Lj} = 0 \quad
{\rm at~} z = z_L ~.
\label{M2fermionBasis2}
\end{align}
Note
\begin{align}
&\hat{\cal S}_{R1} \hat{\cal C}_{L1}  -\hat{\cal S}_{R2} \hat{\cal C}_{L2} 
 = (1 + \alpha_+^2) (  S_R^{\hat c} C_L^{\hat c} - \alpha_+^2   S_L^{\hat c}C_R^{\hat c}) ~.
 \label{M2fermionIdentity2}
\end{align}

As $c \go -c$, $\alpha_\pm \go - \alpha_\mp$ so that
\begin{align}
\hat{\cal C}_{Rj}(z;\lambda, - c, \tilde m) &= \alpha_-^2  \hat{\cal C}_{Lj}(z;\lambda,  c, \tilde m) ~, \cr
\hat{\cal C}_{Lj}(z;\lambda, - c, \tilde m) &= \alpha_-^2  \hat{\cal C}_{Rj}(z;\lambda,  c, \tilde m) ~, \cr
\hat{\cal S}_{Rj}(z;\lambda, - c, \tilde m) &= - \alpha_-^2  \hat{\cal S}_{Lj}(z;\lambda,  c, \tilde m) ~, \cr
\hat{\cal S}_{Lj}(z;\lambda, - c, \tilde m) &= - \alpha_-^2  \hat{\cal S}_{Rj}(z;\lambda,  c, \tilde m) ~.
\label{M2fermionIdentity3}
\end{align}
Further, as $\tilde m \go - \tilde m$, $\alpha_\pm \go - \alpha_\pm$ and
\begin{align}
\hat{\cal C}_{R/Lj}(z;\lambda,  c, -\tilde m) &=  (-1)^{j-1} \,  \hat{\cal C}_{R/Lj}(z;\lambda,  c, \tilde m) ~, \cr
\hat{\cal S}_{R/Lj}(z;\lambda,  c, -\tilde m) &=  (-1)^{j-1} \,  \hat{\cal S}_{R/Lj}(z;\lambda,  c, \tilde m) ~.
\label{M2fermionIdentity4}
\end{align}
In the $\tilde m \go 0$ limit
\begin{align}
&\hat{\cal C}_{R/L1 }(z;\lambda,  c, 0) = C_{R/L1 }(z;\lambda,  c)~, ~~
\hat{\cal S}_{R/L1 }(z;\lambda,  c, 0) = S_{R/L1 }(z;\lambda,  c)~, \cr
&\hat{\cal C}_{R/L2 }(z;\lambda,  c, 0) = \hat{\cal S}_{R/L2 }(z;\lambda,  c, 0) = 0 ~.
\label{M2fermionIdentity5}
\end{align}

Two types of boundary conditions appear at $z=z_L$.

\vskip 5pt
\noindent
\underline{Type A: ~$(N_{+R}, N_{-R}, N_{+L}, N_{-L})= (+,-,-,+)$}

When parity assignment  at $y=L$ for $(N_{+R}, N_{-R}, N_{+L}, N_{-L})$ is  $(+,-,-,+)$,
boundary conditions at $z=z_L$ become
\begin{align}
& D_-(c)N_{+R}=0 ~,~~ N_{+L}=0 ~, \cr
&N_{-R}=0 ~, ~~ D_-(c) N_{-L}=0 ~,
\label{M2fermionBC1}
\end{align}
which leads to the conditions for the parameters in (\ref{M2fermionSol1}) and (\ref{M2fermionSol2})
\begin{align}
\left\{
 \begin{array}{l}
  a\alpha_+ + a'\alpha_- =0,\\
  b-b'=0.\\
 \end{array}
 \right.
\end{align}
It follows that solutions can be written as
\begin{align}
&\begin{pmatrix}   N_{+R} \cr N_{+L} \cr N_{-R} \cr N_{-L} \end{pmatrix} 
= \tilde a \begin{pmatrix}  \hat{\cal C}_{R1}(z; \lambda, c, \tilde m) \cr  \hat{\cal S}_{L1}(z; \lambda, c, \tilde m) \cr
- \hat{\cal S}_{L2}(z; \lambda, c, \tilde m) \cr  \hat{\cal C}_{R2}(z; \lambda, c, \tilde m) \end{pmatrix}
+ \tilde b \begin{pmatrix}  \hat{\cal C}_{R2}(z; \lambda, c, \tilde m) \cr  \hat{\cal S}_{L2}(z; \lambda, c, \tilde m) \cr
- \hat{\cal S}_{L1}(z; \lambda, c, \tilde m) \cr  \hat{\cal C}_{R1}(z; \lambda, c, \tilde m) \end{pmatrix} ,
 \label{M2fermionWave1}
\end{align}
where $\tilde a = a$ and $\tilde b = b/\alpha_+$ are arbitrary constants.

If $N$'s have the same parity assignment at $y=0$ as that at $y=L$, then
(\ref{M2fermionBC1}) must be satisfied at $z=1$ as well. Substituting (\ref{M2fermionWave1})
into (\ref{M2fermionBC1}) and evaluating the conditions at $z=1$, one finds
\begin{align}
\begin{pmatrix} \hat{\cal S}_{L1} & \hat{\cal S}_{L2} \cr \hat{\cal S}_{L2} & \hat{\cal S}_{L1} \end{pmatrix}
\begin{pmatrix} \tilde a \cr \tilde b \end{pmatrix} = 0
\label{M2fermionBC2}
\end{align}
where $ \hat{\cal S}_{L1}  =  \hat{\cal S}_{L1} (1; \lambda, c, \tilde m)$ etc..
The mass spectrum is determined by
\begin{align}
\hat{\cal S}_{L1}^2    -  \hat{\cal S}_{L2}^2  = 0 ~.
\label{M2fermionSpectrum1} 
\end{align}


\vskip 5pt
\noindent
\underline{Type B: ~$(N_{+R}, N_{-R}, N_{+L}, N_{-L})= (-,+,+,-)$}

When parity assignment  at $y=L$ for $(N_{+R}, N_{-R}, N_{+L}, N_{-L})$ is  $(-,+,+,-)$,
boundary conditions at $z=z_L$ become
\begin{align}
&N_{+R}=0 ~, ~~D_+(c) N_{+L}=0 ~, \cr
&D_+ (c) N_{-R}=0 ~, ~~N_{-L}=0 ~.
\label{M2fermionBC3}
\end{align}
This leads to
\begin{align}
\left\{
\begin{array}{l}  a+a'=0,\\  b\alpha_+-b'\alpha_-=0.\\ \end{array}
\right.
\end{align}
It follows that solutions can be written as
\begin{align}
&\begin{pmatrix}   N_{+R} \cr N_{+L} \cr N_{-R} \cr N_{-L} \end{pmatrix} 
= \tilde a \begin{pmatrix}  \hat{\cal S}_{R2}(z; \lambda, c, \tilde m) \cr  \hat{\cal C}_{L2}(z; \lambda, c, \tilde m) \cr
 \hat{\cal C}_{L1}(z; \lambda, c, \tilde m) \cr  - \hat{\cal S}_{R1}(z; \lambda, c, \tilde m) \end{pmatrix}
 + \tilde b  \begin{pmatrix}  \hat{\cal S}_{R1}(z; \lambda, c, \tilde m) \cr  \hat{\cal C}_{L1}(z; \lambda, c, \tilde m) \cr
 \hat{\cal C}_{L2}(z; \lambda, c, \tilde m) \cr - \hat{\cal S}_{R2}(z; \lambda, c, \tilde m) \end{pmatrix} ,
 \label{M2fermionWave2}
\end{align}
where $\tilde a = a/\alpha_+$ and $\tilde b = b$ are arbitrary constants.

If $N$'s have the same parity assignment at $y=0$ as that at $y=L$, then
(\ref{M2fermionBC3}) must be satisfied at $z=1$ as well. 
Substituting (\ref{M2fermionWave2})
into (\ref{M2fermionBC3}) and evaluating the conditions at $z=1$, one finds
\begin{align}
\begin{pmatrix} \hat{\cal S}_{R2} & \hat{\cal S}_{R1} \cr \hat{\cal S}_{R1} & \hat{\cal S}_{R2} \end{pmatrix}
\begin{pmatrix} \tilde a \cr \tilde b \end{pmatrix} = 0 ~. 
\label{M2fermionBC4}
\end{align}
The mass spectrum is determined by
\begin{align}
\hat{\cal S}_{R1}^2    -  \hat{\cal S}_{R2}^2   = 0 ~.
\label{M2fermionSpectrum3} 
\end{align}

\section{Majorana fermions}
\label{Sec:Majorana}

We summarize the notation  adopted in the present paper concerning  Majorana fermions
in four dimensions. Dirac matrices are
\begin{align}
&\{ \gamma^\mu , \gamma^\nu \} = 2 \eta^{\mu \nu} ~,~~ \eta^{\mu \nu} = {\rm diag} (-1, 1,1,1)~, \cr
\noalign{\kern 5pt}
&\gamma^\mu = \begin{pmatrix} &\sigma^\mu \cr \bar \sigma^\mu & \end{pmatrix} , ~~
\begin{pmatrix} \sigma^\mu \cr \bar \sigma^\mu \end{pmatrix} = (\pm I_2 ,  \vec \sigma ) ~,~~
\gamma^5 = \begin{pmatrix} I_2 & \cr & - I_2 \end{pmatrix} .
\end{align}
We define $\overline{\psi} = i \psi^\dagger \gamma^0$. 
Charge conjugation is given by $\psi^C = U_C (\overline{\psi })^t$ where
$U_C \gamma^{\mu t} U_C^\dagger = - \gamma^\mu$.
In our representation
\begin{align}
U_C = i e^{i \delta_c} \begin{pmatrix} \sigma^2 & \cr & \sigma^2 \end{pmatrix}, ~~
\psi = \begin{pmatrix} \xi \cr \eta \end{pmatrix} ~ \go ~
\psi^C = \begin{pmatrix} \eta^c \cr - \xi^c \end{pmatrix} = 
e^{i \delta_c} \begin{pmatrix} \sigma^2 \eta^* \cr - \sigma^2 \xi^* \end{pmatrix} .
\label{Uc1}
\end{align}
Note $(\psi^C)^C = \psi$ whereas $(\eta^c)^c = - \eta$ and $(\xi^c)^c = - \xi$.
It follows that
\begin{align}
&\overline{\psi_1} \psi_2 = - i \eta_1^\dagger \xi_2 + i \xi_1^\dagger \eta_2 = \overline{\psi_2^C} \psi_1^C ~, \cr
\noalign{\kern 5pt}
&\overline{\psi_1} \gamma^\mu \dd_\mu \psi_2 
= - i \eta_1^\dagger \sigma^\mu \dd_\mu \eta_2 + i \xi_1^\dagger \bar \sigma^\mu \dd_\mu \xi_2 ~,  \cr
\noalign{\kern 5pt}
&- i \eta_1^\dagger \sigma^\mu \dd_\mu \eta_2 
= -i \dd_\mu \eta_2^c {}^\dagger \bar \sigma^\mu \eta_1^c
\sim  i \eta_2^{c \dagger} \bar \sigma^\mu \dd_\mu \eta_1^c ~, 
\label{majorana1}
\end{align}
and so on.

In (\ref{neutrinoWave1}) we have introduced wave functions of mass eigenstates satisfying
\begin{align}
&\bar{\sigma}^\mu\partial_\mu f_{\pm R}(x)= m f_{\pm L}(x) ~,~~ 
{\sigma}^\mu\partial_\mu f_{\pm L}(x) =m f_{\pm R}(x) ~,\cr
\noalign{\kern 5pt}
&f_{\pm L}(x)^c = e^{i\delta_C}\sigma^2 f_{\pm L}(x)^* = \pm f_{\pm R}(x)  ~.
\label{majoranaWave1}
\end{align}
Explicit forms of $ f_{\pm L/R}(x) $ are given, for modes propagating in the $x_3$-direction with
$\vec p = (0,0,p)$, by
\begin{align}
&f_{+L}^{(1)} = \mfrac{1}{\sqrt{2E}} 
\begin{pmatrix} \sqrt{E+p} \, e^{-iEt + ip x_3} \cr \noalign{\kern 3pt}
e^{i\delta_c} \sqrt{E-p} \, e^{iEt - ip x_3} \end{pmatrix} , ~
f_{+R}^{(1)}  = \mfrac{1}{\sqrt{2E}} 
\begin{pmatrix} - i\sqrt{E-p} \, e^{-iEt + ip x_3} \cr \noalign{\kern 3pt}
i e^{i\delta_c} \sqrt{E+p} \, e^{iEt - ip x_3} \end{pmatrix} ,  \cr
\noalign{\kern 5pt}
&f_{+L}^{(2)} = \mfrac{1}{\sqrt{2E}} 
\begin{pmatrix} \sqrt{E+p} \, e^{iEt - ip x_3} \cr \noalign{\kern 3pt}
- e^{i\delta_c} \sqrt{E-p} \, e^{- iEt + ip x_3} \end{pmatrix} , ~
f_{+R}^{(2)}  = \mfrac{1}{\sqrt{2E}} 
\begin{pmatrix}  i\sqrt{E-p} \, e^{ iEt - ip x_3} \cr \noalign{\kern 3pt}
i e^{i\delta_c} \sqrt{E+p} \, e^{- iEt + ip x_3} \end{pmatrix} ,  \cr
\noalign{\kern 5pt}
&f_{-L}^{(1)} = \mfrac{1}{\sqrt{2E}} 
\begin{pmatrix} \sqrt{E+p} \, e^{-iEt + ip x_3} \cr \noalign{\kern 3pt}
- e^{i\delta_c} \sqrt{E-p} \, e^{iEt - ip x_3} \end{pmatrix} , ~
f_{-R}^{(1)}  = \mfrac{1}{\sqrt{2E}} 
\begin{pmatrix} - i\sqrt{E-p} \, e^{-iEt + ip x_3} \cr \noalign{\kern 3pt}
- i e^{i\delta_c} \sqrt{E+p} \, e^{iEt - ip x_3} \end{pmatrix} ,  \cr
\noalign{\kern 5pt}
&f_{-L}^{(2)} = \mfrac{1}{\sqrt{2E}} 
\begin{pmatrix} \sqrt{E+p} \, e^{iEt - ip x_3} \cr \noalign{\kern 3pt}
 e^{i\delta_c} \sqrt{E-p} \, e^{- iEt + ip x_3} \end{pmatrix} , ~
f_{-R}^{(2)}  = \mfrac{1}{\sqrt{2E}} 
\begin{pmatrix}  i\sqrt{E-p} \, e^{ iEt - ip x_3} \cr \noalign{\kern 3pt}
- i e^{i\delta_c} \sqrt{E+p} \, e^{- iEt + ip x_3} \end{pmatrix}.
\label{majoranaWave2}
\end{align}
Here $E = \sqrt{p^2 + m^2}$.


\section{Dark fermions}
\label{Sec:DarkFermion}

In addition to the quark and lepton multiplets we introduce dark fermion multiplets in the bulk,
which give relevant contributions to the effective potential $V_\eff (\theta_H)$ to
induce the electroweak symmetry breaking by the Hosotani mechanism.  
They naturally appear from grand unified theory.

\subsection{$Q_{\rm EM}=  \frac{2}{3}, - \frac{1}{3}$:   $( \Psi_{({\bf 3,4})} \equiv \Psi_F )$}

The bulk mass parameter of this multiplet, $c_F$, is assumed to satisfy $|c_F| < \onehalf$.
$\Psi_F $ satisfies boundary condition (\ref{darkFBC2}).  There are no zero modes.  The spectrum
is vector-like.  $(F_1, F_1')$ in Table~\ref{Tab:parity} forms a pair analogous to $(u, u')$ pair,
whereas $(F_2, F_2')$  to $(d, d')$ pair.  
Both pairs satisfy, in the twisted gauge,  the equations similar to Eq.~(\ref{upEq1})
with $c_Q$ replaced by $c_F$.

With the boundary conditions at $y=L$ taken into account, mode functions can be written as
\begin{align}
\begin{pmatrix}  \tilde{\check F}{}_{1R} \cr  \tilde{\check F}{}_{1R}' \end{pmatrix} 
= \begin{pmatrix}  \alpha_{F} S_R (z,\lambda,c_F) \cr \alpha_{F'} C_R (z,\lambda,c_F) \end{pmatrix} f_R(x) , ~
\begin{pmatrix}  \tilde{\check F}{}_{1L} \cr  \tilde{\check F}{}_{1L}' \end{pmatrix}
= \begin{pmatrix}  \alpha_F C_L (z,\lambda,c_F) \cr \alpha_{F'} S_L (z,\lambda,c_F) \end{pmatrix} f_L(x) .
\label{DarkFWave1}
\end{align}
The boundary conditions at $z=1$ are flipped, however, and we have
$D_- \check{F_1}_{R}=0$ and $ \check{F}_{1R}'=0$ there to find
\begin{align}
K_F \begin{pmatrix} \alpha_F \cr  \alpha_{F'} \end{pmatrix} = 
\begin{pmatrix} \cos \onehalf \theta_H C_L^{F} & -i \sin \onehalf \theta_H  S_L^{F} \cr
 -i \sin \onehalf \theta_H  S_R^{F} & \cos \onehalf \theta_H C_R^{F} \end{pmatrix} 
 \begin{pmatrix} \alpha_F\cr  \alpha_{F'} \end{pmatrix}  = 0 ~.
\end{align}
Here $S_{L/R}^{F} = S_{L/R}(1,\lambda,c_F)$ etc..
$\det K_F=0$ leads to the equation determining the spectrum;
\begin{align}
S_L^{F}S_R^{F}+\cos^2\frac{\theta_H}{2}=0 ~.
\label{DFspectrum1}
\end{align}
There are no light modes for $|c_F| < \onehalf$ and small $\theta_H$.
The spectrum of the $(F_2, F_2')$ pair is also given by (\ref{DFspectrum1}).

\subsection{$Q_{\rm EM}= \pm 1$: $E^{\pm}, \hat E^{\pm}$ 
$( \Psi_{({\bf 1,5})}^{\pm})$}

In general $\Psi_{({\bf 1,5})}^+$ and $\Psi_{({\bf 1,5})}^-$ may have
different bulk mass parameters $c_{V^+}$ and $c_{V^-}$.
For charged particles $E^\pm$, equations of motion are given by
\begin{align}
&-k D_-(c_{V^+})\check{E}_{R}^{+} +\sigma^\mu\partial_\mu\check{E}_{L}^{+}
 -\frac{m_{V}^*}{z}\check{E}_{R}^{-} =0 ~, \cr
\noalign{\kern 5pt}
&\overline{\sigma}^\mu\partial_\mu\check{E}_{R}^{+}
-k D_+(c_{V^+})\check{E}_{L}^{+}-\frac{m_{V}}{z}\check{E}_{L}^{-} = 0 ~, \cr
\noalign{\kern 5pt}
&-k D_-(c_{V^-})\check{E}_{R}^{-} +\sigma^\mu\partial_\mu\check{E}_{L}^{-}
 -\frac{m_{V}^*}{z}\check{E}_{R}^{+} =0 ~, \cr
\noalign{\kern 5pt} 
&\overline{\sigma}^\mu\partial_\mu\check{E}_{R}^{-}
 -k D_+(c_{V^-})\check{E}_{L}^{-}  -\frac{m_{V}}{z}\check{E}_{L}^{+} =0 ~.
\label{DFEeq1}
\end{align}
$E^+$ and $E^-$ couple with each other through the mass $m_V$.
Boundary conditions are given by
$\check E_R^+ = D_+ (c_{V^+}) \check E_L^+ =0$ and 
$ D_- (c_{V^-}) \check E_R^- =\check E_L^- =0$ at  $z=1, z_L$.

Mode functions can be easily found for $c_{V^+} = \pm c_{V^-}$.
They are summarized in Appendix B.3.  We quote the results there.
We note that the same result is obtained for $\hat E^\pm$ as for $E^\pm$.


\subsubsection*{Case I:  $c_{V^+}=c_{V^-}=c_V$}

We denote $\tilde m_V = m_V/k$. The boundary condition is Type B.
Mode functions are given by (\ref{MfermionWave2});

\begin{align}
&\begin{pmatrix}   \check E_R^+ \cr \check E_L^+ \cr \check E_R^- \cr \check E_L^- \end{pmatrix} 
= a \begin{pmatrix}  {\cal S}_{R2}(z; \lambda, c_V, \tilde m_V) \cr  {\cal C}_{L2}(z; \lambda, c_V, \tilde m_V) \cr
 {\cal C}_{R1}(z; \lambda, c_V, \tilde m_V) \cr  {\cal S}_{L1}(z; \lambda, c_V, \tilde m_V) \end{pmatrix}
 + b  \begin{pmatrix}  {\cal S}_{R1}(z; \lambda, c_V, \tilde m_V) \cr  {\cal C}_{L1}(z; \lambda, c_V, \tilde m_V) \cr
 {\cal C}_{R2}(z; \lambda, c_V, \tilde m_V) \cr  {\cal S}_{L2}(z; \lambda, c_V, \tilde m_V) \end{pmatrix} ,
 \label{DFEWave1}
\end{align}
where $a,b$ are arbitrary constants.
The expression is valid both in the original gauge and in the twisted gauge, as these fields do
not couple to $\theta_H$ at the tree level.
The spectrum is determined by (\ref{MfermionSpectrum3});
\begin{align}
{\cal S}_{L1}^{V}{\cal S}_{R1}^{V}  -{\cal S}_{L2}^{V}{\cal S}_{R2}^{V} = 0
\label{DFEspectrum1}
\end{align}
where ${\cal S}_{L1}^{V} = {\cal S}_{L1} (1; \lambda, c_V, \tilde m_V) $ etc..

\subsubsection*{Case II: $c_{V^+}=-c_{V^-}=c_V$}

In this case mode functions are given by (\ref{M2fermionWave2});

\begin{align}
&\begin{pmatrix}  \check E_R^+ \cr \check E_L^+ \cr \check E_R^- \cr \check E_L^- \end{pmatrix} 
= a \begin{pmatrix}  \hat{\cal S}_{R2}(z; \lambda, c_V, \tilde m_V) \cr 
 \hat{\cal C}_{L2}(z; \lambda, c_V, \tilde m_V) \cr
 \hat{\cal C}_{L1}(z; \lambda, c_V, \tilde m_V) \cr  
 - \hat{\cal S}_{R1}(z; \lambda, c_V, \tilde m_V) \end{pmatrix}
 +  b  \begin{pmatrix}  \hat{\cal S}_{R1}(z; \lambda, c_V, \tilde m_V) \cr  
 \hat{\cal C}_{L1}(z; \lambda, c_V, \tilde m_V) \cr
 \hat{\cal C}_{L2}(z; \lambda, c_V, \tilde m_V) \cr 
 - \hat{\cal S}_{R2}(z; \lambda, c_V, \tilde m_V) \end{pmatrix} ,
\label{DFEWave2}
\end{align}
where $a, b$  are arbitrary constants. The spectrum is determined by (\ref{M2fermionSpectrum3});
\begin{align}
(\hat{\cal S}_{R1}^V)^2 - (\hat{\cal S}_{R2}^V)^2 = 0 
\label{DFEspectrum2}
\end{align}
where $\hat{\cal S}_{R1}^V = \hat{\cal S}_{R1} (1; \lambda, c_V, \tilde m_V) $ etc..

\subsection{$Q_{\rm EM}=0$:  $N^{\pm}, \hat{N}^{\pm}, S^{\pm}$ $( \Psi_{({\bf 1,5})}^{\pm})$}

$N^{\pm}$,  $\hat{N}^{\pm}$ and $S^{\pm}$ couple with each other through $\theta_H$.
Equations of motion in the original gauges are 
\begin{align}
-k\hat{D}_-(c_{V^\pm}) \begin{pmatrix}  \check{\hat{N}}_{R}^{\pm} \cr
\check{N}_{R}^{\pm} \cr \check{S}_{R}^{\pm} \end{pmatrix} 
+\sigma^\mu \partial_\mu \begin{pmatrix}  \check{\hat{N}}_{L}^{\pm} \cr
\check{N}_{L}^{\pm} \cr \check{S}_{L}^{\pm} \end{pmatrix} 
- \frac{m_V^*}{z} \begin{pmatrix}  \check{\hat{N}}_{R}^{\mp} \cr
\check{N}_{R}^{\mp} \cr \check{S}_{R}^{\mp} \end{pmatrix} &= 0 ~, \cr
\noalign{\kern 5pt}
\overline{\sigma}^\mu\partial_\mu \begin{pmatrix}  \check{\hat{N}}_{R}^{\pm} \cr
\check{N}_{R}^{\pm} \cr \check{S}_{R}^{\pm} \end{pmatrix} 
- k\hat{D}_+(c_{V^\pm})  \begin{pmatrix}  \check{\hat{N}}_{L}^{\pm} \cr
\check{N}_{L}^{\pm} \cr \check{S}_{L}^{\pm} \end{pmatrix} 
-\frac{m_V}{z} \begin{pmatrix}  \check{\hat{N}}_{L}^{\mp} \cr
\check{N}_{L}^{\mp} \cr \check{S}_{L}^{\mp} \end{pmatrix} &= 0 ~.
\label{EoMneutralF3}
\end{align}
Note $\hat{D}_\pm (c)$ is given by (\ref{derivativeD}).

The relation between the original and twisted gauges are given by
$\Psi_{({\bf 1,5})}^{\pm}= \Omega (z) \widetilde{\Psi}_{({\bf 1,5})}^{\pm}$,
where $\Omega (z) = e^{i\theta(z)T_{45}}$, so that
\begin{align}
&\psi_3 = \widetilde{\psi}_3 ~, ~~
\begin{pmatrix}  \psi_4 \cr \psi_{5} \end{pmatrix} =
\begin{pmatrix}  \cos\theta(z) &\sin\theta(z) \cr  -\sin\theta(z) &\cos\theta(z) \end{pmatrix} , \cr
\noalign{\kern 5pt}
&\psi_{3}^{\pm}=\frac{i}{\sqrt{2}} (\hat{N}^{\pm}+N^{\pm} ) ~, ~~
\psi_{4}^{\pm}=\frac{1}{\sqrt{2}}  (\hat{N}^{\pm}-N^{\pm} ) ~,~~
\psi_{5}^{\pm}=S^{\pm} , 
\label{TwistedVector1}
\end{align}
and threfore
\begin{align}
&\begin{pmatrix}  \check{\hat{N}}^{\pm} \cr \check{N}^{\pm} \cr \check{S}^{\pm} \end{pmatrix} 
= \overline{\Omega} (z) \begin{pmatrix}  \tilde{\check{\hat{N}}}{}^{\pm} \cr 
\tilde{\check{N}}{}^{\pm} \cr \tilde{\check{S}}{}^{\pm} \end{pmatrix} , \cr
\noalign{\kern 5pt}
&\overline{\Omega} (z)  = V \begin{pmatrix} 1 & 0 &0 \cr 0 & \cos \theta(z) & \sin\theta (z) \cr
0 & - \sin \theta (z) &\cos \theta (z)\end{pmatrix} V ^{-1} , \cr
\noalign{\kern 5pt}
&V = V^{-1} = \begin{pmatrix} \frac{1}{\sqrt{2}} &  \frac{1}{\sqrt{2}} & 0 \cr 
\frac{1}{\sqrt{2}} &  -\frac{1}{\sqrt{2}} & 0 \cr 0 & 0& 1 \end{pmatrix}.
\label{TwistedVector2}
\end{align}
It follows that
\begin{align}
\hat{D}_- (c_{V^\pm}) \begin{pmatrix}  \check{\hat{N}}_{R}^{\pm} \cr
\check{N}_{R}^{\pm} \cr \check{S}_{R}^{\pm} \end{pmatrix}  =
\overline{\Omega} (z) D_- (c_{V^\pm}) \begin{pmatrix}  \tilde{\check{\hat{N}}}{}_R^{\pm} \cr 
\tilde{\check{N}}{}_R^{\pm} \cr \tilde{\check{S}}{}_R^{\pm} \end{pmatrix} , 
\label{TwistedVector3}
\end{align}
and so on.
Boundary conditions in the original gauge are 
\begin{align}
&\check{\hat{N}}_{R}^{+} = \hat{D}_+(c_{V^+})\check{\hat{N}}_{L}^{+} 
= \hat{D}_-(c_{V^-})\check{\hat{N}}_{R}^{-} = \check{\hat{N}}_{L}^{-} =0 ~, \cr
\noalign{\kern 5pt}
& \check{N}_{R}^{+} = \hat{D}_+(c_{V^+})\check{N}_{L}^{+} 
= \hat{D}_-(c_{V^-})\check{N}_{R}^{-} = \check{N}_{L}^{-} = 0 ~, \cr
\noalign{\kern 5pt}
&\hat{D}_-(c_{V^+})\check{S}_{R}^{+} =  \check{S}_{L}^{+} 
= \check{S}_{R}^{-}=\hat{D}_+(c_{V^-})\check{S}_{L}^{-} =  0~, 
\label{DFneutralBC1}
\end{align}
at both $z=1$ and $z=z_L$.

\subsubsection*{Case I:  $c_{V^+}=c_{V^-}=c_V$}

The boundary conditions in the twisted gauge at $z=z_L$ are obtained
from (\ref{DFneutralBC1}) by replacing $\hat D_\pm (c)$ by $D_\pm (c)$.
Mode functions of $N$ and $\hat N$ fields are given by (\ref{MfermionWave2}),
whereas those of $S$ field by (\ref{MfermionWave1});
\begin{align}
\begin{pmatrix} \widetilde{\check{\hat{N}}}{}_{R}^{+} \cr \widetilde{\check{\hat{N}}}{}_{L}^{+} \cr
\widetilde{\check{\hat{N}}}{}_{R}^{-} \cr  \widetilde{\check{\hat{N}}}{}_{L}^{-} \end{pmatrix}
&=a_{\hat{N}} \begin{pmatrix}   {\cal S}_{R2}(z; \lambda, c_V, \tilde m_V) \cr
{\cal C}_{L2}(z;\lambda, c_V, \tilde m_V) \cr  {\cal C}_{R1}(z;\lambda, c_V, \tilde m_V) \cr
{\cal S}_{L1}(z;\lambda, c_V, \tilde m_V) \end{pmatrix}
 +b_{\hat{N}} \begin{pmatrix}  {\cal S}_{R1}(z;\lambda, c_V, \tilde m_V) \cr
{\cal C}_{L1}(z;\lambda, c_V, \tilde m_V) \cr  {\cal C}_{R2}(z;\lambda, c_V, \tilde m_V) \cr
{\cal S}_{L2}(z;\lambda, c_V, \tilde m_V) \end{pmatrix} , \cr
\noalign{\kern 5pt}
\begin{pmatrix} \widetilde{\check{N}}{}_{R}^{+} \cr \widetilde{\check{N}}{}_{L}^{+} \cr
\widetilde{\check{N}}{}_{R}^{-} \cr  \widetilde{\check{N}}{}_{L}^{-} \end{pmatrix}
&=a_{N} \begin{pmatrix}   {\cal S}_{R2}(z; \lambda, c_V, \tilde m_V) \cr
{\cal C}_{L2}(z;\lambda, c_V, \tilde m_V) \cr  {\cal C}_{R1}(z;\lambda, c_V, \tilde m_V) \cr
{\cal S}_{L1}(z;\lambda, c_V, \tilde m_V) \end{pmatrix}
 +b_{N} \begin{pmatrix}  {\cal S}_{R1}(z;\lambda, c_V, \tilde m_V) \cr
{\cal C}_{L1}(z;\lambda, c_V, \tilde m_V) \cr  {\cal C}_{R2}(z;\lambda, c_V, \tilde m_V) \cr
{\cal S}_{L2}(z;\lambda, c_V, \tilde m_V) \end{pmatrix} , \cr
\noalign{\kern 5pt}
\begin{pmatrix} \widetilde{\check S}{}_{R}^{+} \cr \widetilde{\check S}{}_{L}^{+} \cr
\widetilde{\check S}{}_{R}^{-} \cr \widetilde{\check S}{}_{L}^{-} \end{pmatrix} 
&=a_{S} \begin{pmatrix}  {\cal C}_{R1}(z;\lambda, c_V, \tilde m_V) \cr 
{\cal S}_{L1}(z;\lambda, c_V, \tilde m_V) \cr  {\cal S}_{R2}(z;\lambda, c_V, \tilde m_V) \cr
{\cal C}_{L2}(z;\lambda, c_V, \tilde m_V) \end{pmatrix} 
 +b_{S} \begin{pmatrix}  {\cal C}_{R2}(z;\lambda, c_V, \tilde m_V) \cr
{\cal S}_{L2}(z;\lambda, c_V, \tilde m_V) \cr {\cal S}_{R1}(z;\lambda, c_V, \tilde m_V) \cr
{\cal C}_{L1}(z;\lambda, c_V, \tilde m_V) \end{pmatrix} , 
\label{WaveDFN1} 
\end{align}
where $\tilde m_V = m_V/k$ and 
$a_{\hat{N}}$, $b_{\hat{N}}$, $a_{N}$, $b_{N}$, $a_{S}$, $b_{S}$ are arbitrary parameters. 

We insert (\ref{WaveDFN1} ) into the boundary conditions (\ref{DFneutralBC1}) at $z=1$.
With the aid of (\ref{TwistedVector2}) and (\ref{TwistedVector3}) one finds that
\begin{align}
&K_N \, 
\begin{pmatrix} a_{\hat{N}} \cr a_{N} \cr a_S \cr  b_{\hat{N}} \cr   b_{N} \cr b_S \end{pmatrix} =0 ~, ~~
K_N  = \begin{pmatrix} V & 0 \cr 0 & V \end{pmatrix}
\begin{pmatrix} A & B \cr C & D \end{pmatrix} \begin{pmatrix} V & 0 \cr 0 & V \end{pmatrix} , \cr
\noalign{\kern 10pt}
&A= \begin{pmatrix} {\cal S}_{R2}^V & 0 &0  \cr 
0& c_H{\cal S}_{R2}^V & s_H{\cal C}_{R1}^V \cr
0&-s_H{\cal C}_{L2}^V & c_H{\cal S}_{L1}^V \end{pmatrix} , ~~
B= \begin{pmatrix} {\cal S}_{R1}^V & 0 &0 \cr
0& c_H{\cal S}_{R1}^V & s_H{\cal C}_{R2}^V \cr
0&-s_H{\cal C}_{L1}^V & c_H{\cal S}_{L2}^V \end{pmatrix} ,  \cr
\noalign{\kern 5pt}
&C= \begin{pmatrix} {\cal S}_{L1}^V & 0 &0  \cr
0& c_H{\cal S}_{L1}^V & s_H{\cal C}_{L2}^V \cr
0&-s_H{\cal C}_{R1}^V & c_H{\cal S}_{R2}^V \end{pmatrix} , ~~
D = \begin{pmatrix}  {\cal S}_{L2}^V & 0 &0 \cr
0& c_H{\cal S}_{L2}^V & s_H{\cal C}_{L1}^V \cr
0&-s_H{\cal C}_{R2}^V & c_H{\cal S}_{R1}^V \end{pmatrix}
\label{DFneutralBC2}
\end{align}
where $c_H = \cos \theta_H$, $s_H = \sin \theta_H$  
and ${\cal S}_{L1}^V = {\cal S}_{L1} (1;\lambda, c_V, \tilde m_V)$ etc..
The spectrum is determined by $\det K_N =0$;
\begin{align}
&\det K_N = \det
\begin{pmatrix}  {\cal S}_{R2}^V&{\cal S}_{R1}^V \cr {\cal S}_{L1}^V&{\cal S}_{L2}^V \end{pmatrix}
\det \begin{pmatrix}
c_H{\cal S}_{R2}^V& s_H{\cal C}_{R1}^V& c_H{\cal S}_{R1}^V& s_H{\cal C}_{R2}^V \cr
-s_H{\cal C}_{L2}^V& c_H{\cal S}_{L1}^V& -s_H{\cal C}_{L1}^V& c_H{\cal S}_{L2}^V \cr
c_H{\cal S}_{L1}^V& s_H{\cal C}_{L2}^V& c_H{\cal S}_{L2}^V& s_H{\cal C}_{L1}^V \cr
-s_H{\cal C}_{R1}^V& c_H{\cal S}_{R2}^V& -s_H{\cal C}_{R2}^V& c_H{\cal S}_{R1}^V \end{pmatrix} \cr
\noalign{\kern 10pt}
&= ({\cal S}_{L1}^V{\cal S}_{R1}^V  -{\cal S}_{L2}^V{\cal S}_{R2}^V )
\Big\{ c_H^4 ({\cal S}_{L1}^V{\cal S}_{R1}^V -{\cal S}_{L2}^V{\cal S}_{R2}^V )^2
 +s_H^4 ({\cal C}_{L1}^V{\cal C}_{R1}^V -{\cal C}_{L2}^V{\cal C}_{R2}^V )^2 \cr
\noalign{\kern 5pt}
&\quad
+s_H^2c_H^2 ({\cal C}_{R1}^V{\cal S}_{L2}^V -{\cal C}_{R2}^V{\cal S}_{L1}^V )^2
+s_H^2c_H^2 ({\cal C}_{L1}^V{\cal S}_{R2}^V  -{\cal C}_{L2}^V{\cal S}_{R1}^V )^2 \cr
\noalign{\kern 5pt}
&\quad
+2s_H^2c_H^2 ({\cal C}_{L1}^V{\cal S}_{L1}^V  -{\cal C}_{L2}^V{\cal S}_{L2}^V )
({\cal C}_{R1}^V{\cal S}_{R1}^V  -{\cal C}_{R2}^V{\cal S}_{R2}^V ) \Big\} ~.
\label{DFneutralSpectrum1}
\end{align}

\subsubsection*{Case II: $c_{V^+}= - c_{V^-}=c_V$}

The boundary conditions (\ref{DFneutralBC1}) become
\begin{align}
&\check{\hat{N}}_{R}^{+} = \hat{D}_+(c_{V})\check{\hat{N}}_{L}^{+} 
= \hat{D}_+ (c_{V})\check{\hat{N}}_{R}^{-} = \check{\hat{N}}_{L}^{-} =0 ~, \cr
\noalign{\kern 5pt}
& \check{N}_{R}^{+} = \hat{D}_+(c_{V})\check{N}_{L}^{+} 
= \hat{D}_+(c_{V})\check{N}_{R}^{-} = \check{N}_{L}^{-} = 0 ~, \cr
\noalign{\kern 5pt}
&\hat{D}_-(c_{V})\check{S}_{R}^{+} =  \check{S}_{L}^{+} 
= \check{S}_{R}^{-}=\hat{D}_- (c_{V})\check{S}_{L}^{-} =  0~, 
\label{DFneutralBC3}
\end{align}
at both $z=1$ and $z=z_L$.
Mode functions of $N$ and $\hat N$ fields are given by (\ref{M2fermionWave2}),
whereas those of $S$ field by (\ref{M2fermionWave1});
\begin{align}
\begin{pmatrix} \widetilde{\check{\hat{N}}}{}_{R}^{+} \cr \widetilde{\check{\hat{N}}}{}_{L}^{+} \cr
\widetilde{\check{\hat{N}}}{}_{R}^{-} \cr  \widetilde{\check{\hat{N}}}{}_{L}^{-} \end{pmatrix}
&=a_{\hat{N}} \begin{pmatrix}   \hat{\cal S}_{R2}(z; \lambda, c_V, \tilde m_V) \cr
\hat{\cal C}_{L2}(z;\lambda, c_V, \tilde m_V) \cr  \hat{\cal C}_{L1}(z;\lambda, c_V, \tilde m_V) \cr
-\hat{\cal S}_{R1}(z;\lambda, c_V, \tilde m_V) \end{pmatrix}
 +b_{\hat{N}} \begin{pmatrix}  \hat{\cal S}_{R1}(z;\lambda, c_V, \tilde m_V) \cr
\hat{\cal C}_{L1}(z;\lambda, c_V, \tilde m_V) \cr  \hat{\cal C}_{L2}(z;\lambda, c_V, \tilde m_V) \cr
- \hat{\cal S}_{R2}(z;\lambda, c_V, \tilde m_V) \end{pmatrix} , \cr
\noalign{\kern 5pt}
\begin{pmatrix} \widetilde{\check{N}}{}_{R}^{+} \cr \widetilde{\check{N}}{}_{L}^{+} \cr
\widetilde{\check{N}}{}_{R}^{-} \cr  \widetilde{\check{N}}{}_{L}^{-} \end{pmatrix}
&=a_{N} \begin{pmatrix}   \hat{\cal S}_{R2}(z; \lambda, c_V, \tilde m_V) \cr
\hat{\cal C}_{L2}(z;\lambda, c_V, \tilde m_V) \cr  \hat{\cal C}_{L1}(z;\lambda, c_V, \tilde m_V) \cr
-\hat{\cal S}_{R1}(z;\lambda, c_V, \tilde m_V) \end{pmatrix}
 +b_{N} \begin{pmatrix}  \hat{\cal S}_{R1}(z;\lambda, c_V, \tilde m_V) \cr
\hat{\cal C}_{L1}(z;\lambda, c_V, \tilde m_V) \cr  \hat{\cal C}_{L2}(z;\lambda, c_V, \tilde m_V) \cr
- \hat{\cal S}_{R2}(z;\lambda, c_V, \tilde m_V) \end{pmatrix} , \cr
\noalign{\kern 5pt}
\begin{pmatrix} \widetilde{\check{S}}{}_{R}^{+} \cr \widetilde{\check{S}}{}_{L}^{+} \cr
\widetilde{\check{S}}{}_{R}^{-} \cr  \widetilde{\check{S}}{}_{L}^{-} \end{pmatrix}
&=a_{S} \begin{pmatrix}  \hat{\cal C}_{R1}(z;\lambda, c_V, \tilde m_V) \cr 
\hat{\cal S}_{L1}(z;\lambda, c_V, \tilde m_V) \cr  - \hat{\cal S}_{L2}(z;\lambda, c_V, \tilde m_V) \cr
\hat{\cal C}_{R2}(z;\lambda, c_V, \tilde m_V) \end{pmatrix} 
 +b_{S} \begin{pmatrix}  \hat{\cal C}_{R2}(z;\lambda, c_V, \tilde m_V) \cr
\hat{\cal S}_{L2}(z;\lambda, c_V, \tilde m_V) \cr - \hat{\cal S}_{L1}(z;\lambda, c_V, \tilde m_V) \cr
\hat{\cal C}_{R1}(z;\lambda, c_V, \tilde m_V) \end{pmatrix} , 
\label{WaveDFN2} 
\end{align}
where 
$a_{\hat{N}}$, $b_{\hat{N}}$, $a_{N}$, $b_{N}$, $a_{S}$, $b_{S}$ are arbitrary parameters. 

We insert (\ref{WaveDFN2}) into the boundary conditions (\ref{DFneutralBC3}) at $z=1$.
This time we have, instead of (\ref{DFneutralBC2}), 
\begin{align}
&K_N  = \begin{pmatrix} V & 0 \cr 0 & V \end{pmatrix}
\begin{pmatrix} \hat A & \hat B \cr \hat C & \hat D \end{pmatrix} 
\begin{pmatrix} V & 0 \cr 0 & V \end{pmatrix} , \cr
\noalign{\kern 10pt}
&\hat A= \begin{pmatrix} \hat{\cal S}_{R2}^V & 0 &0  \cr 
0& c_H \hat{\cal S}_{R2}^V & s_H \hat{\cal C}_{R1}^V \cr
0&-s_H \hat{\cal C}_{L2}^V & c_H \hat{\cal S}_{L1}^V \end{pmatrix} , ~~
\hat B= \begin{pmatrix} \hat{\cal S}_{R1}^V & 0 &0 \cr
0& c_H \hat{\cal S}_{R1}^V & s_H \hat{\cal C}_{R2}^V \cr
0&-s_H \hat{\cal C}_{L1}^V & c_H \hat{\cal S}_{L2}^V \end{pmatrix} ,  \cr
\noalign{\kern 5pt}
&\hat C= \begin{pmatrix} \hat{\cal S}_{R1}^V & 0 &0  \cr
0& c_H \hat{\cal S}_{R1}^V & - s_H \hat{\cal C}_{R2}^V \cr
0&-s_H \hat{\cal C}_{L1}^V & - c_H \hat{\cal S}_{L2}^V \end{pmatrix} , ~~
\hat D = \begin{pmatrix}  \hat{\cal S}_{R2}^V & 0 &0 \cr
0& c_H \hat{\cal S}_{R2}^V & - s_H \hat{\cal C}_{R1}^V \cr
0&-s_H \hat{\cal C}_{L2}^V & - c_H \hat{\cal S}_{L1}^V \end{pmatrix}
\label{DFneutralBC4}
\end{align}
where  $\hat{\cal S}_{L1}^V = \hat{\cal S}_{L1} (1;\lambda, c_V, \tilde m_V)$ etc..
The spectrum is determined by
\begin{align}
&\det K_N = \Big\{ (\hat{\cal S}_{R1}^V)^2-(\hat{\cal S}_{R2}^V)^2 \Big\} \cr
\noalign{\kern 5pt}
&\hskip .5cm \times
\Big\{\cos^4\theta_H \left[ (\hat{\cal S}_{L1}^V)^2-(\hat{\cal S}_{L2}^V)^2\right]
 \left[ (\hat{\cal S}_{R1}^V)^2-(\hat{\cal S}_{R2}^V)^2\right] \cr
\noalign{\kern 5pt}
&\hskip 1.cm
+\sin^4\theta_H \left[ (\hat{\cal C}_{L1}^V)^2-(\hat{\cal C}_{L2}^V)^2 \right]
 \left[ (\hat{\cal C}_{R1}^V)^2-(\hat{\cal C}_{R2}^V)^2\right] \cr
\noalign{\kern 5pt}
&\hskip 1.cm
+2\sin^2\theta_H\cos^2\theta_H
\left(\hat{\cal S}_{L1}^V \hat{\cal S}_{R1}^V-\hat{\cal S}_{L2}^V\hat{\cal S}_{R2}^V\right)
\left(\hat{\cal C}_{L1}^V \hat{\cal C}_{R1}^V-\hat{\cal C}_{L2}^V \hat{\cal C}_{R2}^V\right) \cr
\noalign{\kern 5pt}
&\hskip 1.cm
-2\sin^2\theta_H\cos^2\theta_H
\left(\hat{\cal S}_{L1}^V \hat{\cal S}_{R2}^V-\hat{\cal S}_{L2}^V \hat{\cal S}_{R1}^V\right)
\left(\hat{\cal C}_{L1}^V \hat{\cal C}_{R2}^V- \hat{\cal C}_{L2}^V \hat{\cal C}_{R1}^V\right)  \Big\} = 0 ~.
\label{DFneutralSpectrum2}
\end{align}

\vskip 1.cm

\def\jnl#1#2#3#4{{#1}{\bf #2},  #3 (#4)}

\def\Zphys{{\em Z.\ Phys.} }
\def\jssc{{\em J.\ Solid State Chem.\ }}
\def\jpsJ{{\em J.\ Phys.\ Soc.\ Japan }}
\def\ptps{{\em Prog.\ Theoret.\ Phys.\ Suppl.\ }}
\def\PTP{{\em Prog.\ Theoret.\ Phys.\  }}
\def\PTEP{{\em Prog.\ Theoret.\ Exp.\  Phys.\  }}
\def\JMP{{\em J. Math.\ Phys.} }
\def\NPB{{\em Nucl.\ Phys.} B}
\def\NP{{\em Nucl.\ Phys.} }
\def\PLB{{\it Phys.\ Lett.} B}
\def\PL{{\em Phys.\ Lett.} }
\def\PRL{\em Phys.\ Rev.\ Lett. }
\def\PRB{{\em Phys.\ Rev.} B}
\def\PRD{{\em Phys.\ Rev.} D}
\def\PRe{{\em Phys.\ Rep.} }
\def\AP{{\em Ann.\ Phys.\ (N.Y.)} }
\def\RMP{{\em Rev.\ Mod.\ Phys.} }
\def\ZPC{{\em Z.\ Phys.} C}
\def\SCI{\em Science}
\def\CMP{\em Comm.\ Math.\ Phys. }
\def\MPLA{{\em Mod.\ Phys.\ Lett.} A}
\def\IJMPA{{\em Int.\ J.\ Mod.\ Phys.} A}
\def\IJMPB{{\em Int.\ J.\ Mod.\ Phys.} B}
\def\EPJC{{\em Eur.\ Phys.\ J.} C}
\def\PR{{\em Phys.\ Rev.} }
\def\JHEP{{\em JHEP} }
\def\JCAP{{\em JCAP} }
\def\cmp{{\em Com.\ Math.\ Phys.}}
\def\JPA{{\em J.\  Phys.} A}
\def\JPG{{\em J.\  Phys.} G}
\def\NJP{{\em New.\ J.\  Phys.} }
\def\CQG{\em Class.\ Quant.\ Grav. }
\def\ATMP{{\em Adv.\ Theoret.\ Math.\ Phys.} }
\def\ibid{{\em ibid.} }
\def\ChP{{\em Chin.Phys.}C}


\renewenvironment{thebibliography}[1]
         {\begin{list}{[$\,$\arabic{enumi}$\,$]}  
         {\usecounter{enumi}\setlength{\parsep}{0pt}
          \setlength{\itemsep}{0pt}  \renewcommand{\baselinestretch}{1.2}
          \settowidth
         {\labelwidth}{#1 ~ ~}\sloppy}}{\end{list}}

\leftline{\Large \bf References}



\end{document}